\documentclass[twocolumn]{aa}

\usepackage{natbib}
\usepackage{graphicx}
\usepackage{txfonts}
\usepackage{hyperref}

\begin{document} 
\title{The nature of the companion in the Wolf-Rayet system EZ Canis Majoris}
\titlerunning{The low-mass companion of EZ CMa}
  \author{G. Koenigsberger\inst{1}  
     \and
          W. Schmutz\inst{2} 
          }
  \institute{Instituto de Ciencias F\'isicas, Universidad Nacional Aut\'onoma de M\'exico, Ave. Universidad S/N, Cuernavaca, 62210 Morelos, M\'exico\\
             \email{gloria@icf.unam.mx, gloria@astro.unam.mx}
        \and
            Physikalisch-Meteorologisches Observatorium Davos and World Radiation Center,
              Dorfstrasse 33, CH-7260 Davos Dorf, Switzerland\\
              \email{werner.schmutz@pmodwrc.ch}
             }

  \date{Received ; accepted }

\abstract{
{\it Context:} EZ Canis Majoris is a classical Wolf-Rayet star whose binary nature has been debated for decades. It was recently modeled as an  eccentric binary with a periodic brightening at periastron of the emission originating in a shock heated zone near the companion.\\ 
{\it Aims: } The focus of this paper is to further test the binary model and to constrain the nature of the unseen close companion by searching for emission arising in the shock-heated region.\\
{\it Methods:} 
We analyze over 400 high resolution  the International Ultraviolet Explorer spectra obtained between 1983 and 1995 and {\em XMM-Newton} observations obtained in 2010. The light curve and radial velocity (RV) variations were fit with the eccentric binary model and the orbital elements  were constrained.\\
{\it Results:} We find  RV variations in the primary emission lines with a semi-amplitude K$_1\sim$30 km\,s$^{-1}$ in 1992 and 1995, and  a second set of emissions with an anti-phase RV curve with K$_2\sim$150 km\,s$^{-1}$.  The simultaneous model fit to the RVs and the light curve yields the orbital elements for each epoch.  Adopting a Wolf-Rayet mass M$_1\sim$20~M$_\odot$ leads to M$_2\sim$3-5~M$_\odot$, which implies that the companion could be a late B-type star. The eccentric ($e$=0.1) binary model also explains the hard X-ray light curve obtained by {\em XMM-Newton} and the fit to these data indicates that the duration of maximum is shorter than the typical exposure times.  \\
{\it Conclusions:} The anti-phase RV variations of two emission components and the simultaneous fit to the RVs and the light curve are concrete evidence in favor of the binary nature of EZ Canis Majoris. The assumption that the emission from the shock-heated region closely traces the orbit of the companion is less certain, although it is feasible because the companion is significantly heated by the WR radiation field and impacted by the WR wind.    \\
}

\maketitle
%
\section{Introduction}

EZ Canis Majoris (EZ Cma, HD\,50896,  WR6 \citet{vanderHucht2001}) is one of the brightest  Wolf-Rayet (WR) stars that is accessible from the Northern Hemisphere. It is a classical WR star, possessing a strong and fast wind in which the abundances of He and N are enriched compared to typical main sequence stars in the Galaxy.  The high ionization stage of the emission lines that are present in its spectrum point to a high effective temperature, and lead to its classification as a nitrogen-rich, high ionization WR star, WN4. It is one of over 400 WR stars known in our Galaxy,\footnote{http://pacrowther.staff.shef.ac.uk/WRcat/index.php} objects that are considered to be in a short-lived, late evolutionary state, soon to end their life in a supernova event. 

Gaia observations indicate that EZ CMa  is located at a distance of  d$\sim$2.27\, kpc and they confirm its large distance from the galactic plane, z$\sim$380\, pc \citep{2020MNRAS.tmp...34R}.  It is surrounded by an X-ray emitting HII region, with abundances indicative of material ejected by the WR wind  \citep{2012ApJ...755...77T}.


For close to 80 years, the possible binary nature of EZ CMa has been a topic of debate, with early observations suggesting periodicities in the range of 1-13~d \citep{1948PASP...60..383W,  Ross1961, 1967PASP...79...57K,1975A&A....44..219L,1974PASP...86..767S}.  A stable period $\sim$3.7\,d was finally shown to exist \citep{1978MmSAI..49..453F, Firmani_etal1980}, and subsequently confirmed by numerous investigations as summarized in \citet{1992ApJ...397..277R}, \citet{1996A&AS..119...37D}, and \citet{Georgiev_etal1999}. The unseen companion was proposed to be a low-mass object, possibly a neutron star, as evolutionary scenarios call for the existence of such systems \citep{1976IAUS...73...35V}.  However, it was soon found that, although  the 3.7\,d period is always present in the data, the phase-dependent variability is not coherent over timescales longer than a couple of weeks \citep{1989ApJ...343..426D, 1992ApJ...397..277R}. This conclusion was contested by \citet{Georgiev_etal1999}, who found that the variations in the \ion{N}{V}\,$\lambda\lambda 4603$, $4621$ P~Cygni absorptions appeared to show coherent variability over a 15 year timeframe. Thus, the debate has continued.  Recently, \citet[][henceforth SK19]{SK19} successfully fit a binary model to the high-precision photometric observations obtained by the BRITE-Constellation satellite system \citep{Moffat2018}.  The basic conclusion of \citet{SK19} is that the lack of coherence over long timescales can be attributed  to a rapidly precessing eccentric orbit.

In the model presented by SK19, the wind of the WR star collides upon a  nondegenerate, low-mass companion forming a hot and bright shock region which gives rise to excess emission.  The presence of such a companion was previously suggested by  \citet{Skinner_etal2002} based on the X-ray spectral energy distribution and luminosity. The intensity of the excess emission  depends on the orbital separation, being brightest at periastron which is when the companion enters deeper into the WR wind.   Thus, times at which overall brightness increases are associated with periastron passage.  At times of conjunction, a portion of the bright  shock-heated region is occulted by the star that lies closer to us.  This leads to eclipses, as observed in the BRITE data, and gives times of conjunction. Our model was inspired by the results obtained for the WR+O binary system $\gamma^2$ Velorum  where both the emission from the colliding wind and its eclipses were identified \citep{Lamberts_etal2017, Richardson_etal2017}.  In the case of EZ CMa, however, instead of two colliding winds we assume only one wind which produces the shock near the companion's surface.

In this paper we apply the same analysis as in SK19 to X-ray and ultraviolet (UV) data of HD\,50896 with the aim of further testing the model and constraining the nature of the unseen companion.  In Section 2 we describe the observational material and the measurements. In Section 3 we summarize the method of analysis which is then applied to the X-ray light curve obtained by {\em XMM-Newton}.  In Section 4 we analyze four sets of {\it International Ultraviolet Explorer (IUE)} observations to constrain the radial velocity curves of the 3.7\,d orbit.  The results are discussed in Section 5 and the conclusions summarized in Section 6. Supporting material is provided in the appendix, which is available in the electronic version of the paper.

\section{Observational material}

The observational material has been in its majority reported previously and is summarized in Table \ref{table_observations}.  It consists of observations obtained with {\em XMM-Newton} in 2010 and with the {\it International Ultraviolet Explorer (IUE)} in 1983, 1988, 1992, and 1995. 

The {\em XMM-Newton} data were obtained with EPIC and are fully described in \citet{Oskinova_etal2012}. The {\it IUE} spectra were retrieved from the {\it INES} data base\footnote{Available at: sdc.cab.inta-csic.es\/cgi-ines\/IUEdbsMY}.  The Short Wavelength Prime (SWP) spectra cover the wavelength range of $\sim$1180 - 1970 \AA\ and were obtained with the high resolution grating  with exposure times in the range of 200 to 240 seconds. These spectra are described in  \citet{Willis_etal1989}, \citet{1993A&A...267..447S}, \citet{1995ApJ...452L..57S} and \citet{1995ApJ...452L..53M}.
All IUE spectra are flux-calibrated with the INES IUE data processing pipeline.  No correction  for reddening has been applied.

The dominant observational effect caused by the collision of the WR wind with the companion is the production of line emission \citep{{Lamberts_etal2017}, {Richardson_etal2017}}.  Thus, for the analysis of IUE spectra, we chose spectral bands centered on some of the principal emission and absorption features and computed the total flux contained within them as follows:

\begin{equation}
F_{band}=\int_{\lambda_1}^{\lambda_2}F_\lambda d\lambda\;,
\end{equation}

\noindent where $\lambda_1$, $\lambda_2$ are the initial and final wavelengths of the band, $F_\lambda$ is the flux given in ergs cm$^{-2}$ s$^{-1}$ \AA$^{-1}$, and d$\lambda$ is the wavelength increment which for the SWP data is 0.05 \AA.  The advantage of using $F_\lambda$ (as opposed to $F_\lambda-F_c$, with $F_c$ a continuum level) lies in that it avoids the challenges inherent to determining a continuum level which in UV spectra is very uncertain due to the density of  emission lines.

For the velocity measurements, we obtained the flux-weighted average wavelength, defined as:

\begin{equation}
\lambda_c=\frac{\int_{\lambda_1} ^{\lambda_2} \lambda~F_\lambda~d\lambda}{\int_{\lambda_1}^{\lambda_2}F_\lambda~d\lambda}
\end{equation}

\noindent with $\lambda_1<  \lambda <  \lambda_2$.
The choice of this method is based on the fact that  the time-varying asymmetry of emission lines leads to different velocity values depending on the method that is used (gaussian/voigt/lorentz fit; or  bisector of flux average or bisector of the extension of the line wings), and on whether the entire line or only a portion is measured. Hence, it provides a uniform and automated method to measure the large number of IUE spectra.  The disadvantage of the method is that when the variable component is weak and superposed on a stronger, less variable emission, the RVs of the weaker feature are diluted.   Thus, in a second pass, the variable superposed features were measured interactively with  Gaussian fits.

The wavelength bands that were tested for flux variability are listed in Table~\ref{table_UVwaveBands}. They were chosen to cover portions of emission and absorption lines.  In addition, we chose a spectral band (F$_c$1840) which is relatively free of emission lines in order to characterize the continuum variability.

\begin{table}[htb]
\caption{Summary of data sets. \label{table_observations}}
\begin{tabular}{llllrl}
\hline
\hline
Set  & year   & JD start  &   JD end  &    Num     & Ref \\
\hline
IUE  &  1983  & 45580.974   & 45587.890 &    28     &  1 \\  
IUE  &  1988  & 47502.556   & 47507.685 &   130     &  2    \\  
IUE  &  1992  & 48643.473   & 48648.598 &   148     &  3        \\  
IUE  &  1995  & 49730.917   & 49746.574 &   146     &  3,4  \\   
{\em XMM-Newton}     &  2010 & 55480.62     & 55507.35 & ....  &  5,6     \\  
\hline
\hline
\end{tabular}
\tablebib{(1) \citet{Willis_etal1989}; 
(2) \citet{1993A&A...267..447S}; 
(3) \citet{St-Louis_etal1995}; 
(4) \citet{Massa_etal1995};
(5) \citet{Oskinova_etal2012}; 
(6) \citet{Ignace_etal2013}.
}      
\end{table}

\begin{table}[htb]
\caption{UV wavelength bands. \label{table_UVwaveBands}}          
\begin{tabular}{lllll}
\hline
\hline
Name& $\lambda_0$ &$\lambda_1$ & $\lambda_2$&Ion \\  
\hline
SVe      &1197.34& 1194.5 & 1199.6& \ion{S}{V} \\   
1226e    &1226.01& 1217.6 & 1231.6& \ion{Fe}{VI}$^b$ \\ 
NVe      &1242.82& 1237.0 & 1248.5&  \ion{N}{V} \\  
1258e    &1258.30& 1257.2 & 1259.6&  \ion{Fe}{VI}\\ 
1270e    &1269.73& 1267.6 & 1272.3& \ion{Fe}{VI}$^d$ \\  
1270n    &1269.10& 1267.6 & 1271.1&\ion{Fe}{VI}$^e$\\ 
1371e    &1371.30& 1369.0 & 1381.0& \ion{Fe}{V}$^a$\\ 
NIV]     &1486.50& 1478.0 & 1497.0& \ion{N}{IV}] \\   
CIVe     &1550.85& 1545.4 & 1560.5& \ion{C}{IV}  \\ 
HeIIa    &1640.42& 1630.5 & 1633.5& \ion{He}{II} \\  
HeIIe    &1640.42& 1634.2 &  1651.8&\ion{He}{II} \\ 
NIVa     &1718.55& 1709.0 & 1713.0& \ion{N}{IV}   \\ 
NIVe     &1718.55& 1711.5 & 1736.0& \ion{N}{IV} \\  
NIVn     &1718.55& 1714.0 & 1728.0& \ion{N}{IV}  \\  
FeVIpsudo &1270.00& 1250.0& 1290.0&\ion{Fe}{VI}$^f$ \\ 
FeVpsudo&1450.00& 1430.0 & 1470.0&\ion{Fe}{V}$^g$ \\ 
F$_c$1840   & ---  & 1828.3 & 1845.5&  cont    \\   
\hline
\hline
\end{tabular}
\tablefoot{
The abbreviations used  in Column 1 consists of the name of the ion followed by a letter that indicates: 
"{\it a}" P Cyg absorption component; "{\it e}" emission component; "{\it n}";
"{\it psudo}" densely packed lines forming a pseudo continuum. The reference wavelength (in \AA) refers either to the laboratory wavelength of the principal ion listed in column 3, or the center of a blend, or the midpoint of  continuum bands.\\
\tablefoottext{a}{The combined contribution of Fe V lines dominates over that of O~V;}
\tablefoottext{b}{FeVI 1222.82, 1223.97;}        
\tablefoottext{d}{FeV 1269.78, NIV 1270.27, 1272.16, FeVI 1271.10, 1272.07;}
\tablefoottext{f}{FeV1 pseudo-continuum composed of densely packed Fe VI lines}
\tablefoottext{g}{FeV pseudo-continuum composed of densely packed Fe V lines.}
}
\end{table}

\section{The binary model and the X-ray light curve}

\subsection{Description of the model}

The binary star model that was introduced in SK19 is based on the following premises: 1) the WR star wind collides with its unseen companion and the resulting shock region emits radiation in wavelengths ranging from X-rays to radio; 2) the intensity of the emitted radiation is proportional to the wind density, (which decreases with orbital separation) so in the eccentric orbit, a larger flux is emitted at periastron than at apastron; 3) the companion is a low mass object that does not possess a stellar wind; 4) since the shock emission is constrained to a relatively small volume and this volume lies very close to the secondary star, portions of the enhanced emitting region can be eclipsed not only by the WR star but also by the low-mass companion. The eclipses are treated as Gaussian-shaped dips in the light curve, with the duration and the strength of the eclipse as free parameters. 

In more detail,  we compute the intensity variations as a function of time of a localized region that orbits the WR star under the assumption that the variation is caused by an increase or decrease of the energy density. Neglecting the variation in wind velocity, this reduces to variations in the wind particle density, $n_e$.  The process responsible for the localized nature of the emission is the collision of the WR wind on the unseen companion.  Because of the eccentric orbit, the value of $n_e$ varies over the orbital cycle and therefore so does the emission intensity arising in the shock,  which is assumed to follow a d$^{-2}$ relation, with d the orbital separation.\footnote{It has been pointed out by an anonymous referee that Eq.\,10 in \citet{Stevens_etal1992} implies that the distance dependence of the X-ray luminosity should be linear. The assumption of the cited equation is that the geometry of the colliding winds is scale-free, so the volume of shock-heated gas scales with the distance to the power of 3. We note that in the case of EZ CMa the expected scenario is not that of a wind-wind collision but that of a wind colliding with a main sequence star. Thus, we anticipate that the hydrodynamic simulations of \citet{Ruffert1994} for wind accretions are more adequate to guide the expectations. Model FL of \citet{Ruffert1994} might represent the best match, in which the size of the shock region is given by the size of the star. The input energy to the shocked material is then proportional to the density of the impacting wind, which scales with $d^{-2}$.}  

This computation yields a light curve having a maximum at the time of periastron.  However, the orbital precession causes the times of periastron to advance (or recede) compared to a constant anomalistic period, P$_A$.  Our binary model performs a fit to the observations by iterating over P$_A$  and $\dot{\omega}$ (the rate of change of the argument of periastron). The possibility of eclipses is allowed by what we refer to as "attenuation". This means that for each ($P_A$, $\dot{\omega}$), the times of conjunction are computed and the light curve is modified to take into account the possible occultation or attenuation of a portion of the emitting region. Furthermore, the model simultaneously computes the radial velocity of the localized emitting region and of the WR star. Thus, for the fit to be acceptable, the times of conjunction (determined from eclipses) must coincide with times of observed radial velocity curve zero-crossing.  

It is important to note that we do not compute a spectrum. Instead we modulate a mean state which we set to unity, which, for example, for the X-ray spectrum, could be the two-temperature  optically thin plasma model derived by \citet{Skinner_etal2002}. The model yields the expected time-dependent modulation of the emitted intensity which is then compared with the observed variability. The observed data that are used in this process are normalized to their mean value. Thus, the variations are described in terms of fractional departure from the average values.  The goodness-of-fit is evaluated through  summing up the differences between observations and fit curve squared.  The fitting procedure yields the times of periastron passage which, in turn,  provide the anomalistic period (P$_A$) and the argument of periastron $\omega_{per}$.  When simultaneous radial velocity data are available, a theoretical RV curve is constructed and fit together with the light curve fit. This yields the sidereal period (P$_S$). 

Each observation epoch must be fit independently because the timing of the eclipses varies due to the apsidal motion, and there is an additional change of the eclipse timing, which is revealed as an overall quadratic change in the O-C diagram (see Figure 3 of SK19). SK19 interpreted this as a change of the rate of apsidal motion. However, in view of the results reported below, it is more likely that it is the signature of a precession of the system's orbital plane, resulting in a sinusoidal drift of the eclipse timing and in a change of the apparent inclination. The combined effect is that a prediction of the eclipse timing from a linear ephemeris as, for example, by \citet{Antokhin_etal1994}, may deviate by more than 0.3d  (0.1 in phase) after only $\sim$10 to 100\,d, preventing the calculation of a more accurate period by combining two observing epochs, which would be essential for predicting accurately a phase over time intervals of years. Presently, the average sidereal period is not know more accurately than to three digits <$P_S$> $= 3.76 \pm 0.01$\,d, which leads to ephemeris predictions that are off by more than a tenth of the phase after already four months. In a year the uncertainty is on the order of a day and thus, combining observations over a time span more than one year there is the uncertainty that the number count of orbits between two epochs is lost.

\subsection{Model fit to the XMM-Newton observations}

Modeling of the hard X-ray light curve is the ultimate test for the interpretation of the variations in WR6 since this high energy emission must arise in the shock zone and must be attenuated or eclipsed, at least by the WR wind when the collision zone is on the far side of the WR. The orbital eccentricity that was found by SK19 ($e$=0.1) should lead to a $\sim$20\% increase in the hard X-ray flux around the time of periastron, a prediction that is tested in this section.

It is important to keep in mind that in the case of a collision zone that is embedded in an optically thick wind, its emission may not be visible at all orbital phases. In this case, instead of detecting the enhanced periastron emission, one may instead detect a maximum at the time when the optical depth to the observer is smallest; that is, when the companion and its shock region lie between the WR  and the observer. This effect is particularly important for soft X-rays because of the intrinsically higher opacity of the WR wind to softer X-rays. For this reason,  we aim at reproducing only the hard X-rays as this energy range is least affected by absorption.
Figure 11 of \citet{Huenemoerder_etal2015} shows that for photon energies above 3\,keV (below 4\,\AA) the transmission function approaches unity. As the hardest band (band 4) comprises energies between 2.7 and 7\,keV we conclude that the optical depth effects are relatively small in this band. This is confirmed in that band 3, which comprises energies between 1.7 and 2.7\,keV, behaves similarly to band 4.

Early X-ray observations were reported at L$_X$=3.8$\times$10$^{32}$ erg\,s$^{-1}$ in the 0.2-2.5 keV band \citep{1994Ap&SS.221..321W}.  \citet{Skinner_etal2002} reported a similar X-ray luminosity, but also found the presence of harder X-ray emission which they concluded could only originate in a wind collision zone.   \citet{Oskinova_etal2012} reported L$_X\simeq$8$\times$10$^{32}$ ergs\,s$^{-1}$ in the 0.3-12kev band. Fits to the spectrum led these authors to conclude that the X-rays originate very far out in the stellar wind.  Subsequently, \citet{Ignace_etal2013} presented a re-analysis of the same data which led them to conclude that the X-rays are generated in a stationary shock structure, although they favored the corotating interaction region (CIR) scenario over that of a binary colliding wind shock.

\begin{figure}
\centering
\includegraphics[width=0.95\linewidth]{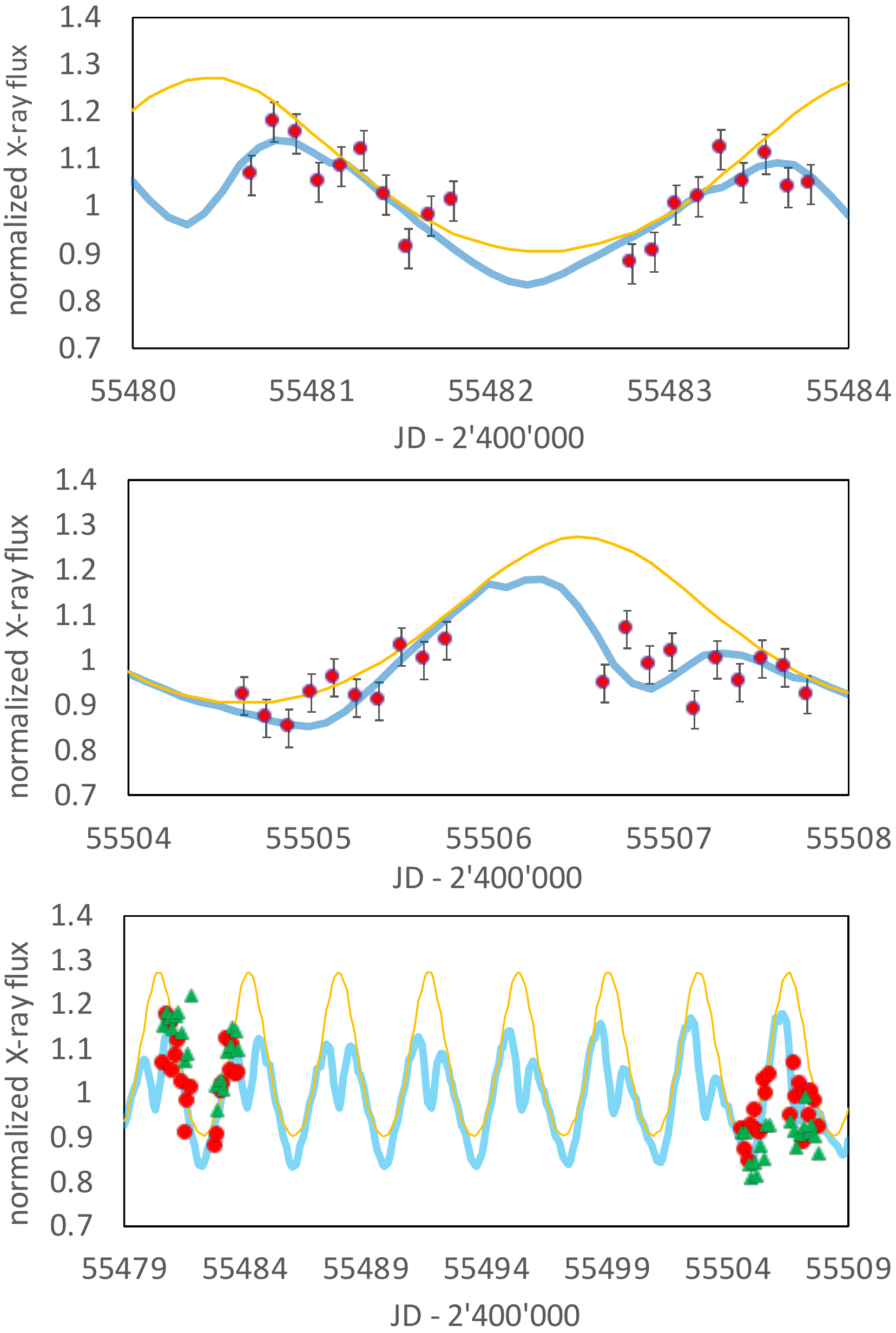}
\caption{
Fit to the hard X-ray (band 4) EPIC observations, which are indicated by the red dots (with error bars only in the top panel).  The uneclipsed photometric light curve from a shocked zone, which is assumed to vary proportionally to the square of the orbital separation, is shown by the orange curve. The modeled attenuated fit including eclipses of the shock zone is shown by the blue curve. Observed band 3 flux (green triangles) is included in the complete view of the fit shown in the bottom panel.
}
\label{fits_XMM}%
\end{figure}

We use the {\em XMM-Newton} EPIC observations that were obtained with four exposures of $\sim$30 hours duration each, on 2010 October 11, 13 and November 4, 6.  Both pairs of observations, seperated by $\sim$1 day, together cover nearly one 3.7\,d period (Oskinova et al. 2012; Ignace et al. 2013).\footnote{The reduced data were kindly provided by Richard Ignace.} Four energy bands were defined in Ignace et al. (2013) as follows: 0.3-0.6 keV (band 1), 0.6-1.7 keV (band 2), 1.7-2.7 keV (band 3)  and 2.7-7.0 keV (band 4). Fig.~\ref{fits_XMM} shows the fit to band 4,  the hardest X-ray band, in which the variation amplitude is largest and presumably, the wind absorption is lowest.

The relative amplitude of band 4 is on the order of $\pm 20$\,\%. The analyses by SK19 yielded for the eccentricity of the orbit $e=0.1$. If we adopt that the flux from the shock zone is varying with the square of the distance between the two objects, we conclude from the observed relative variations that basically 100\,\% of the hard X-ray emission is formed in the central shock. The variation of band 3 is similar to that of band 4, whereas the variations in the softer bands 1 and 2 are much smaller, as expected, since energy in these bands is the sum of freshly heated material close to the central shock plus previously shocked material traveling down stream and emission arising further out in the wind most likely due to global wind instabilities. 

\citet{Huenemoerder_etal2015} analyzed the light curve obtained with the CHANDRA X-ray Observatory during three pointings in 2013. The total count rates varied  by only 5\,\%, which is somewhat smaller but comparable to the variation of the total counts observed by XMM-Newton in 2010 which had a variation of 8\,\%.  Their finding that the variations in the full CHANDRA bandwidth did not correlate with the simultaneous  optical variations  is  consistent  with  the  notion that the soft X-rays are produced primarily in global wind structures, overwhelming  the  variations  occurring in the localized region near the companion. A comparison of the optical variations with the  hard X-ray variations might presumably have yielded a better correlation. 

The fit in Fig.~\ref{fits_XMM} shows two eclipses, one around the minimum of the light curve and a second near the maximum.  The first of these has a duration close to half of the orbit and we attribute it to attenuation by the WR wind of the central collision zone. The effect of the wind optical depth, which is expected to be small for the hard X-ray band, is modeled by a Gaussian curve centered on the conjunction when the WR star is in front. Possible additional absorption by a trailing interaction zone is not taken into account.mThe strength of the eclipse is not well constrained by the observations. In Fig. \ref{fits_XMM}, we have used a relatively moderate absorption of 6.4\,\%, a value which we adapted from the absorption strength calculated from a two-component spectral fit by \citet[][Table 3]{Skinner_etal2002}.  However, a range from almost zero absorption to absorption several factors stronger could equally well be fit because this long duration attenuation basically introduces only an  offset between the uneclipsed  and the eclipsed curves that is nearly constant.

We associate the second eclipse (the one which is superposed on the light curve maximum) with partial occultation by the companion of its shocked region.  This eclipse is computed in the fit, with depth and duration as free parameters.  Although it is better defined by the data than the one near light minimum,  the observations 
between JD\,55506.5 and 55507.5 do not cover the full eclipse duration so the fit parameters are uncertain. This has as consequence that the orbital parameters given in Table~\ref{table_periods} (the anomalistic period between maxima, the sidereal period between eclipses, the epoch of maximum, and the epoch of conjunction) also have significant uncertainties.

The fit indicates that the duration of maximum emission is very short and in particular, shorter than the {\em XMM-Newton} exposure times. This implies that, because of the statistics needed to see a spectrum, there is always much less time (contribution) of the high hard X-ray state than of the low hard X-ray state (eclipsed). In addition the orbit solution shown in Fig. \ref{fits_XMM} reveals that no maximum is fully covered by observations and that the time of expected highest flux is within an eclipse.

\begin{figure}
\includegraphics[width=0.95\linewidth]{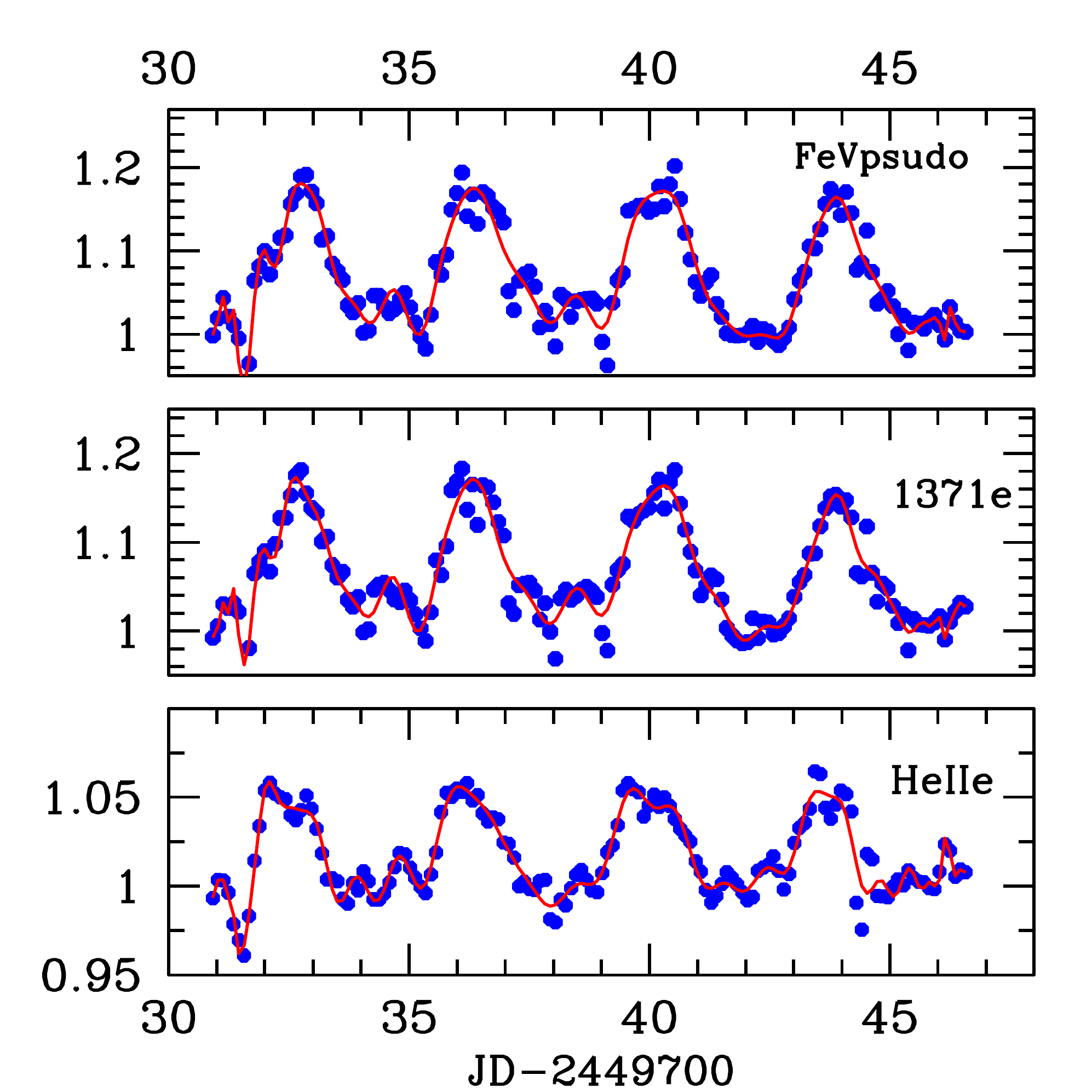}
\caption{UV flux (dots) as a function of JD-2449700 (epoch 1995) in three wavebands illustrating the periodic increase associated with periastron.  Fluxes are normalized to the average value of each band.  The curves show  a Chebyshev 50-order polynomial fit to the measured fluxes.
   }
\label{fluxes1995_days}%
\end{figure}

\begin{figure}
\centering
\includegraphics[width=0.95\linewidth]{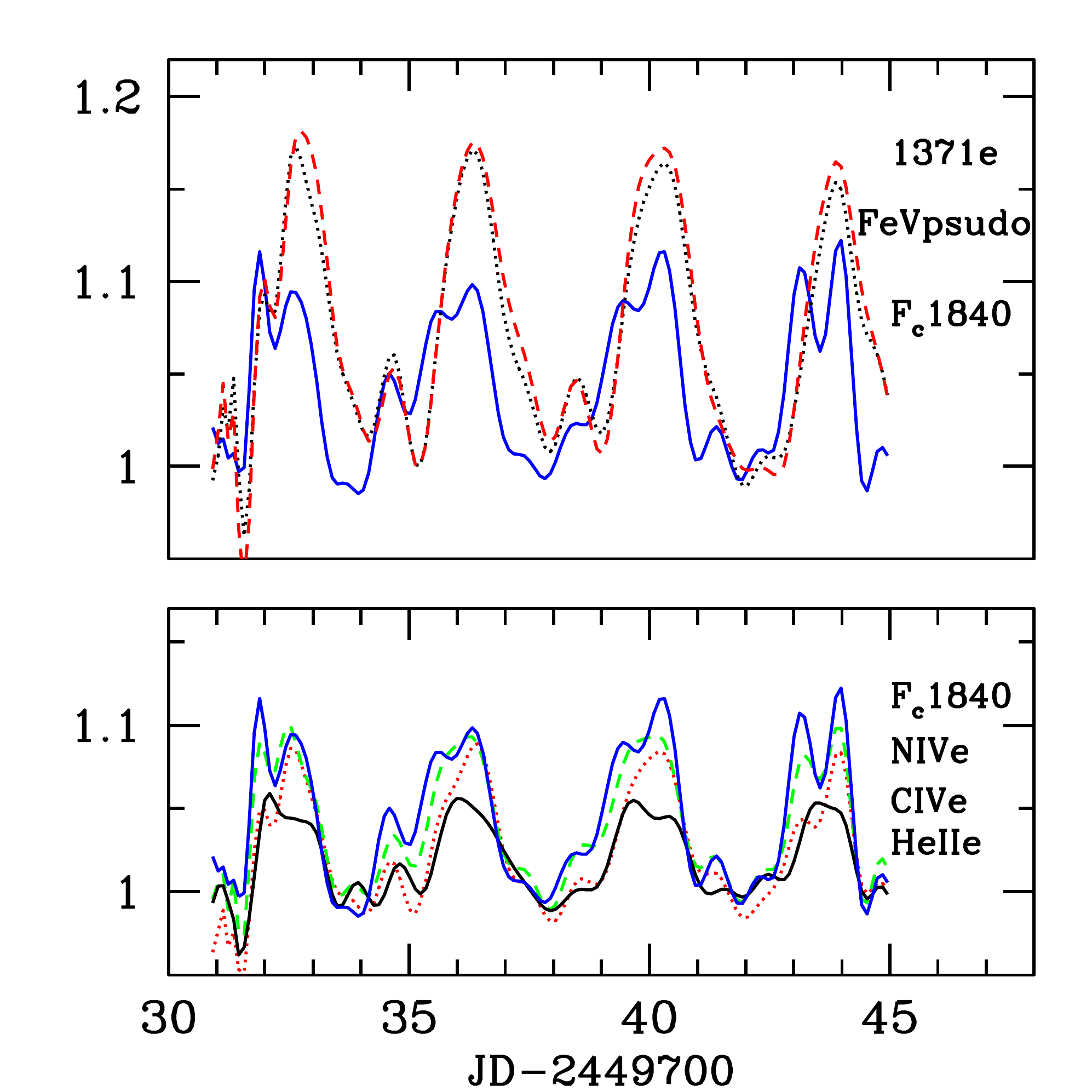}
\caption{Chebyshev 50-order polynomial fit to the UV fluxes showing the phase lag in ascending and descending intensity branches of the following bands:  {\bf Top}:  1371e (red), FeVpsudo (black) and F$_c$1840  (blue). {\bf Bottom}: F$_c$1840 continuum (blue), NIVe (green) CIVe (red) and HeIIe (black).  
   }
\label{cheby1995}%
\end{figure}

\begin{figure}
\includegraphics[width=0.95\linewidth]{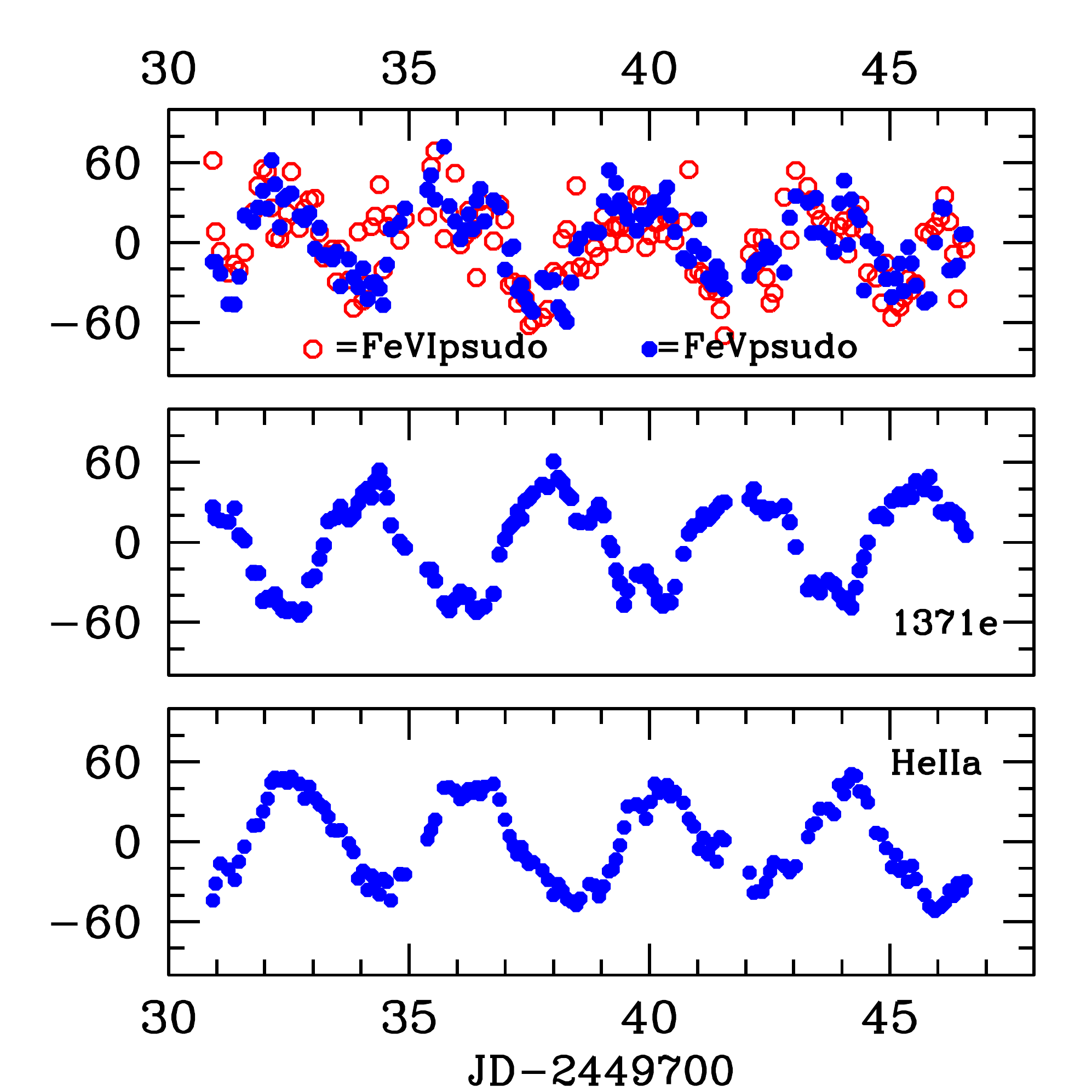}
\caption{Flux-weighted radial velocities obtained from Eq. (2), corrected for their corresponding average values, as a function of JD-2449700 of the following bands: {\bf Top:} FeVpsudo (filled circle) and FeVIpsudo (open circle); {\it Middle:}  1371e; {\bf Bottom:} HeIIa.
}
\label{vels1995_days}
\end{figure}

\begin{figure}
\centering
\includegraphics[width=0.75\linewidth]{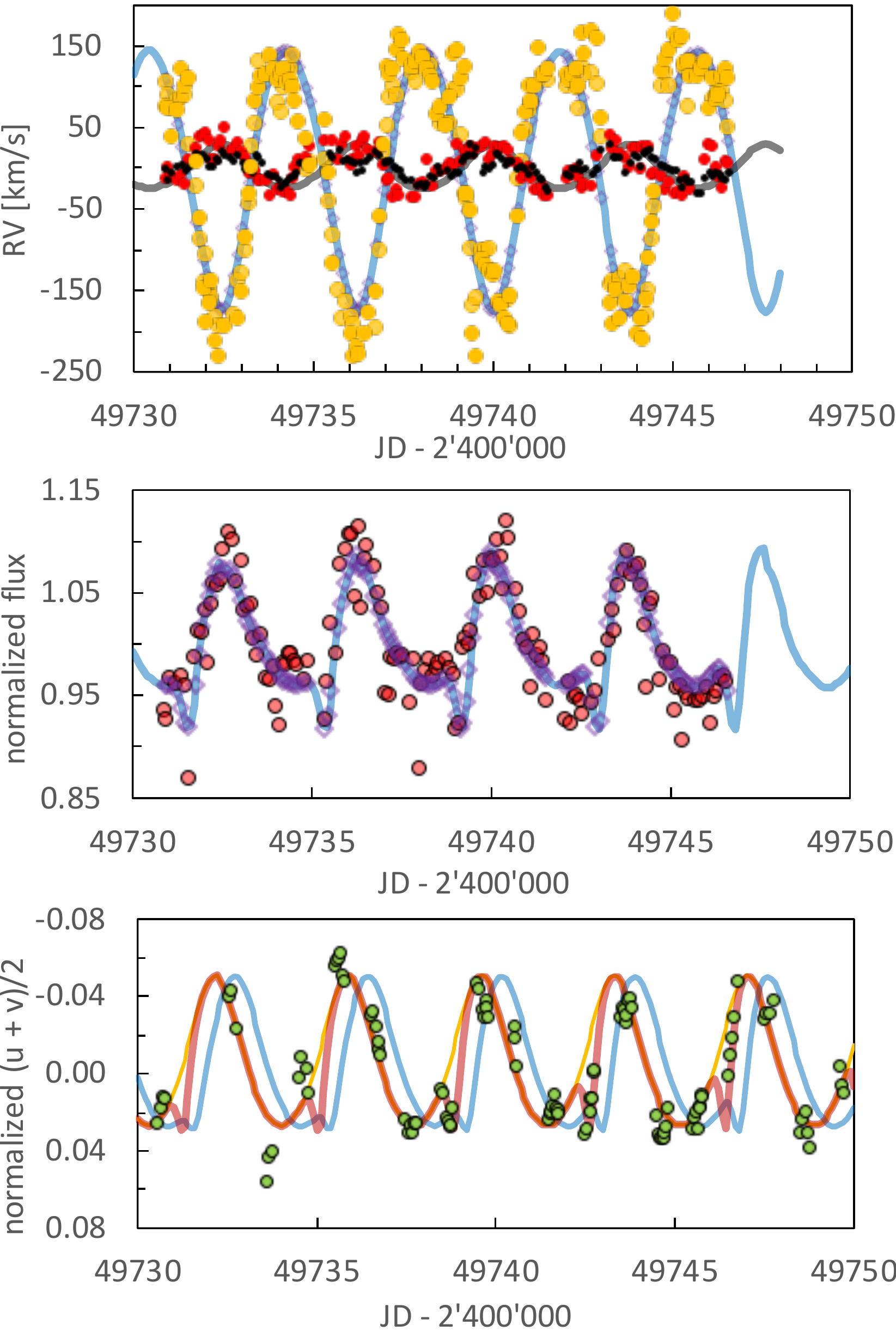}
\caption{Binary model fit to the 1995 IUE data.  {\bf Top:} Flux-weighted mean radial velocities of FeVpsudo and FeVIpsudo (red dots), flux-weighted radial velocities of 1270e (black dots) and Gaussian measured RVs of G1376 (orange dots). The gray and blue curves are the fits to, respectively, the WR star and the less massive object.  {\bf Middle:} Flux variations in the 1371e band (red dots) and the calculated flux variation (blue curve) that is obtained from the simultaneous fit to the RVs that are shown in the top panel.  {\bf Bottom:}  Fit to the 1995  photometric observations of \citet{Morel_etal1997} that were obtained contemporaneously with the IUE-MEGA campaign. The mean of the $u$ and $v$ magnitudes are indicated by {\bf green} dots. The blue curve indicates the photometric variations that were calculated with the parameters obtained by fitting the IUE data given in the two panels above.  The orange shows the results of  an independent fit to the photometric observations. 
  }
\label{fits_1995}%
\end{figure}

\small
\begin{table*}[htb]
\begin{center}
\caption{Times of maxima and dips in the IUE 1995 fluxes \label{times_maxima}}
\begin{tabular}{lllllllllllll}
\hline
\hline
 Band    &  \multicolumn{3}{c}{Cycle 1}&  \multicolumn{3}{c}{Cycle 2} &  \multicolumn{3}{c}{Cycle 3}&  \multicolumn{3}{c}{Cycle 4}\\
                   &   Max  &gfwhm&Dip   & Max   &gfwhm &Dip$^a$ & Max    &gfwhm&Dip  & Max  &gfwhm &Dip \\
\hline
F$_c$1840 & 32.34&1.36&31.56&  36.03&1.62&34.93:&  39.98&1.58 &38.96& 43.60&1.55& 43.54       \\
NIVe      & 32.41&1.19&31.53&  36.07&1.45& ...  &  39.90&1.48 &38.91&  43.64&1.34& 43.56   \\
NIVa      & 32.41&1.37&31.47&  36.10&1.35&35.24:&  39.94&1.39 &38.95&  43.73&1.45& 42.77        \\
HeIIe     & 32.41&1.46&31.55&  36.12&1.27&35.11:&  39.93&1.22 &38.92&  43.65&1.29& 44.37:      \\
CIVe  & 32.60&1.24&31.55&  36.29&1.23&35.30:&  40.09&1.33 &39.18&  43.73&1.26& 44.35       \\
1371e  & 32.74&1.05&31.59&  36.34&1.07&35.28:&  40.22&1.16 &39.07&  43.88&0.89& ....        \\
FeVpseudo& 32.78&1.14&31.57&  36.38&1.28&35.30:&  40.18&1.30 &39.09&  43.89&1.22& 42.69:      \\
HeIIa     & 32.98&1.29&31.52&  36.74&1.50&35.34:&  40.50&1.33 &39.2:&  44.38&1.53&           \\
\hline
\hline
\end{tabular}
\tablefoot{
Times are in units of JD-2449700.
Times of maxima,  Gaussian full width at half maximum (gfwhm) and dips were measured with Gaussian fits to the flux curves.
\tablefoottext{a}{The uncertain measurements are noted with a ``:'' and in Cycle 2, the uncertainty is because the times were interpolated over missing data.}
\tablefoottext{b}{Other sets of apparent dips are seen in NIVa as follows: 1) superposed on the maximum at days 32.28,36.20:,39.85, 43.71;  and 2) near minima  at days 34.10, 37.94:, 41.72, 45.54.}
}
\end{center}
\end{table*}

\begin{figure}
\includegraphics[width=0.45\linewidth]{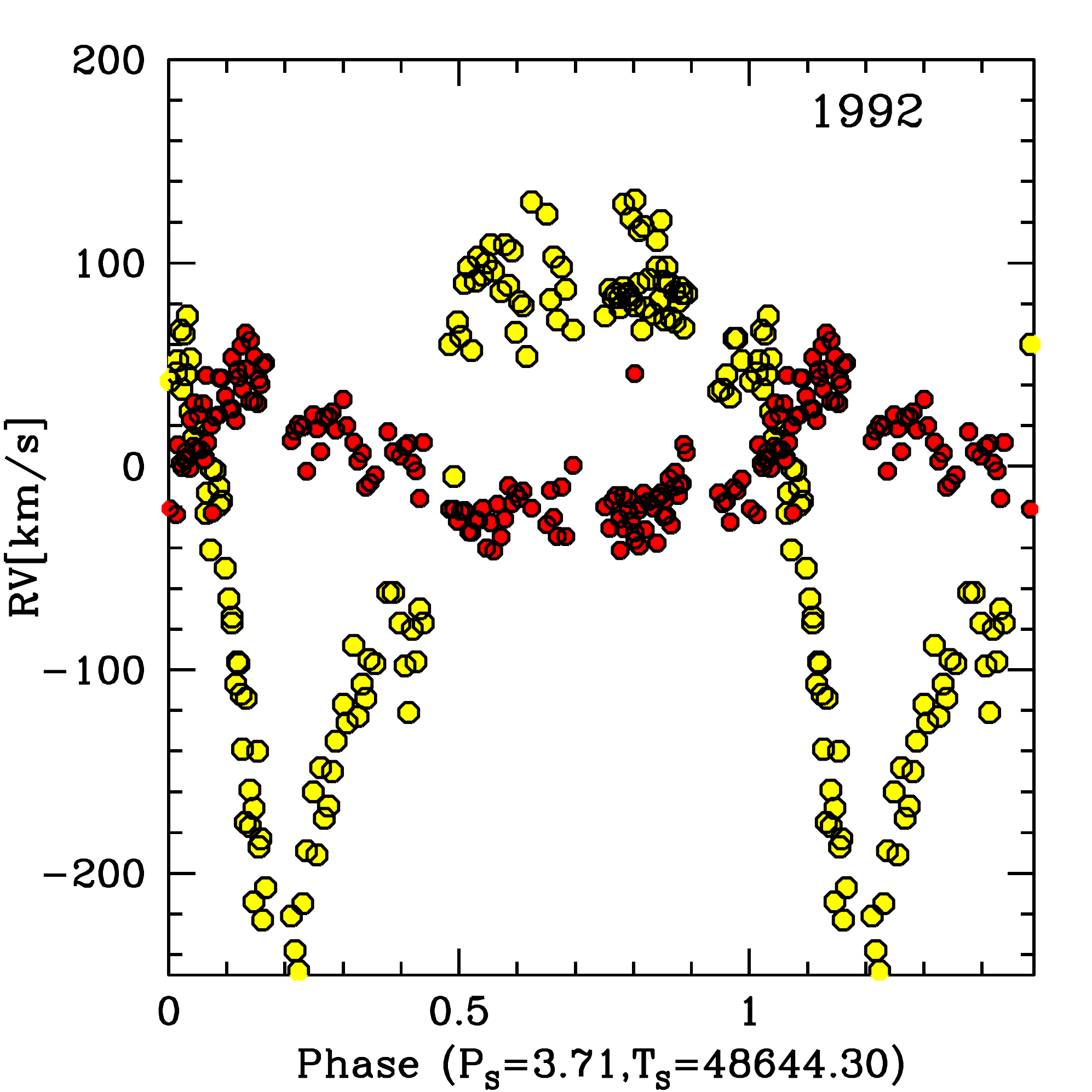}
\includegraphics[width=0.45\linewidth]{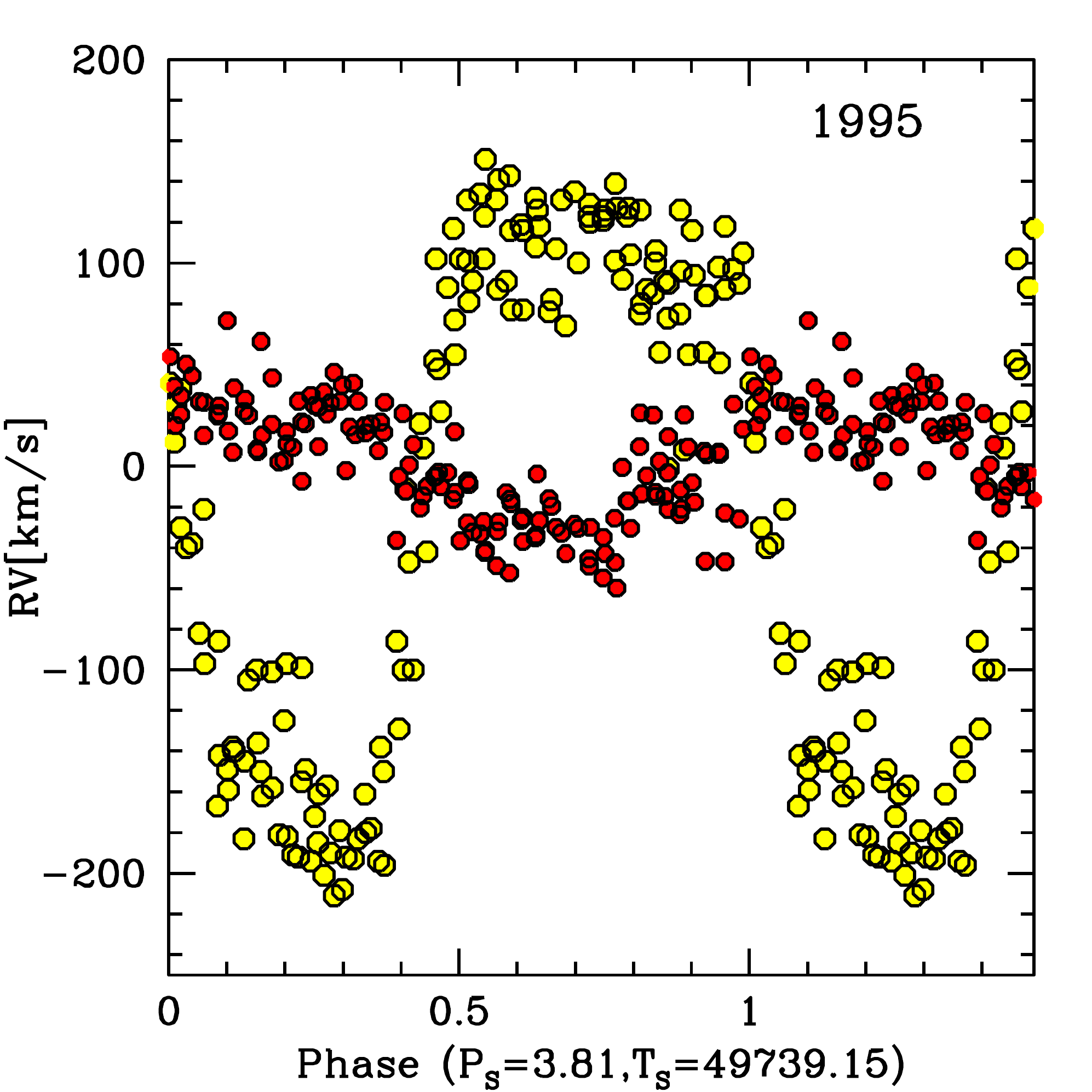}
\caption{Radial velocities as a function of sidereal phase of the G1376 emission peak measured with Gaussian fits (yellow circles) and the flux-weighted velocities of the FeVpsudo band (red circles).  Phase zero corresponds to the conjunction with the WR star in front.  {\bf Left:} Epoch 1992; {\bf Right:} Epoch 1995.
   }
\label{RVcurves2}%
\end{figure}

\section{The binary model applied to the IUE data}

The IUE observations that we analyze consist of SWP spectra obtained sequentially over at least one orbital cycle in the years 1983, 1988, 1992 and 1995.  The four data sets are comprised of 404 spectra.

\citet{Willis_etal1989} analyzed the 28 SWP spectra of 1983 and remarked on their very low level of variability compared to spectra that had been acquired in the timeframe 1978-1980. A slightly higher level of variability was reported by \citet{1993A&A...267..447S} in a 6-day run of observations in 1988, which suggested the presence of variations on a 1\,d timescale. The spectra obtained in 1992 cover $\sim$1.4 orbital cycles and were discussed  in \citet{St-Louis_etal1995} .

The 1995 data set was obtained over 16 consecutive days as part of  the IUE-MEGA campaign (Massa et al. 1995). The spectra display clear line and continuum variability on the 3.7\,d period, which were interpreted by \citet{St-Louis_etal1995} in terms of a global wind structure pattern which could equally well be explained with an ionized cavity around a neutron star companion or a corotating interaction region.  They identified two types of line profiles in the strong emission features corresponding to approximately opposite phases in the 3.7\,d period.  The first, which they termed the ``quiet'' state, is characterized by P Cygni profiles having an absorption edge indicative of a terminal wind speed $\sim$1900\,km~s$^{-1}$.  In the second, termed the ``high velocity'' state, there is enhanced absorption extending out as far as $\sim$2900\,km~s$^{-1}$ together with excess emission at lower speeds.  The 1992 spectra follow the same  pattern as those of 1995.  However, in the 1983 and 1988 data, the high velocity state is absent. 

Morel et al. (1997) revisited these IUE spectra in conjunction with a comprehensive analysis of optical spectra and photometry. They found: a) the presence of extra emission subpeaks that travel across the emission line profiles; b) a correlation between the strength of the emission lines and the stellar brightness; c) a 1 day recurrence time scale for variations; d) at maximum light there is enhanced \ion{He}{I} absorption at high velocities; e)  the \ion{N}{V}~$\lambda 4604$ P~Cyg absorption component disappears as the star brightens.

\subsection{Epoch 1995}

We address the 1995 data set first, since it covers four of the 3.7\,d cycles and thus allows a broader  perspective of the variability cycles.  Using Eq.~(1), we find periodic flux variations on the 5-20\,\% level in most\footnote{The exceptions are the weaker features such as 1197, 1197n, 1226, 1226n, 1258n and 1270n,  and  FeVIpsudo} of the emission bands listed in Table \ref{table_UVwaveBands} and by factors of up to 2.5 in the  P Cygni absorption bands.  The variability is illustrated in Figure~\ref{fluxes1995_days} where the flux in three of the bands are plotted over the four orbital cycles.  The times of maximum, obtained by fitting a Gaussian function to the maxima are listed in Table \ref{times_maxima}. In each cycle, the first to reach maximum are F$_c$1840,  NIVe, NIVa, and HeIIe, followed in progression by CIVe, 1371e, FeVpsudo, and  HeIIa.  The average separation in time between the first and last bands to reach maximum is 0.7\,d. The manner in which the flux of each band rises and declines is shown by the Chebyshev fits in Figure~\ref{cheby1995} whose Gaussian full width at half  maximum (gfwhm)  range from $\sim$1d to 1.6\,d; that is, $\leq$40\% of the 3.7\,d cycle.   We note that the \ion{N}{V} resonance doublet does not show flux variations on the 3.7\,d period.

Many of the flux-weighted radial velocities obtained with Eq.~(2) also display periodic variations. The semi-amplitude for the emission line RVs is $\sim$40 km/s, as illustrated in Figure~\ref{vels1995_days}, where the following bands are plotted:  FeVpsudo, FeVIpsudo,  HeIIa, and 1371e. The fourth of these is clearly in anti-phase with the first three, thus suggesting the existence of a double-line RV curve.  The \ion{N}{V} resonance doublet shows similar RV variations to those of 1371e, but the curve is considerably more noisy.

Comparing Figures~\ref{fluxes1995_days} and \ref{vels1995_days} one can see dips in the fluxes (for example, around day 38) which coincide with the times at which the FeVpsudo RV curve crosses from negative to positive velocities. The dips are assumed to appear when a portion of the shock region is occulted by either the WR or the secondary star;  that is, the dips are  associated with times of conjunction. 

According to the wind collision model, periastron is expected to coincide with maximum line flux. In addition, the  orbital motion of the unseen companion should be evident in the RVs of high ionization lines formed near the vortex of the shock. Thus, the binary model was applied to numerous combinations of flux and radial velocity curves.\footnote{Variable features are plotted in the Appendix.}  In  addition to fitting these curves, the constraint was imposed that the  model also fit the dips in the flux curves.  

The spectral features which yield the most consistent results under the imposed constraints are 1371e and the FeVIpsudo and FeVpsudo bands,\footnote{The flux in these features attains maximum at approximately the same time and their ionization potentials are similar.} and they provide a first set of orbital periods and initial epochs.  It must be noted, however that the flux-weighted RVs give only a lower limit to the variation amplitude.  This is because the bulk of the emission comes from the WR wind, which dilutes the flux-weighted RVs of the weaker shock emission.  Hence, we manually measured the top of the 1371e emission and a neighboring emission at 1362~\AA\ by fitting Gaussians. We refer to these features as G1376 and G1362, respectively. This was performed also for  the absorption feature that lies at $\lambda$1324.8~\AA, henceforth referred to as G1324, a relatively clean P Cygni absorption component.  

The Gauss-fit RVs of both emission lines yield nearly identical RV curves with a semiamplitude $\sim$150 km s$^{-1}$. Repeating the fitting procedure using the flux-weighted RVs of the FeV/FeVIpsudo bands (for the WR orbital motion) and the Gaussian-measured RVs of G1376 (for the companion) yields the values of the periods (anomalistic and sidereal) and corresponding initial epochs that are listed in  Table~\ref{table_periods}. The  fits are shown in Figure~\ref{fits_1995}.  The observed variability of the 1371e shown in this figure leads us to conclude that 33\,\% of its flux arises in the collision zone. The fit also shows that it is partially eclipse when the WR star is in front. The best fit  model parameters  also indicate that the flux is weakly attenuated  when the companion is in front.   The RV curves plotted as a function of sidereal phase in  Figure~\ref{RVcurves2} (right).

The photometric observations in the visual spectral range that were obtained contemporaneously with the 1995 IUE spectra \citep{Morel_etal1997} were also fit with the model and are shown in Figure~\ref{fits_1995}. The resulting ephemerides are consistent with those derived from the IUE flux and radial velocity curves.  However,there is a time-offset between maxima in the visual photometry and the 1371e.  This offset is such that visual maxima occur $\sim$0.5~d earlier than the 1371e maxima, which is in the same sense as  the offset between the $F_c$1840 continuum maximum and the 1371e band. One interpretation for this and the other time offsets is that different zones in the shock and in the neighboring outflowing wind respond differently to the changing wind density encountered around periastron. An additional factor involves the viewing angle to the portion of the companion star's surface which is irradiated by the WR star and predicted to be significantly hotter than the opposite hemisphere  (see Section 4.5) and is expected to contribute to continuum flux when in view. From the observed variability we conclude that 19\,\% of the optical photometric flux is associated with the collision zone.

\subsection{Epoch 1992}
The 1992 data set cover a little over one orbital cycle, but have a higher density of phase coverage that the 1995 data set. We applied  a similar procedure as for the 1995 epoch.  The fits are illustrated in Figure~\ref{fits_1992} and the RV curves as a function of sidereal phase are shown in Figure~\ref{RVcurves2} (left). 
 
Except for a sharper minimum in the 1992 companion's RV curve, the 1992 and 1995 RV curves are very similar.We explored the possibility of a higher eccentricity, $e$=0.4, in 1992 which improved the RV fit (see Fig.~\ref{fits_e0.4_1992}).  However, such a large value for $e$ is ruled out by the fits for the other epochs, which leads us to speculate that the sharp RV minimum may be caused by a wind effect.

\subsection{Epochs 1983 and 1988}

\begin{figure}
\includegraphics[width=0.48\linewidth]{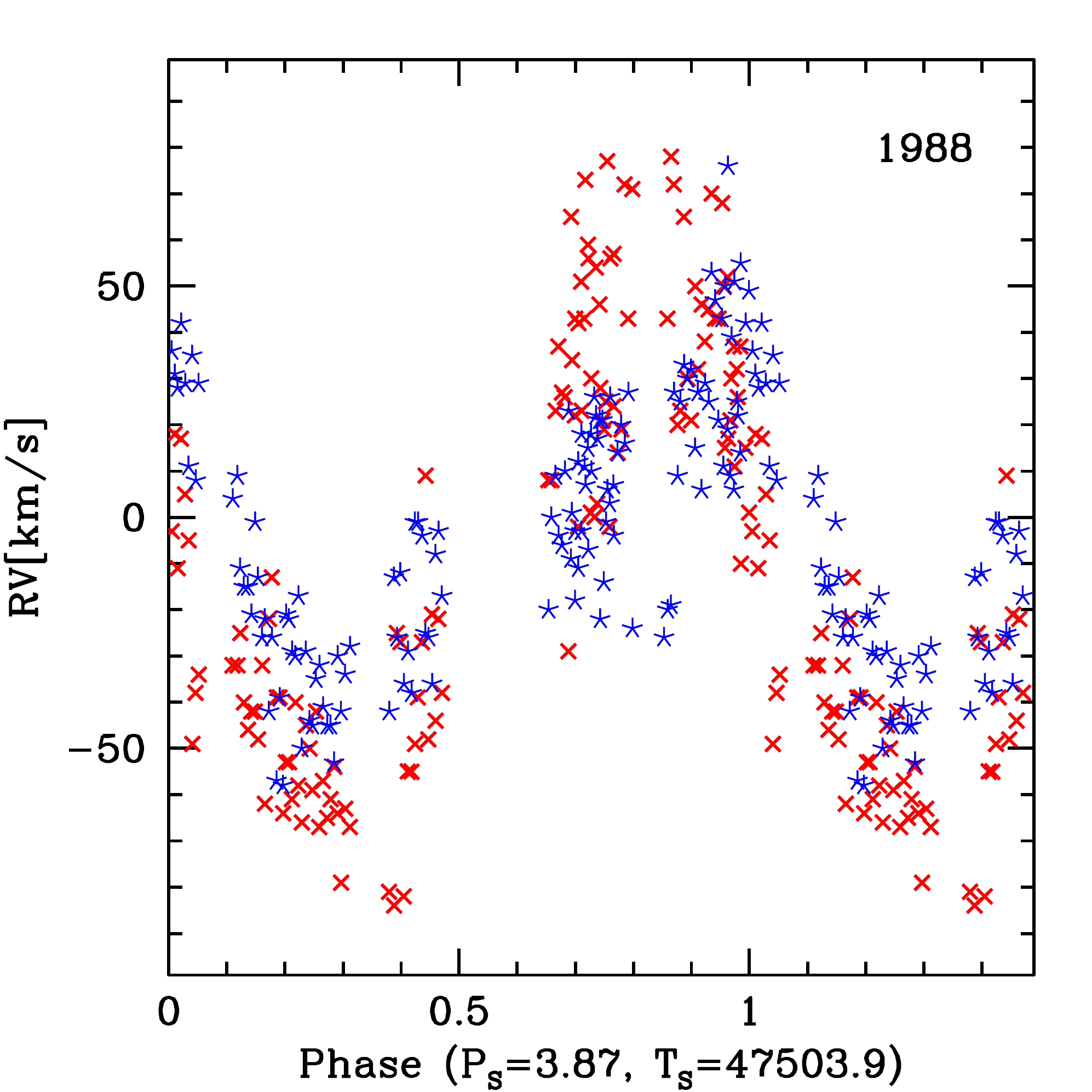}
\includegraphics[width=0.48\linewidth]{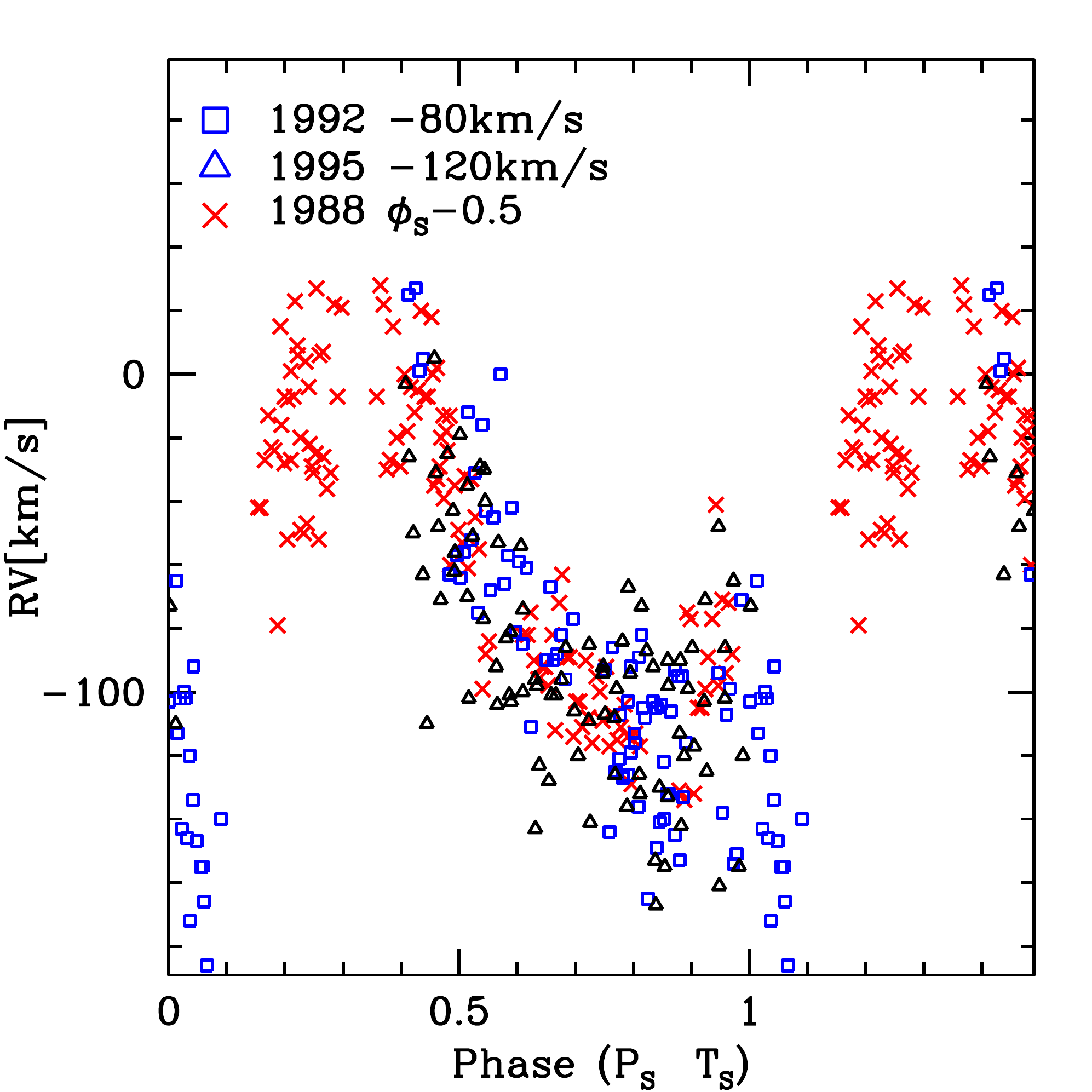}
\caption{{\bf Left:} Epoch 1988 Gaussian measured RVs of the G1376 emission (star) and G1324 absorption (cross) plotted as a function of sidereal orbital phase. RVs are corrected for their corresponding average values.  {\bf Right:} Epoch 1988 RVs of G1324 (cross) unshifted in velocity but shifted by 0.5 in sidereal phase, compared to the same measurements for 1992 (square, shifted by -80 km~s$^{-1}$) and 1995 (triangle; shifted by -120 km~s$^{-1}$).  The abscissa is the sidereal phase for 1992 and 1995, computed with the parameters listed in Table 4 for the corresponding epoch. 
   }
\label{RVcurves3}%
\end{figure}

As noted previously, the 1983 and 1988 spectra display much weaker variability than  present in 1992/1995. Here we focus on the 1988 data because the phase coverage in 1983 is very sparse compared to the other epochs and because the average 1983 spectrum is nearly indistinguishable from that of 1988.  The flux-weighted RVs display notable variations only in  bands  CIVe, HeIIe,  NIV], 1270e, 1270n, FeVpsudo and FeVIpsudo (see Figure \ref{vels_group1}), all with a very small ($<$40 km~s$^{-1}$) semi-amplitude.  Fits to the flux-weighted RVs of the two latter bands and the fluxes in 1270e yield the tentative ephemerides for this epoch listed in Table~\ref{table_periods}. Gaussian measured RVs  yield somewhat larger amplitude variations in the two emission lines analyzed (G1376 and G1362) and in the G1324 absorption. The corresponding RV curves folded with the sidereal ephemeris are plotted in Figure~\ref{RVcurves3} (left). The G1324 absorption vanishes during periastron in the 1992 and 1995 epochs, for which its RVs could not be measured.  At other times, however, it appears to behave similarly to what is observed in 1988, as illustrated in Figure~\ref{RVcurves3} (right) if shifted both in phase and in velocity.  The significance of the phase shift is not clear.  The mean velocity shift, however, could possibly be attributed to a change in the systemic velocity.

\begin{figure*}
\centering
\includegraphics[width=0.48\linewidth]{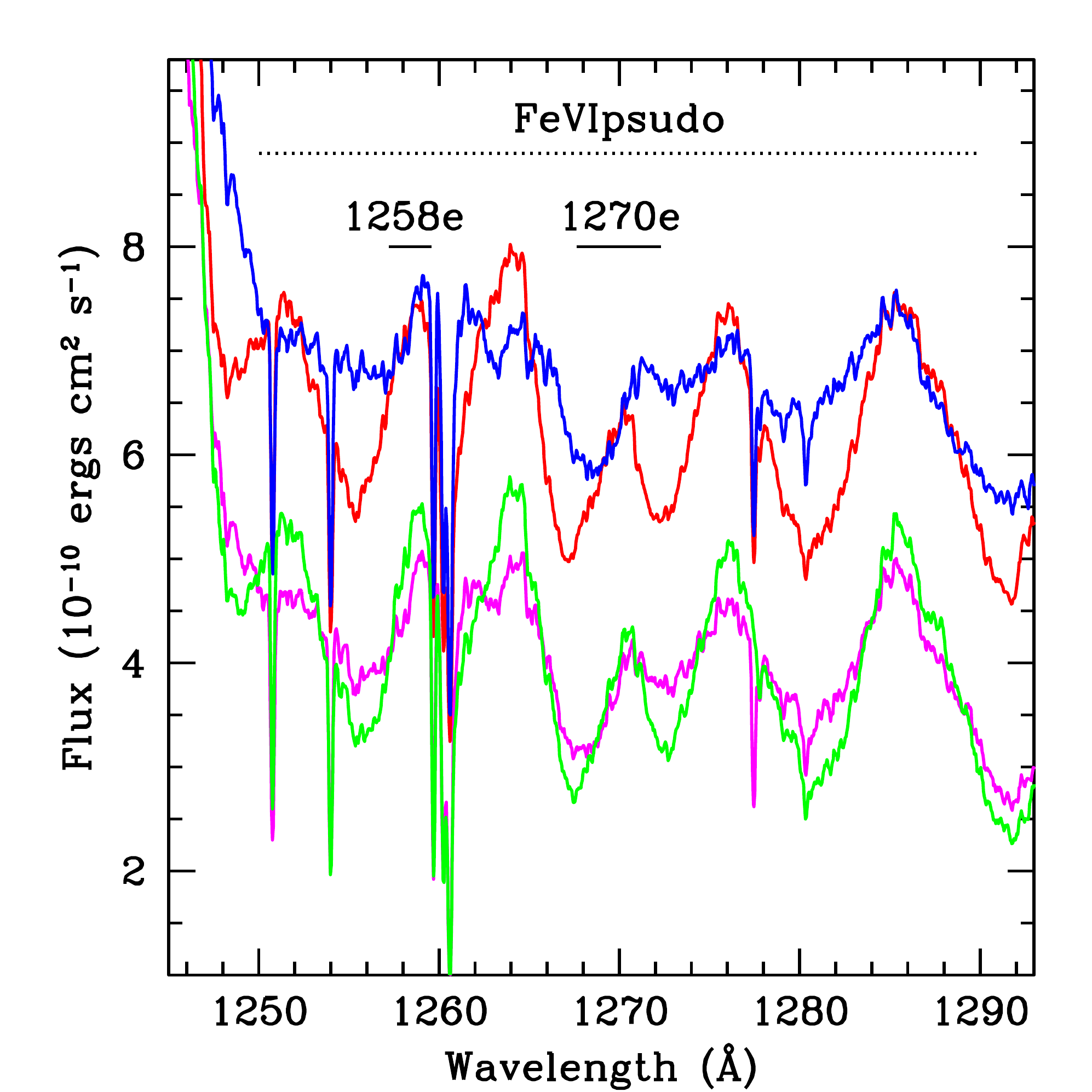}
\includegraphics[width=0.48\linewidth]{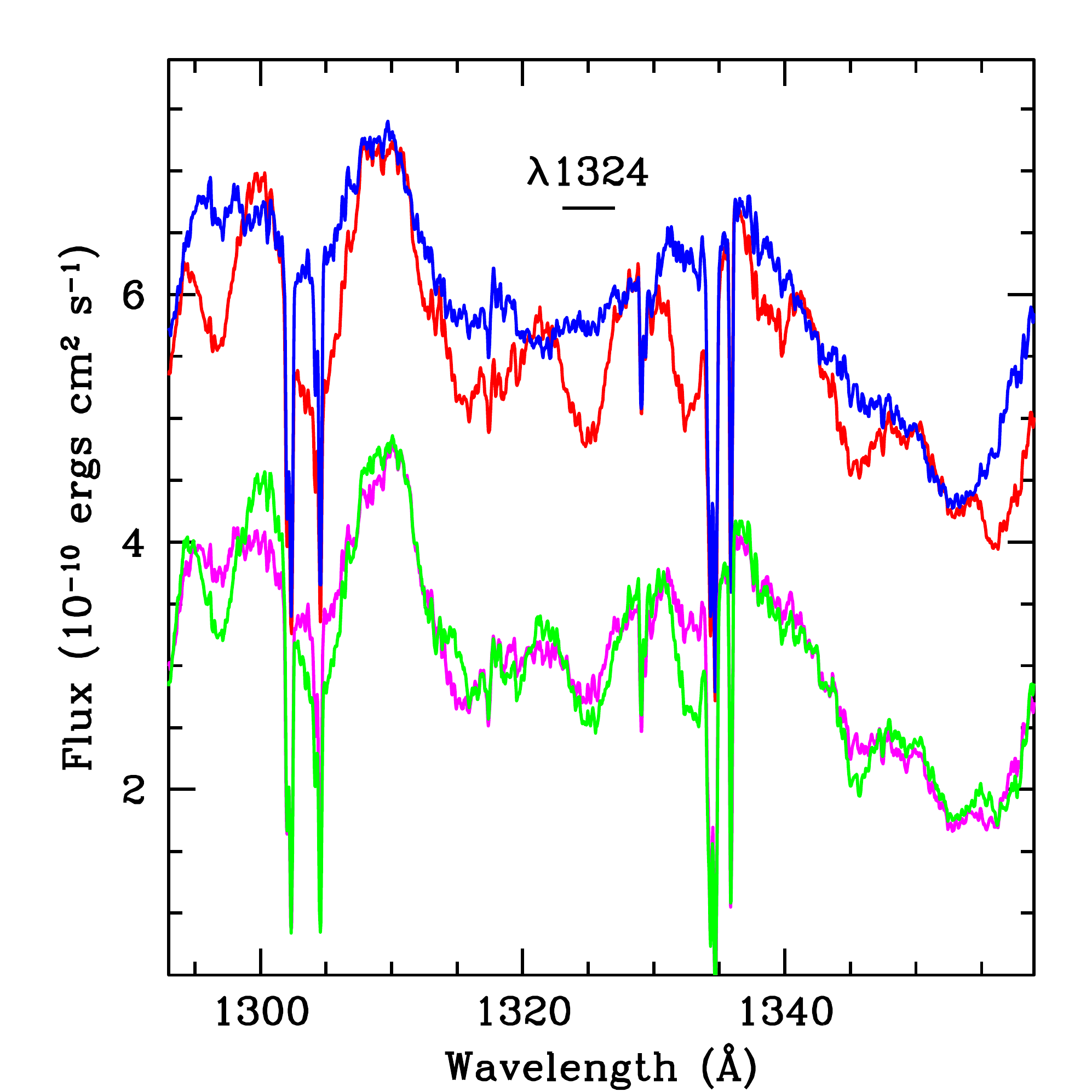}
\includegraphics[width=0.48\linewidth]{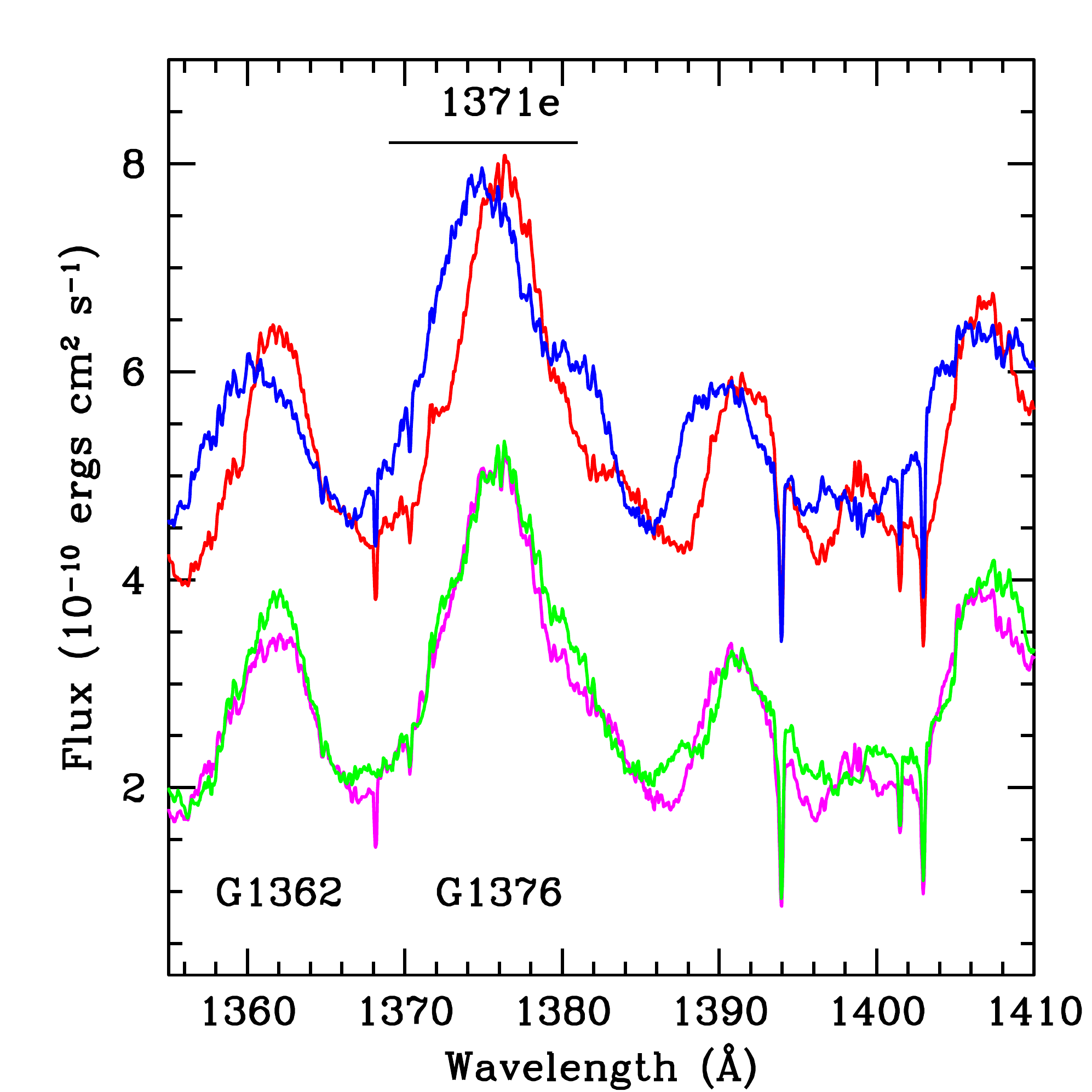}
\includegraphics[width=0.48\linewidth]{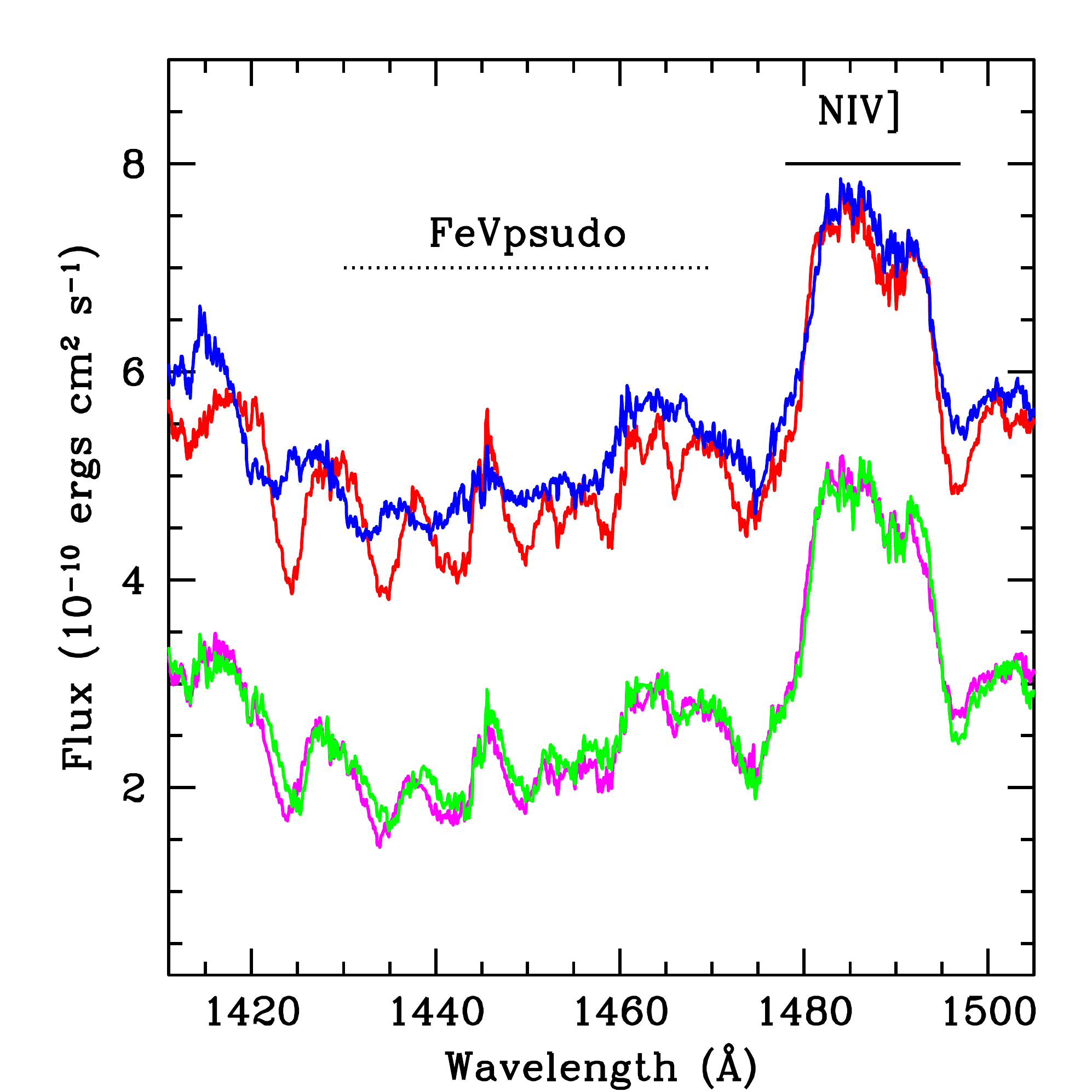}
\caption{Binned spectra of 1995 at four (sidereal) orbital phases.   Low mass companion approaching ($\phi_S$=0.2-0.3; blue) and receding ($\phi_S$0.7-0.8; red).  These spectra are shifted vertically for clarity in the figure.  The spectra corresponding to conjunctions are at $\phi_S$= 0.9-1.1 (WR in front, magenta) and $\phi_S$=0.4-0.6 (green).  Each panel illustrates a different wavelength region. The horizontal lines enclose wavelength regions defining bands that were used for flux measurements.
}
\label{spectra_1995}
\end{figure*}

\begin{figure}
\centering
\includegraphics[width=0.48\linewidth]{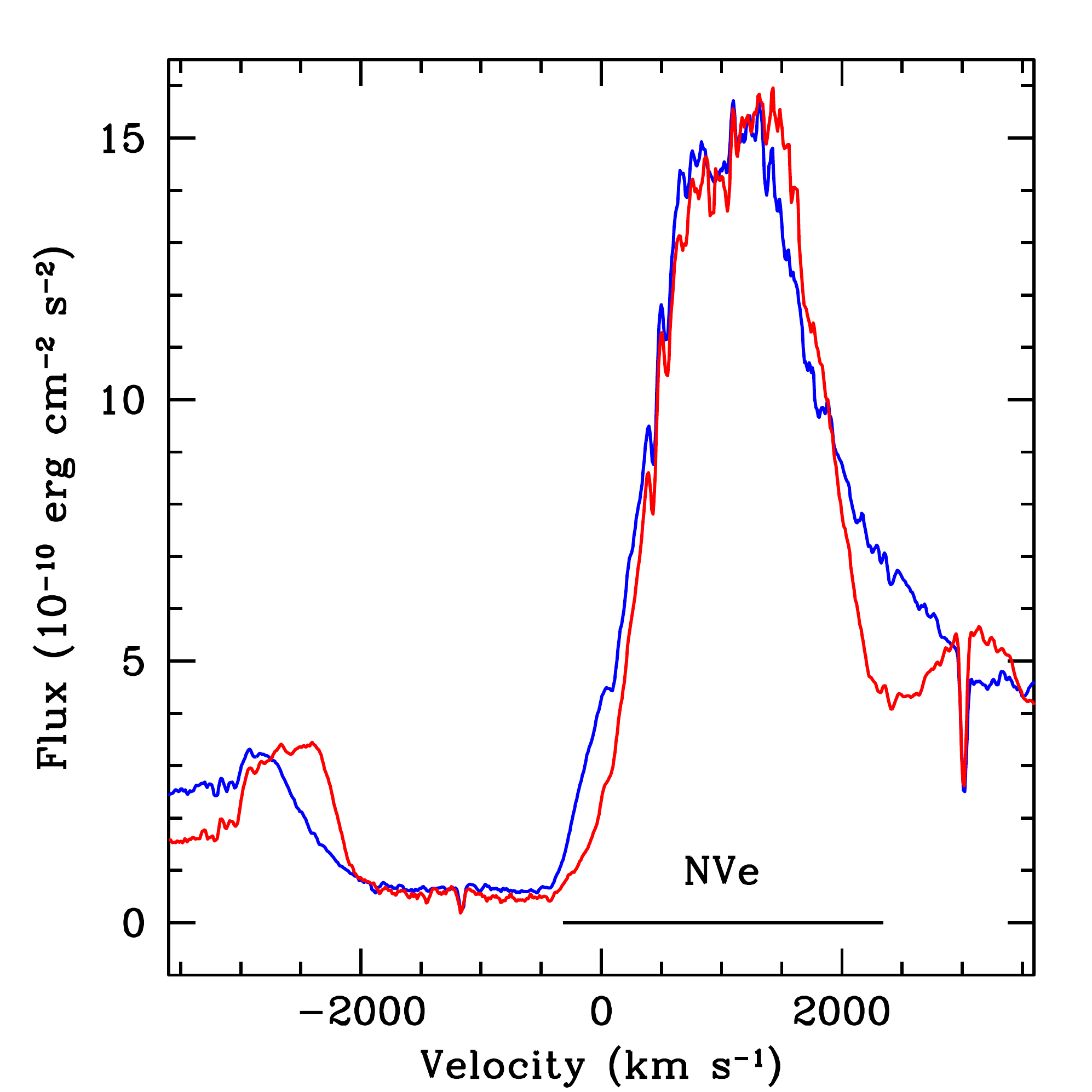}
\includegraphics[width=0.48\linewidth]{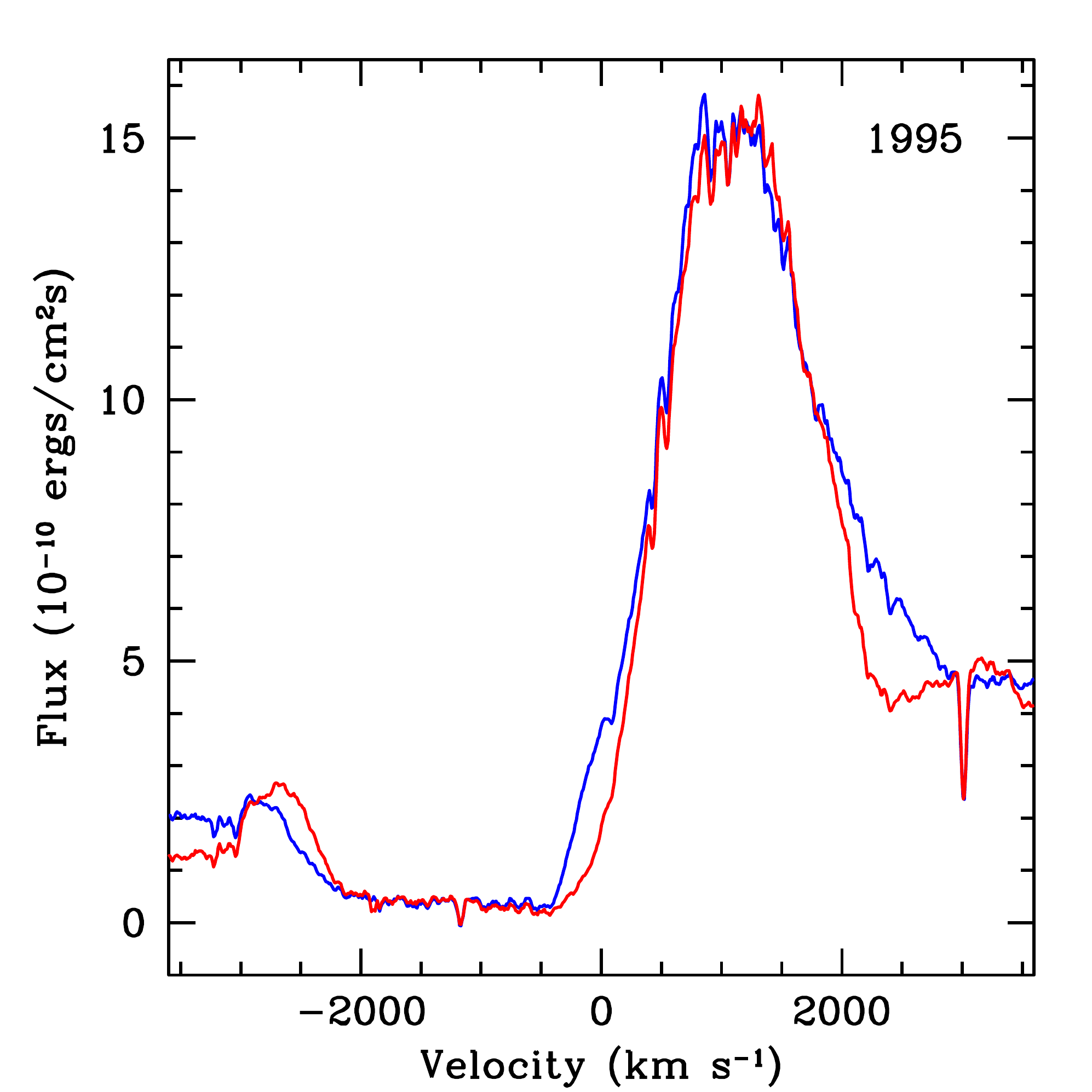}
\includegraphics[width=0.48\linewidth]{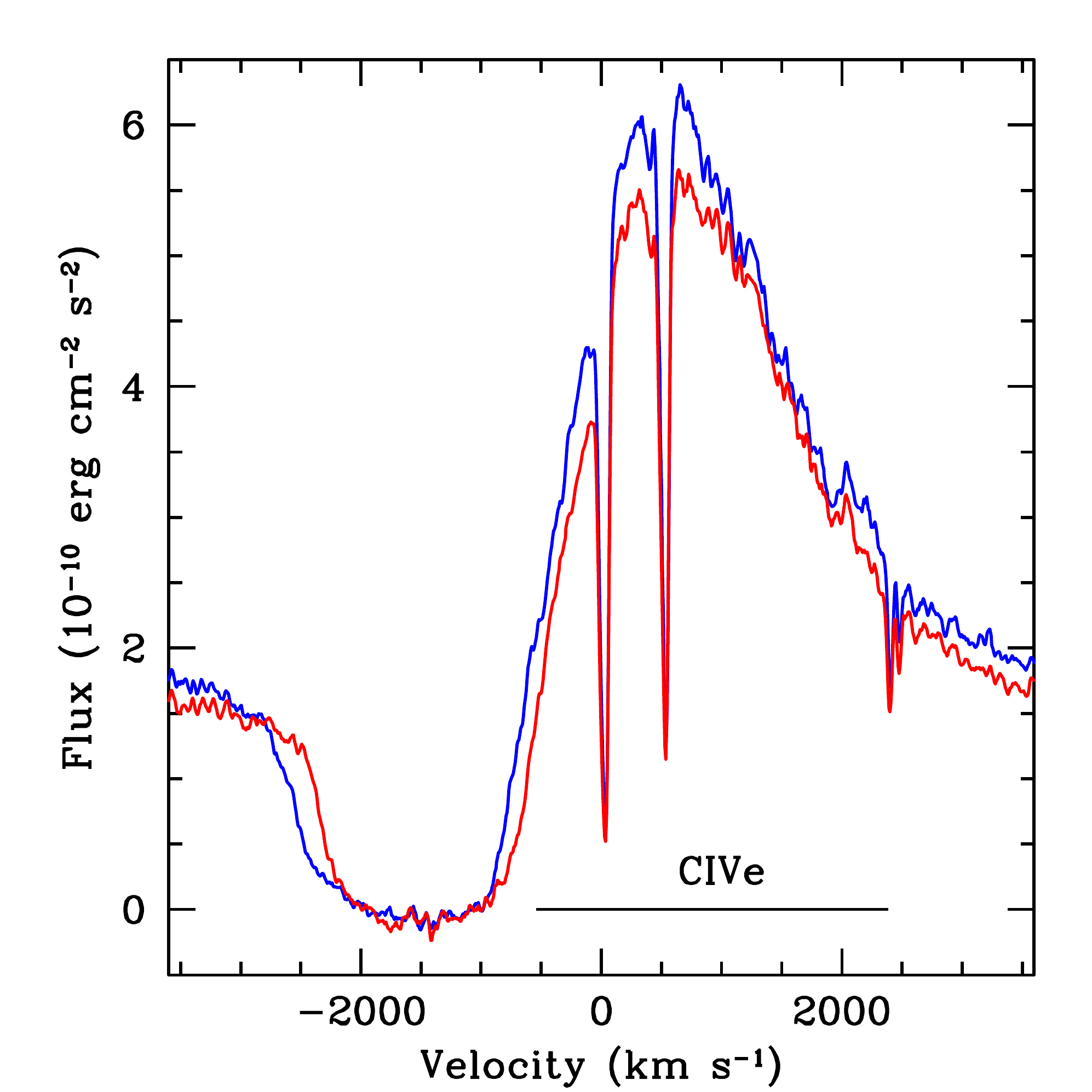}
\includegraphics[width=0.48\linewidth]{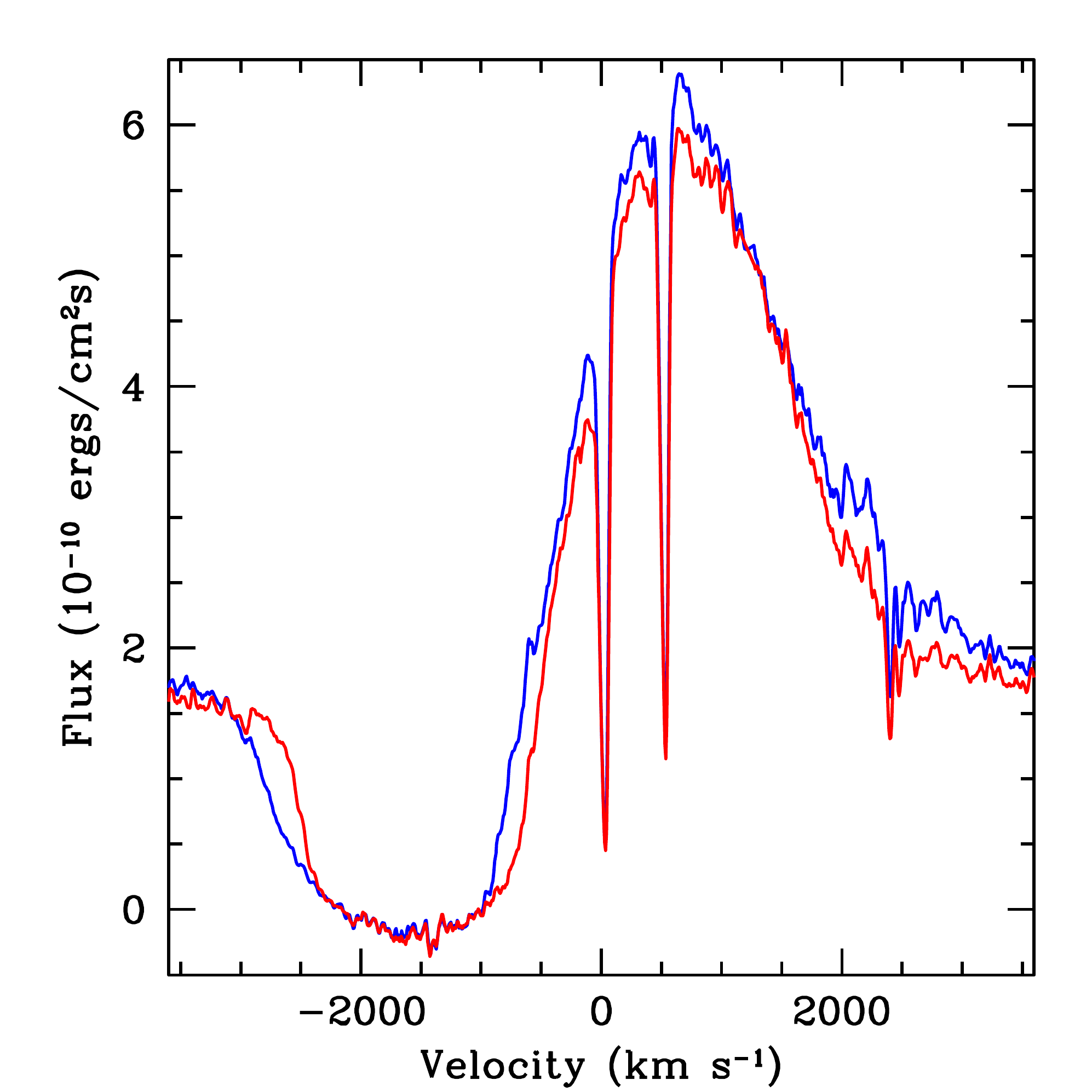}
\includegraphics[width=0.48\linewidth]{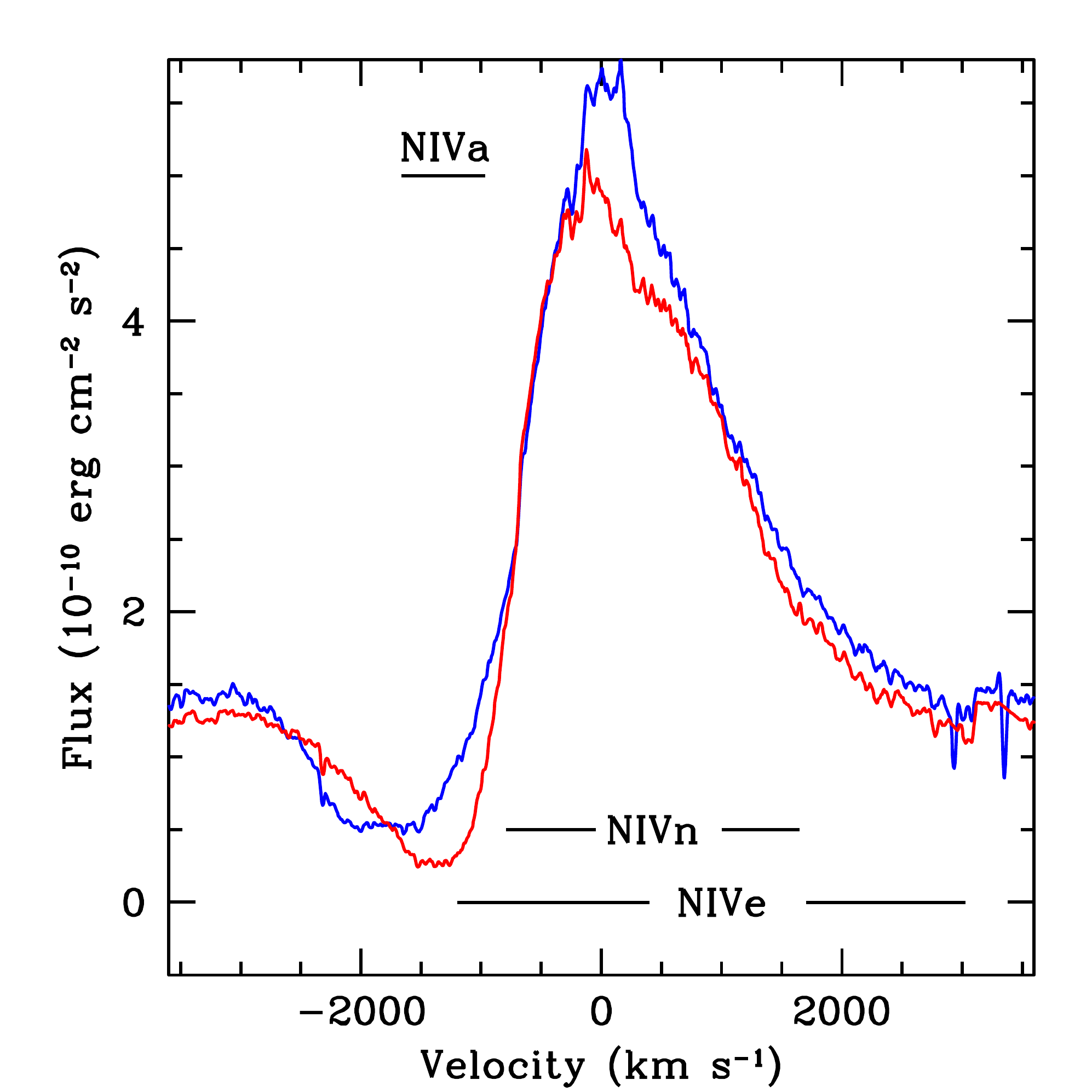}
\includegraphics[width=0.48\linewidth]{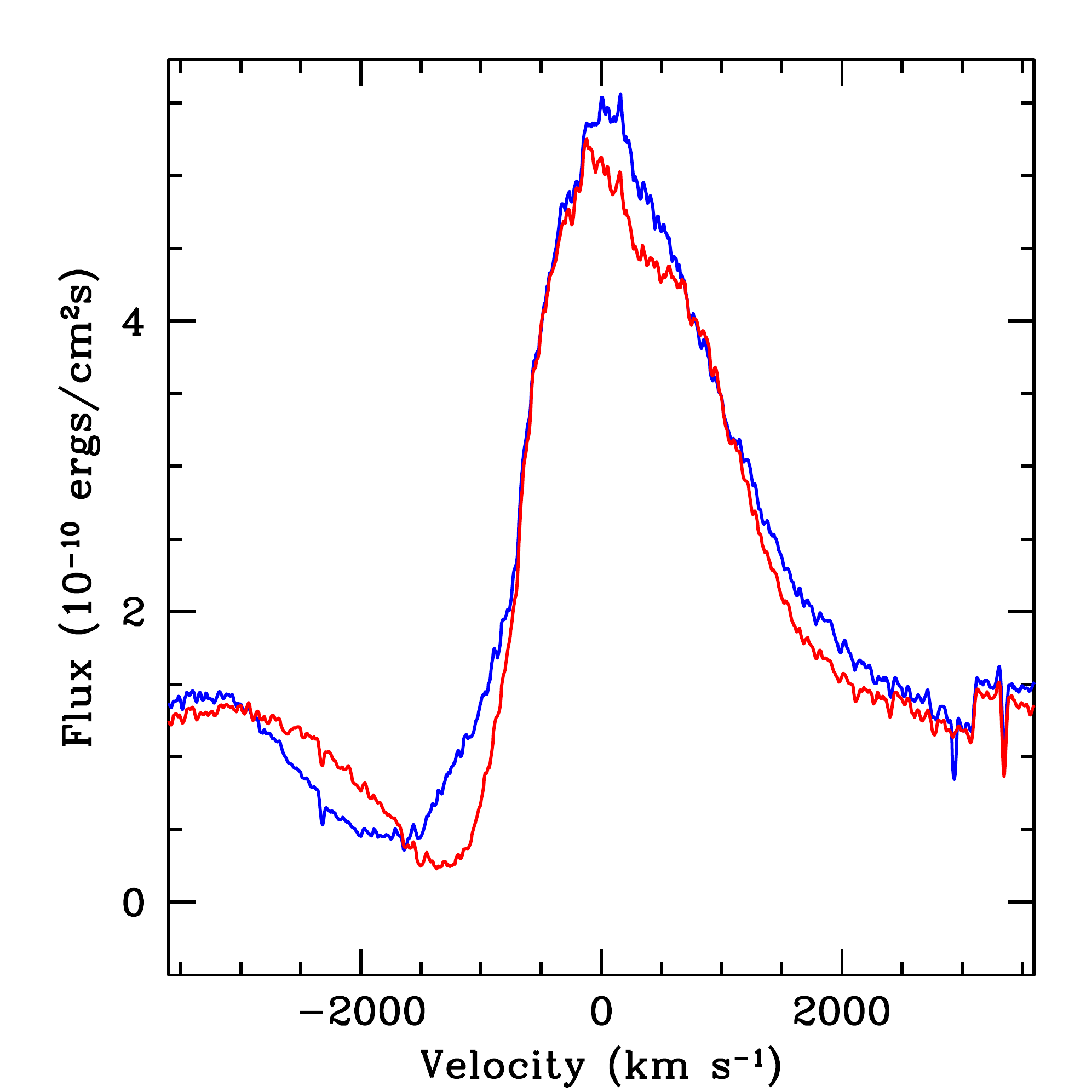}
\caption{Binned spectra at sidereal phases $\phi$=0.2-0.3 (blue) and 0.7-0.8 (red) for epochs 1992 (left) and 1995 (right), showing their very similar phase-dependent line profile variability. The abscissa is in velocity units  centered on the reference wavelengths: {\bf Top:} \ion{N}{V} $\lambda$1238.32; {\bf middle:} \ion{C}{IV} $\lambda$1538.20;  {\bf bottom:} \ion{N}{IV} $\lambda$1718.55.  The corresponding wavelength bands defined for these features and listed in Table 2 are indicated in the left panel with horizontal lines. }
               \label{spectra_phases}%
\end{figure}

\begin{figure}
\centering
\centering
\includegraphics[width=0.50\linewidth]{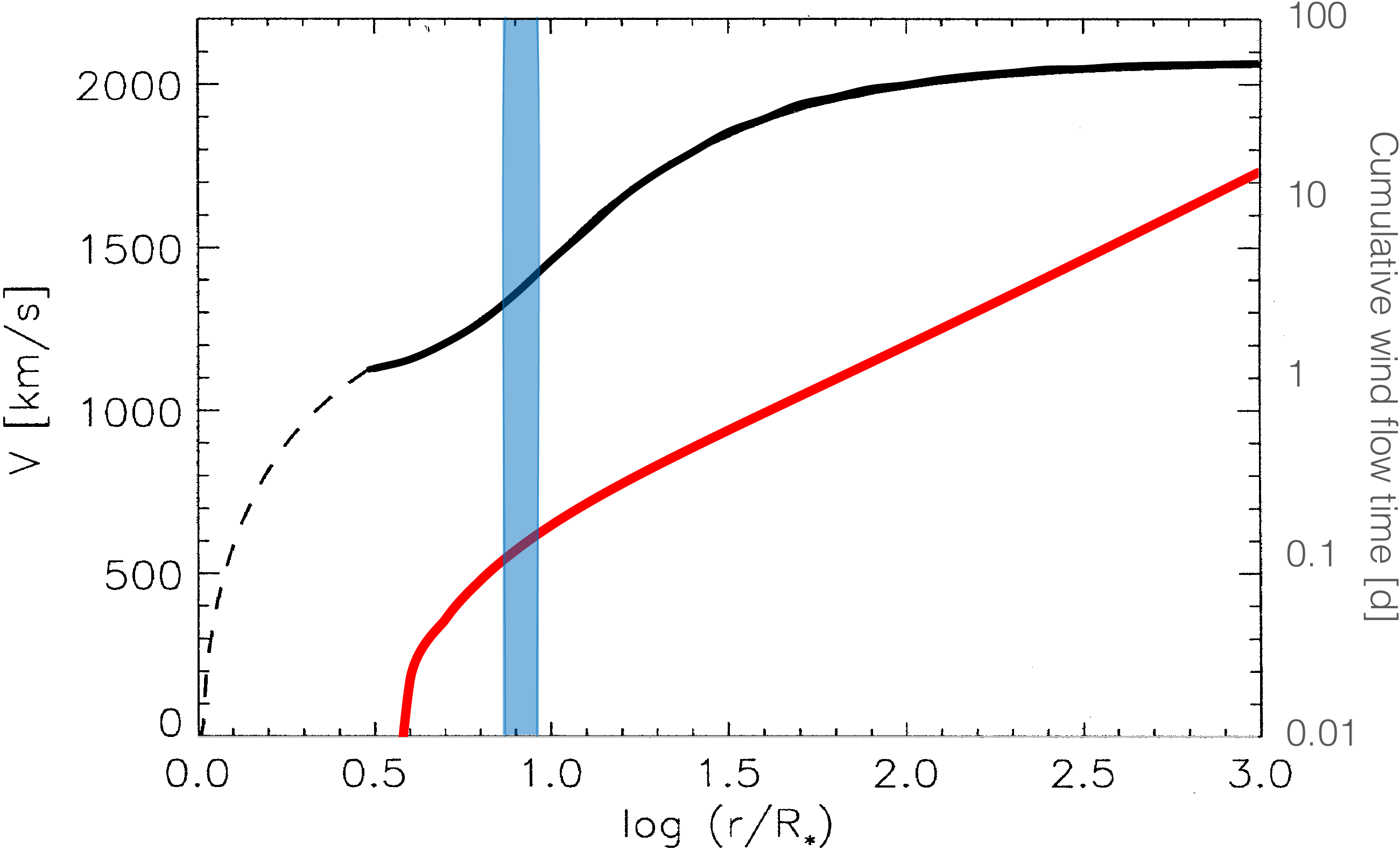}
\caption{The integrated flow time of the wind (axis on the right) from the calculated velocity law shown in Fig 10 of Schmutz (1997). The velocity law is shown in black and the cumulative flow time of the wind in red, starting at the photosphere, where the wind optical depth is $\tau$=1. This corresponds to the model radius 3R$_*$ ($=10$\,R$_\odot$).  The blue rectangle encloses the distance between the two stars at periastron, 26 R$_\odot$  (=0.87 log$_{10}$(r/R$_*$)), and apastron, 32 R$_\odot$  (=0.96 log$_{10}$(r/R$_*$)).}
               \label{velocity_law}%
\end{figure}

\begin{figure}
\centering
\includegraphics[width=\linewidth]{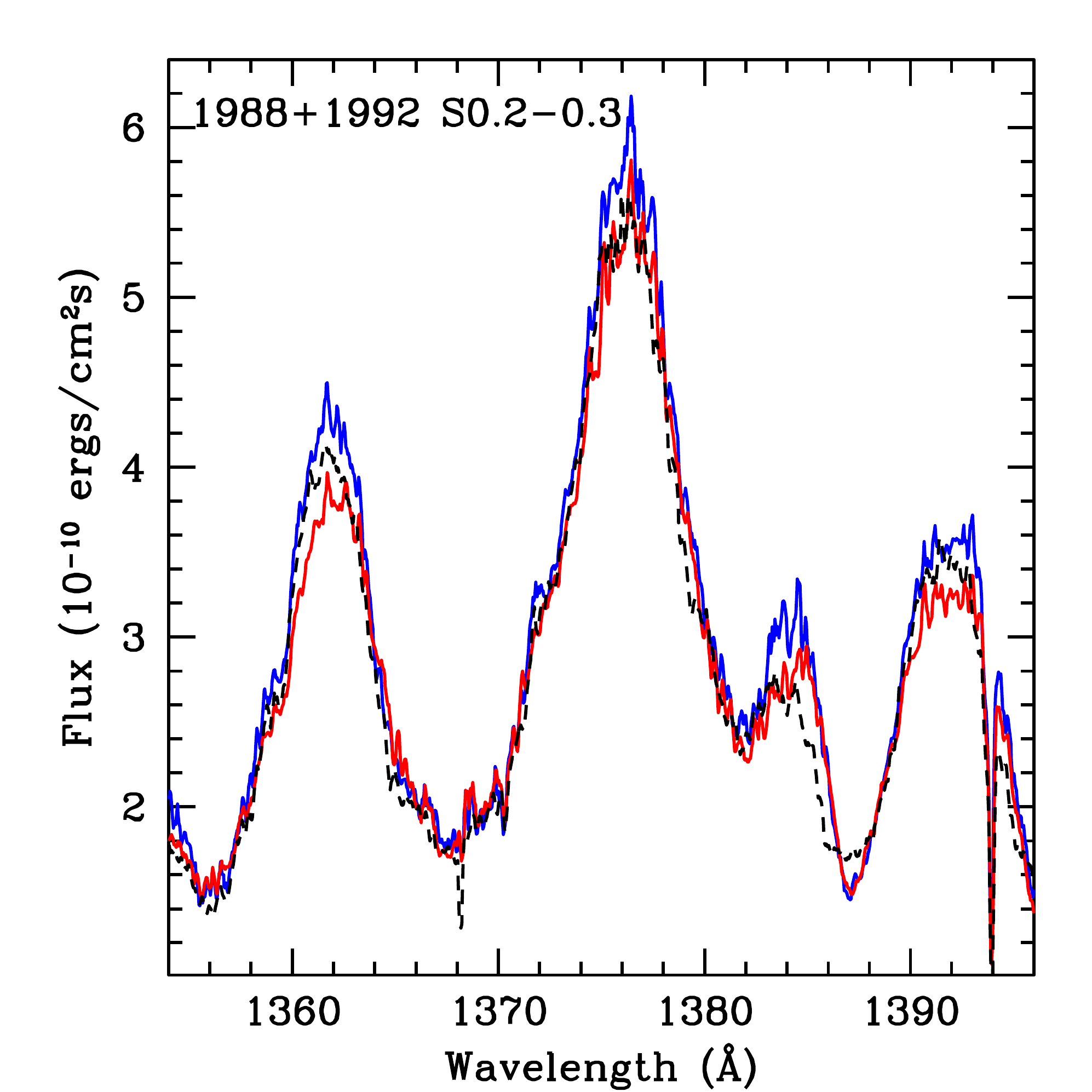}
\caption{Line profiles in the average spectra of 1988 at minimum RV (blue) and maximum RV (red)  (see Fig. 7, left) illustrating their similarity to the apastron spectrum of 1992 (dash).
}
\label{profiles_1988}
\end{figure}

\subsection{Properties of the orbit}

The fits described in the previous sections yield the anomalistic and sidereal periods P$_A$ and P$_S$, respectively, and the precession period $U$ as listed in Table~ \ref{table_periods}. Inspection of the P$_A$ and P$_S$ values shows differences $\leq$0.1\,d that can be accounted for by the short timespans covered by each observing campaign.   Even for the 1995 epoch, which covers five orbital periods, the values of P$_A$ and P$_S$ can only be determined to two significant digits. Thus, the determined  values of P$_S$ are not significantly different from  3.766\,d determined by \citet{Antokhin_etal1994} from three months of uninterupted narrow band photometry in mid February to mid May 1993, and which agrees  with the sidereal period determined by SK19. We tested the sensitivity of our fits to the period by setting the sidereal period $P_\mathrm{s}=3.766$\,d, and varying only the apsidal motion in the fit procedure to get the value of P$_A$. To the eye the resulting fit does not differ from what is shown in Figure \ref{fits_1995} and numerically the agreement is not significantly worse based on the scatter of the measurements. Comparing in Table \ref{table_periods} the two entries for the ephemerides of the IUE data in 1995, provides insight into  the uncertainties of the values. In particular it cannot be excluded that the sidereal period remained stable in all epochs.

The 1983 and 1988 fits suggest a much smaller inclination of  the orbital plane than what we obtain of 1992 and 1995.  A time-dependent variation in the orbital plane inclination is consistent with rapid apsidal motion, and could provide an explanation for the epoch-dependent spectral changes that are discussed in Section 5.

 \subsection{The nature of the close companion}

The orbital parameters obtained from simultaneous fits to the Gauss-measured radial velocities of the peak G1376 and  the fluxes 1371e  are listed in Table \ref{table_orbit}.  The amplitudes obtained from the G1376 RVs and the flux-weighted mean radial velocities of FeVpsudo and FeVIpsudo  are similar for 1992 and 1995. The ratios of the velocities are 4.1 and 5.9 for 1992 and 1995, respectively. If we assume a mass of 20\,M$_\odot$ for the WR star, adopted from \citet{Schmutz1997}\footnote{There are more recent determination of the stellar parameters of EZ CMa, such as, for example, \citet{Hamann_etal2019}. These analyses yield stellar parameters that can be considered consistent regarding the luminosity and thus, yielding a mass on the order as assumed here. However, no recent analyses has been done in such detail as \citet{Schmutz1997}, who has  calculated as part of the spectroscopic analysis the velocity law of the mass loss consistently with the radiation field and the ionization structure.}, then we obtain a mass range from 3.4\,M$_\odot$ to 4.8\,M$_\odot$ for the low-mass companion. As a WR star of 20\,M$_\odot$ stems from an even more massive star with its hydrogen rich envelope being shed, this star is only a few million years old. Thus, consulting the evolutionary calculation tables of \citet{Ekstrom_etal2012} we conclude that the less massive object is likely to be a main sequence B star, the models of which give an effective temperate on the order of 15000\,K, a radius of 2.3\,R$_\odot$, and 240\,L$_\odot$. However, this star cannot be a normal B-type star.

The B-star's atmosphere on the side that is facing the WR star is irradiated by the much more luminous WR companion. Our orbital estimates given in Table 5 yield an average distance of 0.14 AU, which in turn yields a luminosity of about 800\,L$_\odot$ incident onto the B star. The irradiance onto the B-star at the center of the irradiated surface is $\sim$15 times larger than the flux emitted by the unperturbed B star. Its atmosphere thus needs to compensate for the incoming radiation by adjusting its temperature and temperature gradient at optical depth $\tau = 2/3$ to emit the additional flux. A simple black-body estimate indicates that the photosphere  at the central point should look like a  30000\,K atmosphere. The optically thin zone that lies above it is dominated by the WR radiation field that has a still hotter spectral distribution and this should produce an ionization equilibrium comparable to that of the WR star. In addition to the WR radiation, the B star is subjected to the incoming supersonic flow of WR wind that collides upon the already heated atmosphere.  Thus,  the higher layers are formed by the shocked accreting WR wind, which is cooling down from its peak temperatures that produce the X-rays. Hence, the expectation is that the B-star surface layers are extremely hot and can produce emission lines of a similar ionization stage as those in the WR wind. It is therefore quite likely that the features we detect moving in anti-phase to the WR orbital motion are formed just above the B star's photosphere that is facing the WR star and thus do represent the B-star's orbital motion.  This conclusion, however, requires a more detailed study of the irradiated and shocked atmosphere structure. 

In low mass binaries, the response of a nondegenerate star to the irradiation from a very luminous companion has been found to significantly alter its properties. The intense radiation pressure can deform its surface \citep{2002MNRAS.337..431P}, and cause structural changes that can lead to expansion \citep{1991Natur.350..136P} and even an irradiation driven wind \citep{1989ApJ...336..507R,1999ApJ...513..442G}.  These effects can form a synergy with those caused by the incoming WR wind.  For example, even a modest expansion of the B-star would increase the cross-section that is exposed to the WR radiation field as well as that which intercepts the incoming wind. Though highly speculative, such variable interactions may play a role in causing epoch-dependent variations.

\subsection{The third body}

The strongest evidence pointing to a third object in the system is still the rapid apsidal motion. However, changes in the systemic velocity, derived from the orbital solutions to the IUE radial velocities, suggest a possible $K_3 \sim$150 km~s$^{-1}$ for the orbital velocity amplitude of the WR+B-star (Table~\ref{table_orbit}). A similar conclusion may be derived from the shifting systemic velocity obtained from the G1324 absorption (Figure 7) which is in the same sense and with the same amplitude as the possible $K_3$ for the other lines.

A caveat to this interpretation lies in the possible epoch-to-epoch wind structure changes.  A systematically increasing wind speed could shift the G1324 P Cyg absorption to shorter wavelengths and affect the systemic speed of other lines as well.  

\section{Discussion}

\subsection{Phase-dependent line profile variability}

Significant line profile variability is present in the 1992 and 1995 data sets, as already reported by St-Louis et al. (1995), while much lower levels of variability are present in 1983 and 1988. Orbital phase-dependent line profile variations can be an important source of uncertainty for the interpretation of the radial velocity curves, which is why we examine them in detail here. 

The phase-dependent variability in the line profiles of 1992 and 1995 are nearly identical.  One reason why this is the case is that, according to the ephemerides for these epochs, periastron occurs, respectively, 0.29 and 0.25 in phase after the conjunction when the WR is in front.  This means that in both cases, the unseen companion is approaching the observer in its orbit at a time when the wind density at the orbital distance is largest.  Hence, the excess emission is strongest, and it is blue-shifted  because of the companion's radial velocity and  because there is shock-heated WR wind expanding outward with a velocity component approaching the observer.

To illustrate the above, the spectra were averaged into four sidereal phase bins: {\it bin 1} (0.9$\leq \phi \leq$1.1)  corresponds to the WR star in front; {\it bin 3} (0.4$\leq \phi \leq$0.6)  corresponds to the companion in front; {\it bin 2} (0.2$\leq \phi \leq$0.3) corresponds to the companion approaching in its orbit; and {\it bin 4} (0.7$\leq \phi \leq$0.8) corresponds to the companion receding.  It is important to note that the {\it bin 2} phase interval includes periastron. The binned 1995 spectra are plotted in Figure~\ref{spectra_1995} and show the excess emission of the {\it bin 2} spectrum. Specifically, whereas most of the lines in the $\lambda\lambda$1250-1350 spectral region (dominated by \ion{Fe}{VI} transitions) have P Cyg profiles at other phases, at this phase they are filled in by emission. 

The majority of blends in the  $\lambda\lambda$1250-1350 spectral region arise from  transitions between excited levels of \ion{Fe}{VI}, and in the  $\lambda\lambda$1350-1470 region from \ion{Fe}{V}. 
The dominant effect is that the intrinsic WR spectrum in 1992 and 1995, consisting of P Cygni profiles in the \ion{Fe}{V-VI} region, is altered around periastron by the superposition of  emissions that fill in the  absorptions. Excess emission persists in the {\it bin 3} spectrum since at this phase the unseen companion is on the near side of the WR.  Contributing lines are likely the same \ion{Fe}{VI} shifted to shorter wavelengths as they originate in outflowing shock-heated WR wind.

In addition to  \ion{Fe}{V-VI} transitions, lines from even higher ionization species  originating in the shock regions are expected to  strengthen at periastron.  For example, in the spectral regions shown in Figure~\ref{spectra_1995}, one finds that the excess emission at  $\lambda\lambda$ 1265-1268 may include contributions from  \ion{Fe}{VII}, \ion{S}{IV}, \ion{S}{V}, and \ion{Si}{V}.  In $\lambda\lambda$ 1322-1327 there are lines of \ion{S}{IV}, \ion{S}{VI}, and possibly \ion{Si}{IX}.  In $\lambda\lambda$ 1330-1333 possible contributions include \ion{Si}{VI}, \ion{O}{IV}, \ion{S}{IV}, \ion{Mg}{IV}, \ion{Fe}{VII}, \ion{Si}{VII}.   

The \ion{Fe}{V} P Cyg absorption components in the $\lambda\lambda$1350-1420 region are not as filled-in by emission around periastron as is the \ion{Fe}{VI} region.  It is here where a wavelength shift in the emission peaks is most evident  when comparing the  {\it bin 2} and {\it bin 4} spectra. Specifically, the line maxima at  phases $\sim$0.2-0.3 are shifted by $-$350 km/s with respect to those of phases $\sim$0.7-0.8. Thus, in addition to the excess emission that extends out to -1300 km s$^{-1}$, the peak emission of the associated line, near line center, shifts blueward in {\it bin 2} with respect to the opposite phase.  We interpret this emission to arise in the shock region closest to the companion.

The P Cygni profiles of the resonance and other strong lines change notably between the two periapses ({\it bin 2} and  {\it bin 4}),  as already noted by St-Louis et al. (1995) and illustrated in  Figure~\ref{spectra_phases}.  Near periastron ({\it bin 2}, companion approaching), they have much faster blue-shifted absorption edges as well as excess emission on the blue wing at low velocity, which echoes the behavior of the excess emission in the \ion{Fe}{V} and \ion{Fe}{VI} lines.  It is important to note, however, that the  saturated portion of the absorptions (referred to as V$_{black}$) does not change significantly.

An interpretation for the higher speed P Cyg absorptions around periastron is not straightforward. There is a lag between the time when the WR wind  passes the companion's orbit and the time when this material  attains the fastest speed indicated by the excess absorption. This is illustrated with the integrated flow time of the WR wind, using the velocity law given by Schmutz (1997; Figure 10). Our Figure~\ref{velocity_law} shows that the wind velocity is 1300~km~s$^{-1}$ when it reaches the companion's orbit at periastron and 1400~km~s$^{-1}$ at apastron, and that the flow time is 0.1 to 0.2~d to the companion. This is a relatively small fraction of the orbital period. However, for the material that forms the absorptions with shifts greater than 2000~km~s$^{-1}$ the flow time is half an orbit or longer.  This means that, if the fast P Cygni absorptions that are observed at periastron arise in the WR wind, the material causing them could have left the star  at the time of apastron or even at around the time of the previous periastron.  Alternatively, the high speeds may arise in material that is accelerated by shocks, instead of representing a global wind velocity. The accelerating mechanism, however, would have to affect both the approaching and receding wind, since  excess emission in the \ion{He}{II} $\lambda$1640 {\em red wing} is also present at the same time as the excess blue absorption.  Furthermore, there is a correlated change in the shape of the \ion{N}{IV} $\lambda$1718 which also points to additional absorption at faster speeds.

The most notable aspect of the 1988 spectra  is their significantly weaker variability compared to the later epochs.  This is perplexing, but may potentially be understood as a consequence of a slower WR wind during this epoch, as implied by the P Cyg profiles as discussed below.

Also noteworthy is the fact that most of the 1988 spectral features are very similar to those in the apastron ({\it bin 4}) spectra in 1992 and 1995. The similarity is illustrated in Figure~\ref{profiles_1988}, where we compare the average 1988 spectra at RV minimum and at maximum  with the apastron spectrum of 1992. The one major difference between these spectra is in the resonance lines, which in 1988 have less extended P Cygni absorptions than in the later epochs. This is further discussed in Section 5.2.

\begin{figure}
\centering
 \includegraphics[width=\linewidth]{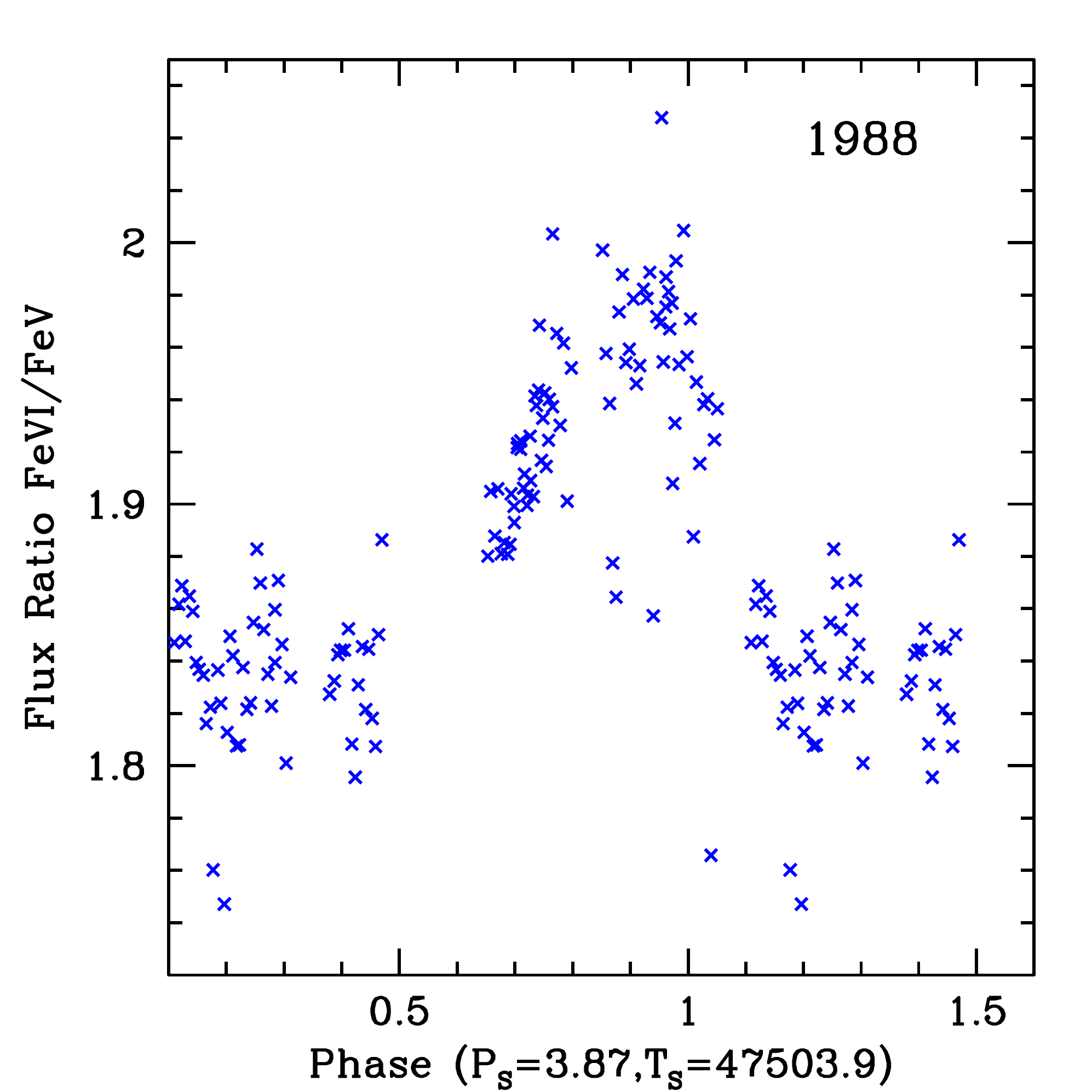}
\caption{Flux ratio FeVIpsudo/FeVpsudo plotted as a function of sidereal phase in the 1988 data, with periastron at $\phi_S \sim$0. 
   }
\label{ratio_1988}%
\end{figure}

The 1988 spectra display a phase-dependent variation in the flux ratio \ion{Fe}{VI}/\ion{Fe}{V} as illustrated in Figure~\ref{ratio_1988}.  The ratio rises toward maximum just prior to the time of periastron. A similar ionization variation is not seen in the 1992 and 1995 epochs, possibly because the shock is stronger and instead of \ion{Fe}{V-VI}, even higher ionization stages dominate.

\begin{figure}
\centering
\includegraphics[width=0.44\linewidth]{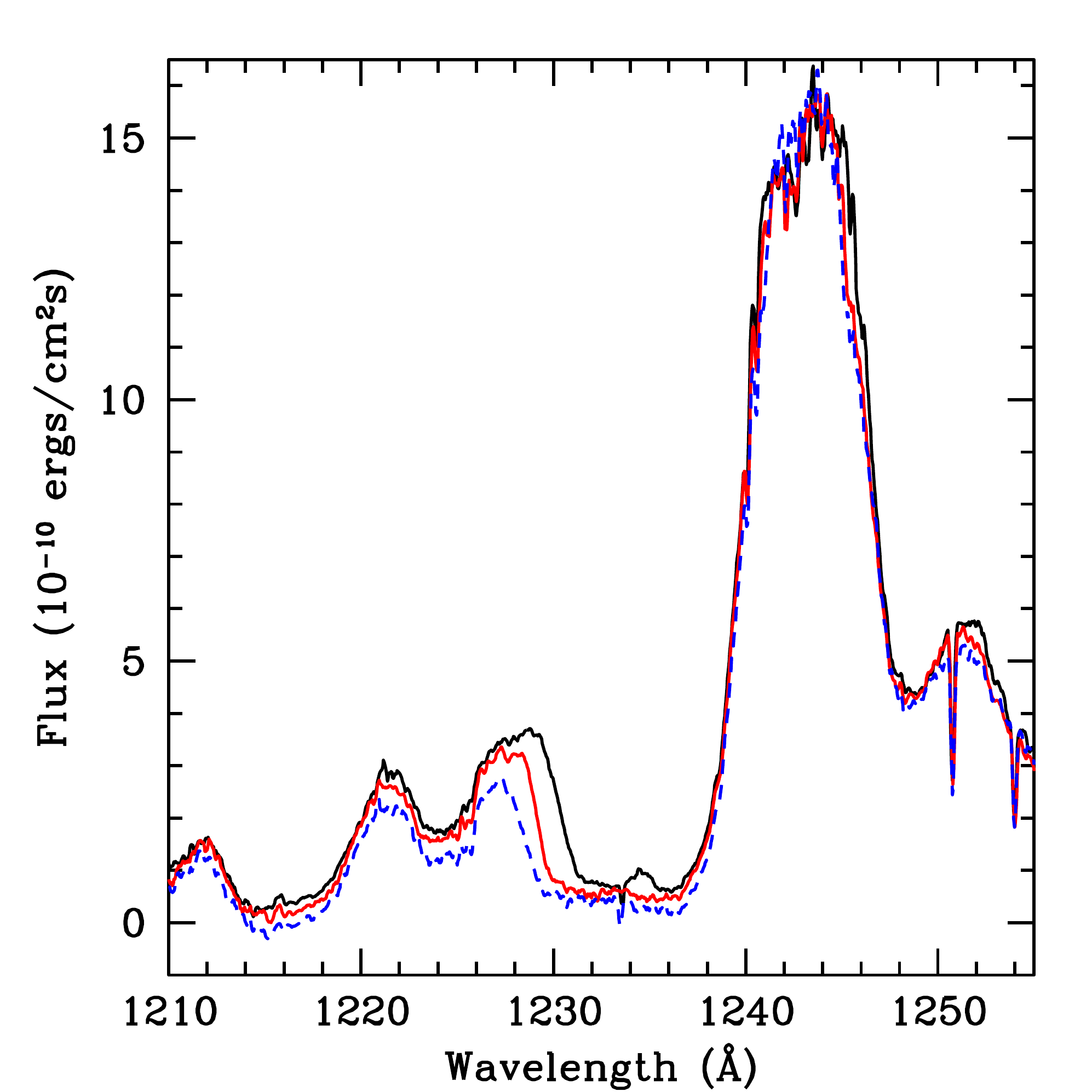}
\includegraphics[width=0.44\linewidth]{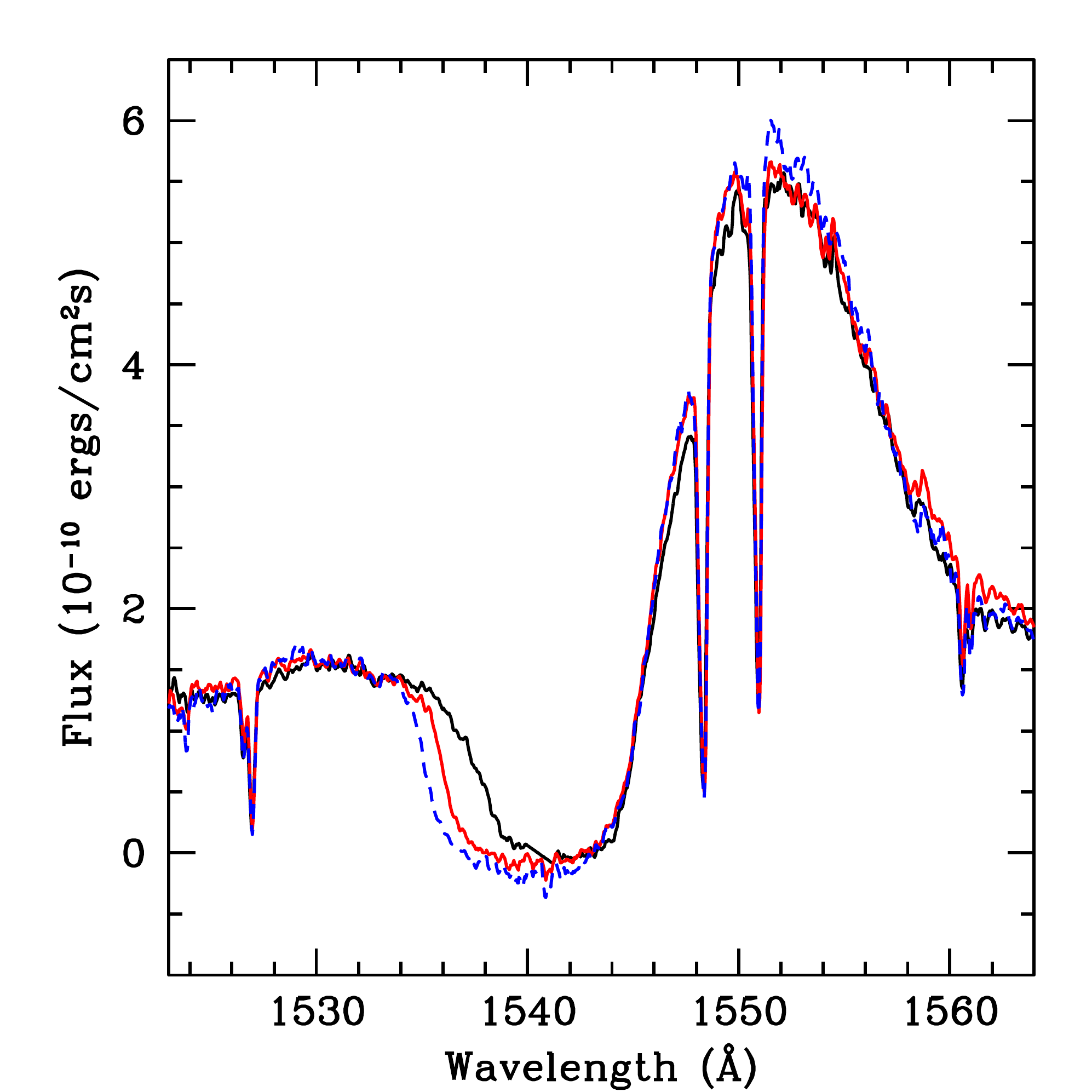}
\includegraphics[width=0.44\linewidth]{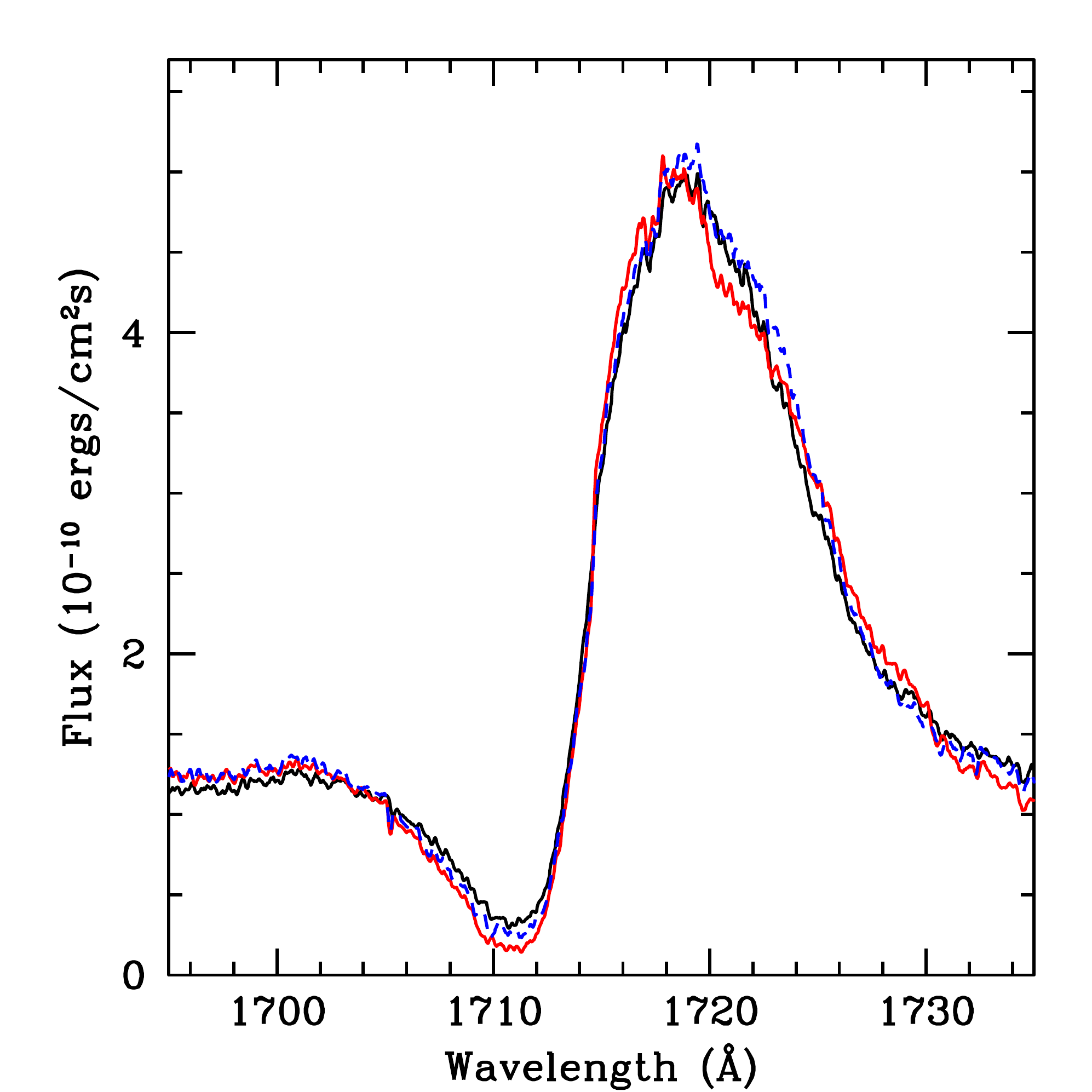}
\includegraphics[width=0.44\linewidth]{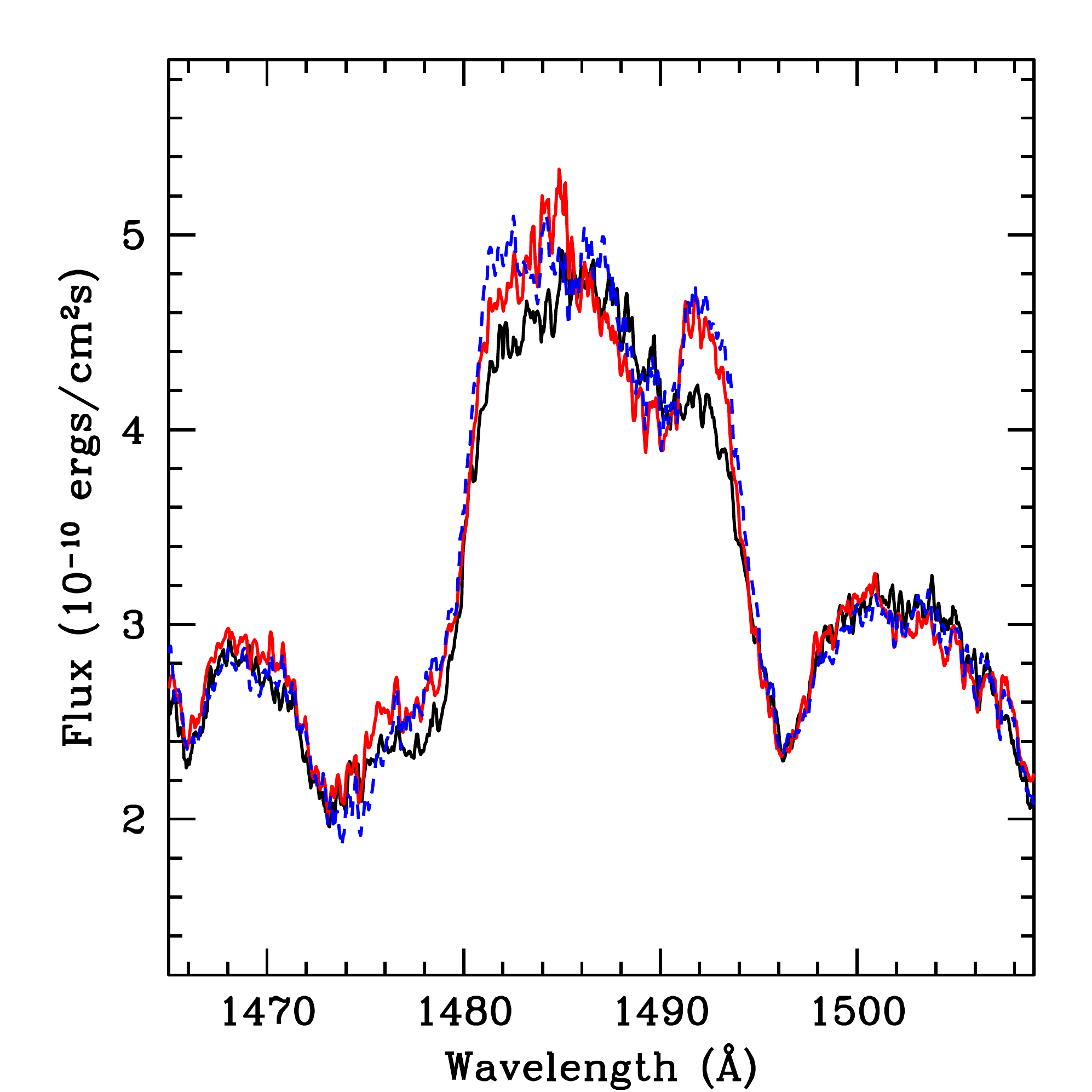}
\caption{Line profiles of \ion{N}{V} and \ion{C}{IV} in the average spectrum of HD\,50896 in 1988 (black) and in the binned spectra at apastron in 1992 (red) and 1995 (blue dash) showing  a systematic increase in the extent of the P Cygni resonance line absorptions which is less evident in \ion{N}{IV}\,$\lambda$1718 with no width change in  \ion{N}{IV}]\,$\lambda$1486.
  }
               \label{spectra_epochs}%
\end{figure}

\subsection{Epoch to epoch variations}

The epoch-dependent changes are quantified in Table \ref{table_UVaverages_stdevs} where we list for the 1983, 1988, 1992 and 1995 epochs the average over all orbital phases flux, flux-weighted velocity, and the corresponding standard deviations for the spectral bands that were measured.  Inspection of this table discloses two notable trends. The first trend is one in which the continuum flux in the $\lambda\lambda$1828-1845 band increased over time. Specifically, its value with respect to that of 1995 was 0.88, 0.90, and 0.99  for 1983, 1988 and 1992, respectively. A similar result was also reported in Morel et al. (1997).

The second trend is that the degree of variability over the 22 bands that were measured also increased over time. The flux standard deviations of Epochs 1983, 1988 and 1992 are, respectively,  0.33, 0.62 and 0.86 that of 1995.  Similarly, the corresponding velocity standard deviations are 0.53, 0.68, and 0.95.

As mentioned above, the 1983 and 1988 spectra are very similar to the {\it bin 4} (around apastron) spectra of 1992 and 1995.  Thus, the increase over time in the fluxes and their standard deviations are dominated by the blue-shifted emissions that appear in the {\it bin 2} (periastron) spectra of the later years. 

Comparing the {\it bin 4} (apastron) spectra of 1992 and 1995 with the average spectra of 1983 and 1988 discloses a single difference: the maximum extent of the \ion{N}{V} and \ion{C}{IV} P Cygni absorptions. These resonance lines are shown in Figure~\ref{spectra_epochs}.  In the later epochs, the entire P Cygni absorption component, including the saturated portion of the profile, is stretched towards shorter wavelengths. The extent of their blue edge increases from -2400~km\,s$^{-1}$ in 1983 and 1988 to -2500~km\,s$^{-1}$  in 1992 and then to -2680~km\,s$^{-1}$ in 1995.  The corresponding V$_{black}$ is  -1840~km\,s$^{-1}$ (1983 and 1988), -2100~km\,s$^{-1}$ (1992) and -2300~km\,s$^{-1}$ (1995).  Such an effect is not present in \ion{He}{II} nor \ion{N}{IV}, which arise from excited transitions.   Thus, the resonance lines suggest the presence in 1992 and 1995 of low density material that lies along the line-of sight to the WR core and that is accelerated to faster speeds than present in the earlier epochs. The slight emission increase in \ion{N}{ IV]} $\lambda$1486 may be another manifestation of the same phenomenon. However, the \ion{N}{IV}\,$\lambda 1718$ P Cyg absorption shows only a weak change from epoch to epoch.  This line is formed nearer to to the base of the wind  than the resonance lines, and it has a smaller opacity so it is less sensitive to processes that occur far out in the wind. 

We conclude by speculating that the faster wind speed in the later epochs results in a stronger feedback in the shock-heated region near the companion which, in turn causes, a larger degree of orbital phase-dependent variability.

\begin{table*}[htb]
\caption{Summary of Ephemerides. \label{table_periods}}
\begin{tabular}{llcccccccc}
\hline
Set   &{Year} &{T$_{Anom}$}  &{P$_A$} &$\dot{\omega}_{per}$& {U}&{deg/P$_A$} & T$_0$(WR in front) & {P$_S$} & $\omega($T$_0)$ \\
\hline
\hline
IUE LW & 1983.675 & 45582.4: &  3.71:: & 0.15:: & -42:: & -32::    & 45582.4:: & 3.4:: &  $271^\circ$  \\ 
IUE SW & 1983.673 & 45581.7: & 3.74: &  -0.15:  &  42:  & 32:   &  45581.8:  & 4.1:  & $259^\circ$ \\  
IUE & 1988.936    & 47504.0  & 3.74: & -0.057:  & 110:  & 12:    &  47503.9:  & 3.87: & $100^\circ$   \\ 
IUE & 1992.061  &  48645.38  & 3.61: & -0.045   & 126   & 10.3 &48644.30  & 3.71: & $207^\circ$  \\   
IUE & 1995.059  &  49740.17  & 3.72  & -0.040   & 157   &  8.5   & 49739.15  & 3.81 & $199^\circ$ \\ 
IUE $P_\mathrm{s}$ fixed & 1995.058 
                &  49740.12  & 3.69  &  -0.037  &  170  & 7.8   & 49739.16  &  {\emph{3.766}} &  $195^\circ$ \\
Morel &1995.057 &  49739.59  & 3.71  & -0.034   & 187 &  7.2   & 49738.83  & 3.79 & $175^\circ$ \\  
{\em XMM-Newton} & 2010.785  &  55484.15  & 3.72  &  -0.034  & 187 & 7.2 & 55482.16    & 3.80 & $100^\circ$ \\ 
BRITE&2015.888  &  57348.14  & 3.625 &  -0.075  & 83.6  & 15.6 &   ....     &3.789  \\
\hline
\end{tabular}

\tablefoot{Columns 3-10 are obtained from the model fits to the observed light curves and radial velocities. T$_{Anom}$ and T$_0$ are, respectively, the time of periastron and the time when the WR is in front of the companion, and are given in units of JD$-$2400000. P$_A$ and P$_S$ are the anomalistic and sidereal periods, respectively and are given in days. The rate of change of the argument of periastron, $\dot\omega_{per}$, is given in radians per day; $U$ is the apsidal period, given in days; {deg/P$_A$} is $\dot\omega_{per}$ given in degrees per anomalistic period; and $\omega(T_0)$ is the argument of periastron at time $T_0$ in degrees measured from elongation. The uncertainties in the values are discussed in section 4.4.
}
\end{table*}

\begin{table*}[htb]
\caption{Summary of orbit parameters. \label{table_orbit}}
\begin{tabular}{lccccc}
\hline
Year & 
$K_\mathrm{WR}$ [km/s]  &
$K_{2^\mathrm{nd}\mathrm{ obj}}$ &
$\gamma$ [km/s] & 
$i^\mathrm{a)}$ & 
M$_{2^\mathrm{nd}\mathrm{ obj}}$ [M$_\odot$] \\
\hline
\hline
1983.67  &     - &     - & 1050  & $<2^\circ$  & - \\ 
1988.936  &  -  &     13: & 1145 & $<2^\circ$ & - \\ %
1992.061  &     33  &    137 & $974$ & $25^\circ$ & 4.8 \\ 
1995.059  &     27  &     159 & $959$ & $28^\circ$ & 3.4 \\ 
\hline
\end{tabular}

$^\mathrm{a)}$ The inclination is calculated from the assumption that $M_\mathrm{WR} = 20$\,M$_\odot$.
\end{table*}

\subsection{CIRs, wind-wind collisions and periastron effects}

Some of the observed variations in EZ CMa and other presumed single sources with strong winds has often been attributed to corotating interaction regions \citep[CIRs,][]{1984ApJ...283..303M}.  These are assumed to arise in an aspherical stellar wind or in active zones near the stellar photosphere of a rotating single star.   As the perturbations propagate outward, they form a  spiral density structure that is embedded in the WR wind \citep{2019MNRAS.489.2873C}. Qualitatively, this spiral structure may be analogous that that which arises from a wind-wind collision and its trailing shock region that propagates outward \citep{2008MNRAS.388.1047P,2009MNRAS.394.1758P}. The dynamics and detailed structure of the spiral depend on the orbital parameters and mass-loss properties of the two stars. 

The line profile variability expected from a wind-wind collision region was modeled by \citet{2009MNRAS.395..962I} for the case of  forbidden emission lines. Although their model is applied to a WR+OB binary, their predicted variations are qualitatively similar to those that  were reported by Firmani et al. (1980) for the \ion{He}{II} $\lambda$\,4686 emission line; that is, it changes from a flat-top to a peak that is either centered or on the blue or the red side of the underlying broad emission. Thus, one could speculate that the high ionization UV lines describe the processes occurring near the companion while the optical lower ionization lines describe those that occur further downstream from the shock.

Associating the CIR phenomenon with an origin in the binary companion would help solve the question surrounding the trigger for the CIR formation in EZ CMa. However, a quantitative analysis is required in order to test whether the observational diagnostics of CIR-like spirals produced in a colliding wind binary  are similar to those in assumed rotating single stars.

Interaction effects play a role in all close binary systems and particularly so when one or both possess a stellar wind.  In addition to eclipse, occultation and wind collision effects, additional phase-dependent phenomena arise when the orbit is eccentric.   In addition to the variations in the wind collision strength discussed in this paper,  type of interaction effect is the tidal excitation of oscillations that are triggered by periastron passage in an eccentric binary \citep{1995ApJ...449..294K,1999RMxAA..35..157M,2011ApJS..197....4W,2019arXiv191108687G}.  Although the eccentricity of EZ CMa is not as large as that of binaries in which such oscillations have been observed, the theory predicts their existence.    
Pulsations have long been suspected to contribute towards enhancing mass-loss rates in massive stars \citep{2007AIPC..948..345T}.  This raises the question of the role that tidally excited oscillations may play in producing a structured stellar wind, particularly in the orbital plane. For example, one may speculate that the larger tidal perturbation amplitude at periastron  leads to a structured wind with alternating slower and faster shells, with the latter catching up with and shocking the former. 

A final question of potential interest concerns the amount of B-star material that might be ablated by the combined effect of irradiation and  wind collision.  The outflowing shocked WR wind would then carry H-rich material it has "lifted" from the B-star, leading to a mix of chemical compositions in the wind. 

\section{Conclusions}

We analyzed X-ray and UV observational data sets that were obtained during 5 different epochs between 1983 and 2010 with the aim of further testing the eccentric binary model for EZ CMa that was proposed by SK19 and with the objective of constraining the properties of the low-mass companion.  In this model, the WR wind is assumed to collide with the low-mass companion forming a very hot and highly ionized region which gives rise to hard X-rays and emission lines.

The observed flux variability in the {\em XMM-Newton} hard X-ray bands is explained by a model in which the energy deposited in the shock is proportional to the wind density at the shock location. The modulation by 0.1 eccentricity with variations proportional 1/d$^2$ explains  what is observed; that is, $\pm 20$\,\% modulation.  In addition, we find that the predicted time over which the emission maximum occurs is shorter than typical exposure times and that the time when the emission is maximum  (that is, at periastron) coincided in these data with an eclipse. Hence,  the phase with the hardest emission may not yet have been  observed.

The extensive sets of IUE observations allow a detailed radial velocity analysis which shows that the strong WR emission lines follow a periodic RV curve with a semi-amplitude K$_1\sim$30 km\,s$^{-1}$ in 1992 and 1995, and that a second set of weaker emissions move in an anti-phase RV curve with K$_2\sim$150 km\,s$^{-1}$.  The simultaneous model fit to the RVs and the light curve yields the orbital elements for each epoch.  Adopting a Wolf-Rayet mass M$_1\sim$20~M$_\odot$ leads to M$_2\sim$3-5~M$_\odot$ which, given the age of the WR star,  corresponds to a late B-type star \citep{2014ApJ...795...82S}.   We argue, however, that it is unlikely to be a normal B-type star because of the large incoming radiative flux from the WR, the additional heating by the shock-produced X-rays, and the accretion of impinging WR wind material.  The combined effect of these processes can significantly alter the B-star envelope structure \citep[][and references therein]{2002ASPC..279..253B} and we speculate that it may lead to the ablation of H-rich B-star  material which is then carried away by the WR wind. 

We speculate that as the mixture of shocked WR wind that flows past the B-star and the ablated B-star material  expands outward, it forms  a spiral structure that  may be interpreted as the corotating interaction regions that have been modeled in previous studies \citep{Ignace_etal2013}.

Considerable orbital-phase dependent variations are observed in the UV spectral line profiles during  1992 and 1995, as already discussed in \citet{St-Louis_etal1995}.  We find that the high velocity state that was identified by these authors corresponds to periastron passage, when the shock-induced line emission is strongest and which occurs in these epochs at the quadrature when the companion is approaching us. We are unable to find a satisfactory explanation for the peculiarly extended P Cygni absorption edges that appear in all the strong lines during this high state. 

A similar high state is absent in the 1988 IUE spectra. All the spectral features of this epoch are nearly identical to the low state spectra of 1992 and 1995, with the exception of the faster wind speed in the latter two epochs as deduced from the saturated portion of resonance P Cyg absorptions.   This suggests that the weaker phase-dependent variability in 1988 may be due to a slower WR wind speed which, in turn, would produce a weaker shock region near the companion. 

The rapidly precessing argument of periastron deduced by SK19 implies the presence of a third object in the system. The systemic velocities obtained from the orbital solutions of the four IUE epochs differ by $\sim$200 km~s$^{-1}$, which would be consistent with an orbit of the WR (+ B-type companion) around the third object.  However, many more observation epochs are needed before this result can be confirmed. 

Although many questions remain unanswered, we consider that the anti-phase RV variations of two emission components and the simultaneous fit to the RVs and the light curve are concrete evidence in favor of the binary nature of EZ Canis Majoris. The assumption that the emission from the shock-heated region traces the orbit of the companion is less certain, given the changing shock strength over the orbital cycle combined with the possible ablation of the companion's envelope,  questions  requiring further investigation.

\begin{acknowledgements}
This paper is based on INES data from the IUE satellite. GK thanks Ken Gayley for helpful discussions and  the Indiana University Astronomy Department for hosting a visit during which much of this paper was completed, and  acknowledges support from CONACYT grant 252499 and UNAM/PAPIIT grant IN103619.
\end{acknowledgements}

\bibliographystyle{aa} 
\bibliography{references} 

\begin{thebibliography}{53}
\expandafter\ifx\csname natexlab\endcsname\relax\def\natexlab#1{#1}\fi

\bibitem[{{Antokhin} {et~al.}(1994){Antokhin}, {Bertrand}, {Lamontagne}, \&
  {Moffat}}]{Antokhin_etal1994}
{Antokhin}, I., {Bertrand}, J.-F., {Lamontagne}, R., \& {Moffat}, A. F.~J.
  1994, \aj, 107, 2179

\bibitem[{{Beer} \& {Podsiadlowski}(2002)}]{2002ASPC..279..253B}
{Beer}, M.~E. \& {Podsiadlowski}, P. 2002, Astronomical Society of the Pacific
  Conference Series, Vol. 279, {Irradiation Effects in Compact Binaries}, ed.
  C.~A. {Tout} \& W.~{van Hamme}, 253

\bibitem[{{Carlos-Leblanc} {et~al.}(2019){Carlos-Leblanc}, {St-Louis},
  {Bjorkman}, \& {Ignace}}]{2019MNRAS.489.2873C}
{Carlos-Leblanc}, D., {St-Louis}, N., {Bjorkman}, J.~E., \& {Ignace}, R. 2019,
  \mnras, 489, 2873

\bibitem[{{Drissen} {et~al.}(1989){Drissen}, {Robert}, {Lamontagne}, {Moffat},
  {St-Louis}, {van Weeren}, \& {van Genderen}}]{1989ApJ...343..426D}
{Drissen}, L., {Robert}, C., {Lamontagne}, R., {et~al.} 1989, \apj, 343, 426

\bibitem[{{Duijsens} {et~al.}(1996){Duijsens}, {van der Hucht}, {van Genderen},
  {Schwarz}, {Linders}, \& {Kolkman}}]{1996A&AS..119...37D}
{Duijsens}, M.~F.~J., {van der Hucht}, K.~A., {van Genderen}, A.~M., {et~al.}
  1996, \aaps, 119, 37

\bibitem[{{Ekstr{\"o}m} {et~al.}(2012){Ekstr{\"o}m}, {Georgy}, {Eggenberger},
  {Meynet}, {Mowlavi}, {Wyttenbach}, {Granada}, {Decressin}, {Hirschi},
  {Frischknecht}, {Charbonnel}, \& {Maeder}}]{Ekstrom_etal2012}
{Ekstr{\"o}m}, S., {Georgy}, C., {Eggenberger}, P., {et~al.} 2012, \aap, 537,
  A146

\bibitem[{{Firmani} {et~al.}(1980){Firmani}, {Koenigsberger}, {Bisiacchi},
  {Moffat}, \& {Isserstedt}}]{Firmani_etal1980}
{Firmani}, C., {Koenigsberger}, G., {Bisiacchi}, G.~F., {Moffat}, A.~F.~J., \&
  {Isserstedt}, J. 1980, \apj, 239, 607

\bibitem[{{Firmani} {et~al.}(1978){Firmani}, {Koenigsberger}, {Bisiacchi},
  {Ruiz}, \& {Solar}}]{1978MmSAI..49..453F}
{Firmani}, C., {Koenigsberger}, G., {Bisiacchi}, G.~F., {Ruiz}, E., \& {Solar},
  A. 1978, \memsai, 49, 453

\bibitem[{{Gayley} {et~al.}(1999){Gayley}, {Owocki}, \&
  {Cranmer}}]{1999ApJ...513..442G}
{Gayley}, K.~G., {Owocki}, S.~P., \& {Cranmer}, S.~R. 1999, \apj, 513, 442

\bibitem[{{Georgiev} {et~al.}(1999){Georgiev}, {Koenigsberger}, {Ivanov},
  {St.-Louis}, \& {Cardona}}]{Georgiev_etal1999}
{Georgiev}, L.~N., {Koenigsberger}, G., {Ivanov}, M.~M., {St.-Louis}, N., \&
  {Cardona}, O. 1999, \aap, 347, 583

\bibitem[{{Guo} {et~al.}(2019){Guo}, {Shporer}, {Hambleton}, \&
  {Isaacson}}]{2019arXiv191108687G}
{Guo}, Z., {Shporer}, A., {Hambleton}, K., \& {Isaacson}, H. 2019, arXiv
  e-prints, arXiv:1911.08687

\bibitem[{{Hamann} {et~al.}(2019){Hamann}, {Gr{\"a}fener}, {Liermann},
  {Hainich}, {Sander}, {Shenar}, {Ramachand ran}, {Todt}, \&
  {Oskinova}}]{Hamann_etal2019}
{Hamann}, W.~R., {Gr{\"a}fener}, G., {Liermann}, A., {et~al.} 2019, \aap, 625,
  A57

\bibitem[{{Huenemoerder} {et~al.}(2015){Huenemoerder}, {Gayley}, {Hamann},
  {Ignace}, {Nichols}, {Oskinova}, {Pollock}, {Schulz}, \&
  {Shenar}}]{Huenemoerder_etal2015}
{Huenemoerder}, D.~P., {Gayley}, K.~G., {Hamann}, W.~R., {et~al.} 2015, \apj,
  815, 29

\bibitem[{{Ignace} {et~al.}(2009){Ignace}, {Bessey}, \&
  {Price}}]{2009MNRAS.395..962I}
{Ignace}, R., {Bessey}, R., \& {Price}, C.~S. 2009, \mnras, 395, 962

\bibitem[{{Ignace} {et~al.}(2013){Ignace}, {Gayley}, {Hamann}, {Huenemoerder},
  {Oskinova}, {Pollock}, \& {McFall}}]{Ignace_etal2013}
{Ignace}, R., {Gayley}, K.~G., {Hamann}, W.~R., {et~al.} 2013, \apj, 775, 29

\bibitem[{{Kuhi}(1967)}]{1967PASP...79...57K}
{Kuhi}, L.~V. 1967, \pasp, 79, 57

\bibitem[{{Kumar} {et~al.}(1995){Kumar}, {Ao}, \&
  {Quataert}}]{1995ApJ...449..294K}
{Kumar}, P., {Ao}, C.~O., \& {Quataert}, E.~J. 1995, \apj, 449, 294

\bibitem[{{Lamberts} {et~al.}(2017){Lamberts}, {Millour}, {Liermann},
  {Dessart}, {Driebe}, {Duvert}, {Finsterle}, {Girault}, {Massi}, {Petrov},
  {Schmutz}, {Weigelt}, \& {Chesneau}}]{Lamberts_etal2017}
{Lamberts}, A., {Millour}, F., {Liermann}, A., {et~al.} 2017, \mnras, 468, 2655

\bibitem[{{Lindgren} {et~al.}(1975){Lindgren}, {Lundstrom}, \&
  {Stenholm}}]{1975A&A....44..219L}
{Lindgren}, H., {Lundstrom}, I., \& {Stenholm}, B. 1975, \aap, 44, 219

\bibitem[{{Massa} {et~al.}(1995{\natexlab{a}}){Massa}, {Fullerton}, {Nichols},
  {Owocki}, {Prinja}, {St-Louis}, {Willis}, {Altner}, {Bolton}, \&
  {Cassinelli}}]{1995ApJ...452L..53M}
{Massa}, D., {Fullerton}, A.~W., {Nichols}, J.~S., {et~al.} 1995{\natexlab{a}},
  \apjl, 452, L53

\bibitem[{{Massa} {et~al.}(1995{\natexlab{b}}){Massa}, {Fullerton}, {Nichols},
  {Owocki}, {Prinja}, {St-Louis}, {Willis}, {Altner}, {Bolton}, {Cassinelli},
  {Cohen}, {Cooper}, {Feldmeier}, {Gayley}, {Harries}, {Heap}, {Henriksen},
  {Howarth}, {Hubeny}, {Kambe}, {Kaper}, {Koenigsberger}, {Marchenko},
  {McCandliss}, {Moffat}, {Nugis}, {Puls}, {Robert}, {Schulte-Ladbeck},
  {Smith}, {Smith}, {Waldron}, \& {White}}]{Massa_etal1995}
{Massa}, D., {Fullerton}, A.~W., {Nichols}, J.~S., {et~al.} 1995{\natexlab{b}},
  \apjl, 452, L53

\bibitem[{{Moffat} {et~al.}(2018){Moffat}, {St-Louis}, {Carlos-Leblanc},
  {Richardson}, {Pablo}, \& {Ramiaramanantsoa}}]{Moffat2018}
{Moffat}, A.~F.~J., {St-Louis}, N., {Carlos-Leblanc}, D., {et~al.} 2018, in 3rd
  BRITE Science Conference, ed. G.~A. {Wade}, D.~{Baade}, J.~A. {Guzik}, \&
  R.~{Smolec}, Vol.~8, 37--42

\bibitem[{{Morel} {et~al.}(1997){Morel}, {St-Louis}, \&
  {Marchenko}}]{Morel_etal1997}
{Morel}, T., {St-Louis}, N., \& {Marchenko}, S.~V. 1997, \apj, 482, 470

\bibitem[{{Moreno} \& {Koenigsberger}(1999)}]{1999RMxAA..35..157M}
{Moreno}, E. \& {Koenigsberger}, G. 1999, \rmxaa, 35, 157

\bibitem[{{Mullan}(1984)}]{1984ApJ...283..303M}
{Mullan}, D.~J. 1984, \apj, 283, 303

\bibitem[{{Oskinova} {et~al.}(2012){Oskinova}, {Gayley}, {Hamann},
  {Huenemoerder}, {Ignace}, \& {Pollock}}]{Oskinova_etal2012}
{Oskinova}, L.~M., {Gayley}, K.~G., {Hamann}, W.~R., {et~al.} 2012, \apjl, 747,
  L25

\bibitem[{{Parkin} \& {Pittard}(2008)}]{2008MNRAS.388.1047P}
{Parkin}, E.~R. \& {Pittard}, J.~M. 2008, \mnras, 388, 1047

\bibitem[{{Parkin} {et~al.}(2009){Parkin}, {Pittard}, {Corcoran}, {Hamaguchi},
  \& {Stevens}}]{2009MNRAS.394.1758P}
{Parkin}, E.~R., {Pittard}, J.~M., {Corcoran}, M.~F., {Hamaguchi}, K., \&
  {Stevens}, I.~R. 2009, \mnras, 394, 1758

\bibitem[{{Phillips} \& {Podsiadlowski}(2002)}]{2002MNRAS.337..431P}
{Phillips}, S.~N. \& {Podsiadlowski}, P. 2002, \mnras, 337, 431

\bibitem[{{Podsiadlowski}(1991)}]{1991Natur.350..136P}
{Podsiadlowski}, P. 1991, \nat, 350, 136

\bibitem[{{Rate} \& {Crowther}(2020)}]{2020MNRAS.tmp...34R}
{Rate}, G. \& {Crowther}, P.~A. 2020, \mnras [\eprint[arXiv]{1912.10125}]

\bibitem[{{Richardson} {et~al.}(2017){Richardson}, {Russell}, {St-Jean},
  {Moffat}, {St-Louis}, {Shenar}, {Pablo}, {Hill}, {Ramiaramanantsoa},
  {Corcoran}, {Hamuguchi}, {Eversberg}, {Miszalski}, {Chen{\'e}}, {Waldron},
  {Kotze}, {Kotze}, {Luckas}, {Cacella}, {Heathcote}, {Powles}, {Bohlsen},
  {Locke}, {Handler}, {Kuschnig}, {Pigulski}, {Popowicz}, {Wade}, \&
  {Weiss}}]{Richardson_etal2017}
{Richardson}, N.~D., {Russell}, C.~M.~P., {St-Jean}, L., {et~al.} 2017, \mnras,
  471, 2715

\bibitem[{{Robert} {et~al.}(1992){Robert}, {Moffat}, {Drissen}, {Lamontagne},
  {Seggewiss}, {Niemela}, {Cerruti}, {Barrett}, {Bailey}, {Garcia}, \&
  {Tapia}}]{1992ApJ...397..277R}
{Robert}, C., {Moffat}, A. F.~J., {Drissen}, L., {et~al.} 1992, \apj, 397, 277

\bibitem[{{Ross}(1961)}]{Ross1961}
{Ross}, L.~W. 1961, \pasp, 73, 354

\bibitem[{{Ruderman} {et~al.}(1989){Ruderman}, {Shaham}, \&
  {Tavani}}]{1989ApJ...336..507R}
{Ruderman}, M., {Shaham}, J., \& {Tavani}, M. 1989, \apj, 336, 507

\bibitem[{{Ruffert}(1994)}]{Ruffert1994}
{Ruffert}, M. 1994, \aaps, 106, 505

\bibitem[{{Schmidt}(1974)}]{1974PASP...86..767S}
{Schmidt}, G.~D. 1974, \pasp, 86, 767

\bibitem[{{Schmutz}(1997)}]{Schmutz1997}
{Schmutz}, W. 1997, \aap, 321, 268

\bibitem[{{Schmutz} \& {Koenigsberger}(2019)}]{SK19}
{Schmutz}, W. \& {Koenigsberger}, G. 2019, \aap, 624, L3

\bibitem[{{Silaj} {et~al.}(2014){Silaj}, {Jones}, {Sigut}, \&
  {Tycner}}]{2014ApJ...795...82S}
{Silaj}, J., {Jones}, C.~E., {Sigut}, T.~A.~A., \& {Tycner}, C. 2014, \apj,
  795, 82

\bibitem[{{Skinner} {et~al.}(2002){Skinner}, {Zhekov}, {G{\"u}del}, \&
  {Schmutz}}]{Skinner_etal2002}
{Skinner}, S.~L., {Zhekov}, S.~A., {G{\"u}del}, M., \& {Schmutz}, W. 2002,
  \apj, 579, 764

\bibitem[{{St. -Louis} {et~al.}(1993){St. -Louis}, {Howarth}, {Willis},
  {Stickland }, {Smith}, {Conti}, \& {Garmany}}]{1993A&A...267..447S}
{St. -Louis}, N., {Howarth}, I.~D., {Willis}, A.~J., {et~al.} 1993, \aap, 267,
  447

\bibitem[{{St-Louis} {et~al.}(1995{\natexlab{a}}){St-Louis}, {Dalton},
  {Marchenko}, {Moffat}, \& {Willis}}]{1995ApJ...452L..57S}
{St-Louis}, N., {Dalton}, M.~J., {Marchenko}, S.~V., {Moffat}, A.~F.~J., \&
  {Willis}, A.~J. 1995{\natexlab{a}}, \apj, 452, L57

\bibitem[{{St-Louis} {et~al.}(1995{\natexlab{b}}){St-Louis}, {Dalton},
  {Marchenko}, {Moffat}, \& {Willis}}]{St-Louis_etal1995}
{St-Louis}, N., {Dalton}, M.~J., {Marchenko}, S.~V., {Moffat}, A.~F.~J., \&
  {Willis}, A.~J. 1995{\natexlab{b}}, \apjl, 452, L57

\bibitem[{{Stevens} {et~al.}(1992){Stevens}, {Blondin}, \&
  {Pollock}}]{Stevens_etal1992}
{Stevens}, I.~R., {Blondin}, J.~M., \& {Pollock}, A.~M.~T. 1992, \apj, 386, 265

\bibitem[{{Toal{\'a}} {et~al.}(2012){Toal{\'a}}, {Guerrero}, {Chu}, {Gruendl},
  {Arthur}, {Smith}, \& {Snowden}}]{2012ApJ...755...77T}
{Toal{\'a}}, J.~A., {Guerrero}, M.~A., {Chu}, Y.~H., {et~al.} 2012, \apj, 755,
  77

\bibitem[{{Townsend}(2007)}]{2007AIPC..948..345T}
{Townsend}, R. 2007, in American Institute of Physics Conference Series, Vol.
  948, Unsolved Problems in Stellar Physics: A Conference in Honor of Douglas
  Gough, ed. R.~J. {Stancliffe}, G.~{Houdek}, R.~G. {Martin}, \& C.~A. {Tout},
  345--356

\bibitem[{{van den Heuvel}(1976)}]{1976IAUS...73...35V}
{van den Heuvel}, E.~P.~J. 1976, in IAU Symposium, Vol.~73, Structure and
  Evolution of Close Binary Systems, ed. P.~{Eggleton}, S.~{Mitton}, \&
  J.~{Whelan}, 35

\bibitem[{{van der Hucht}(2001)}]{vanderHucht2001}
{van der Hucht}, K.~A. 2001, \nar, 45, 135

\bibitem[{{Welsh} {et~al.}(2011){Welsh}, {Orosz}, {Aerts}, {Brown},
  {Brugamyer}, {Cochran}, {Gilliland}, {Guzik}, {Kurtz}, {Latham}, {Marcy},
  {Quinn}, {Zima}, {Allen}, {Batalha}, {Bryson}, {Buchhave}, {Caldwell},
  {Gautier}, {Howell}, {Kinemuchi}, {Ibrahim}, {Isaacson}, {Jenkins}, {Prsa},
  {Still}, {Street}, {Wohler}, {Koch}, \& {Borucki}}]{2011ApJS..197....4W}
{Welsh}, W.~F., {Orosz}, J.~A., {Aerts}, C., {et~al.} 2011, \apjs, 197, 4

\bibitem[{{Willis} {et~al.}(1989){Willis}, {Howarth}, {Smith}, {Garmany}, \&
  {Conti}}]{Willis_etal1989}
{Willis}, A.~J., {Howarth}, I.~D., {Smith}, L.~J., {Garmany}, C.~D., \&
  {Conti}, P.~S. 1989, \aaps, 77, 269

\bibitem[{{Willis} {et~al.}(1994){Willis}, {Schild}, {Howarth}, \&
  {Stevens}}]{1994Ap&SS.221..321W}
{Willis}, A.~J., {Schild}, H., {Howarth}, I.~D., \& {Stevens}, I.~R. 1994,
  \apss, 221, 321

\bibitem[{{Wilson}(1948)}]{1948PASP...60..383W}
{Wilson}, O.~C. 1948, \pasp, 60, 383

\end{thebibliography}

\appendix

\section{Complementary information}

\begin{sidewaystable*}[htb]
\caption{Average flux and velocity values for each IUE wavelength band measured for epochs 1983, 1988, 1992, and 1995. \label{table_UVaverages_stdevs}}
\begin{tabular}{rllrrrrrrrrrrrrrrrr}
\hline
\hline
    &                      &   &\multicolumn{4}{c}{|-------------  1983  -------------|} &\multicolumn{4}{c}{|-------------  1988  -------------|} &\multicolumn{4}{c}{|-------------  1992  -------------|} &\multicolumn{4}{c}{|-------------  1995 -------------|}   \\
$\lambda_0$ &~~~$\lambda_1$~~-~~$\lambda_2$&ID &  v   & sd & flux    & sd &   v   & sdv & flux & sdf & v   & sdv & flux    & sdf & v   & sdv & flux    & sdf    \\
\hline
1197.34& 1194.5-1199.6&  SVe      &  -45 &     9 &  2.441&  0.048 &  -39&   8&  2.503  &0.082&  -48& 12& 2.445 & 0.166&  -28&   21&  2.092&  0.356 \\ 
1197.34& 1195.4-1199.6& SVn       &   38 &     6 &  2.119&  0.044 &   42&   5&  2.179  &0.072&   40&  6& 2.101 & 0.132&   55&   14&  1.818&  0.300 \\ 
1226.01& 1217.6-1231.6& 1226e$^b$ & -172 &    12 &  3.143&  0.085 & -178&  36&  3.222  &0.136& -331& 42& 2.583 & 0.201& -398&   46&  1.960&  0.361 \\ 
1227.80& 1225.6-1230.0& 1226n$^c$ &   15 &     8 &  1.427&  0.033 &   16&  16&  1.448  &0.103&  -66& 30& 1.043 & 0.117& -107&   23&  0.726&  0.145 \\ 
1242.82& 1237.0-1248.5&  NVe     &  115 &     8 & 10.640&  0.158 &  118&   8& 11.120  &0.302& 106 & 11&10.878 & 0.545&  118&   16& 10.478&  1.040 \\ 
1258.30& 1257.2-1259.6&  1258e   &    34 &     3 &  1.293&  0.031 &  33&   3&  1.286  &0.041&   32&  4& 1.208 & 0.059&   33&    5&  1.133&  0.089 \\ 
1269.73& 1267.6-1272.3& 1270e$^d$ &   47 &     6 &  1.706&  0.043 &   50&   8&  1.710  &0.548&   60& 11& 1.703 & 0.113&   66&   13&  1.652&  0.149 \\ 
1269.10& 1267.6-1271.1&1270n$^e$  &   84 &     3 &  1.339&  0.035 &   84&   3&  1.333  &0.038&   80&  7& 1.287 & 0.064&   84&    7&  1.238&  0.097 \\ 
1371.30& 1369.0-1381.0&   1371e$^a$&  898 &     7 &  4.508&  0.095 &  911&   5&  4.547  &0.118&  869& 28& 4.695 & 0.271&  867&   32&  4.685&  0.259 \\ 
1486.50& 1478.0-1497.0& NIV]     &  150 &     7 &  7.374&  0.097 &  149&   5&  7.445  &0.122&  139&  9& 7.884 & 0.226&  148&   12&  8.034&  0.245 \\ 
1550.85& 1545.4-1560.5& CIVe     &  408 &     8 &  5.640&  0.074 &  397&  15&  5.641  &0.112&  386& 11& 6.041 & 0.203&  385&    9&  6.084&  0.214 \\ 
1550.85& 1548.2-1554.2&  CIVn    &  136 &     4 &  2.922&  0.042 &  133&   5&  2.969  &0.060&  128&  4& 3.115 & 0.109&  133&    5&  3.169&  0.091 \\ 
1640.42& 1630.5-1633.5& HeIIa    &-1546 &     7 &  0.275&  0.024 &-1548&  11&  0.264  &0.018&-1499& 22& 0.355 & 0.091&-1503&   30&  0.395&  0.097 \\ 
1640.42& 1644.4-1655.0& HeIIr$^h$& 1387 &     8 &  7.422&  0.169 & 1391&   9&  7.325  &0.239& 1404&  9& 7.813 & 0.317& 1392&   13&  7.654&  0.345 \\ 
1640.42& 1633.0-1637.3& HeIIb$^j$& -824 &     3 &  2.834&  0.101 & -822&   3&  2.920  &0.087& -847& 20& 3.371 & 0.243& -852&   21&  3.176&  0.243 \\ 
1640.42& 1634.2- 1651.8&HeIIe    &  319 &     7 & 20.947&  0.314 &  303&   9& 21.474  &0.467& 305 &  5&22.161 & 0.437&  303&   10& 22.230&  0.530 \\ 
1718.55& 1709.0-1713.0& NIVa     &-1279 &    14 &  0.199&  0.013 &-1288&  17&  0.175  &0.016&-1255& 19& 0.238 & 0.084&-1265&   26&  0.228&  0.085 \\ 
1718.55& 1711.5-1736.0& NIVe     &  642 &     6 &  6.388&  0.080 &  640&   6&  6.491  &0.109&  635&  9& 6.850 & 0.248&  630&   10&  6.792&  0.245 \\ 
1718.55& 1714.0-1728.0& NIVn     &  328 &     6 &  4.974&  0.063 &  330&   5&  5.113  &0.081&  338&  6& 5.351 & 0.148&  344&    6&  5.361&  0.135 \\ 
1718.55& 1715.0-1723.5& NIVnn     &  123 &     6 &  3.575&  0.046 &  122&   4&  3.696  &0.072&  115&  5& 3.819 & 0.113&  125&    6&  3.863&  0.087 \\ 
1270.00& 1250.0-1290.0&FeVIpsudo  & -153 &    11 & 16.200&  0.359 & -152&  19& 16.439  &0.461& -106& 28&16.250 & 0.941&  -83&   30& 15.980&  1.103 \\ 
1450.00& 1430.0-1470.0&FeVpsudo   &  264 &    13 &  8.851&  0.178 &  241&  11&  8.676  &0.256&  293& 26& 9.198 & 0.541&  307&   28&  9.139&  0.552 \\ 
 ----  & 1594.1-1610.0&``cont''  &  --- &   --- &  2.318&  0.042 & ----&----&  2.266  &0.041& ----&-- & 2.440 & 0.098&---- &---- &  2.362&  0.107   \\ 
 ----  & 1828.3-1845.5&``cont''  &  --- &   --- &  1.421&  0.026 & ----&----&  1.442  &0.019&---- &---& 1.600 & 0.067&---- &---- &  1.611&  0.067 \\ 
 ----  & 1750.0-1900.0&``cont''  &  --- &   --- & 13.685&  0.212 & ----&----& 13.649  &0.173&---- &---& 14.999& 0.640&---- &---- & 14.980&  0.647 \\ 
\hline
\hline
\end{tabular}
\tablefoot{
The reference wavelength in column 1  is given in \AA\ and corresponds to the laboratory wavelength for the ion listed in column 3 or to the midpoint of the limits of the continuum bands. Average velocities and corresponding standard deviations (sdv) are in units of km\,s$^{-1}$. Average integrated fluxes over the wavelength band listed in column 2 and corresponding standard deviations (sdf) are in units of 10$^{-9}$\,erg\,cm$^{-1}$\,s$^{-1}$.  Fluxes are not corrected for reddening. ``cont'' indicates possible continuum. 
\\
\tablefoottext{a}{The contribution of Fe V lines dominates over the O V contribution, as evidenced by the red shift of the centroid.}
\tablefoottext{b}{FeVI 1222.82, 1223.97}          
\tablefoottext{c}{FeVI 1228.61, 1228.96; in 1995 it is severely affected by NV P Cyg}
\tablefoottext{d}{FeV 1269.78, NIV 1270.27, 1272.16, FeVI 1271.10, 1272.07}
\tablefoottext{e}{FeV 1269.78, NIV 1270.27, FeVI 1271.10}
\tablefoottext{f}{FeV1 pseudo-continuum composed of densely packed Fe VI lines}
\tablefoottext{g}{FeV pseudo-continuum composed of densely packed Fe V lines}
\tablefoottext{h}{HeIIr is the red wing of the HeII 1640 emission line}
\tablefoottext{h}{HeIIb is the blue wing of the HeII 1640 emission line}
}
\end{sidewaystable*}

\begin{figure*}
\centering
\includegraphics[width=0.19\linewidth]{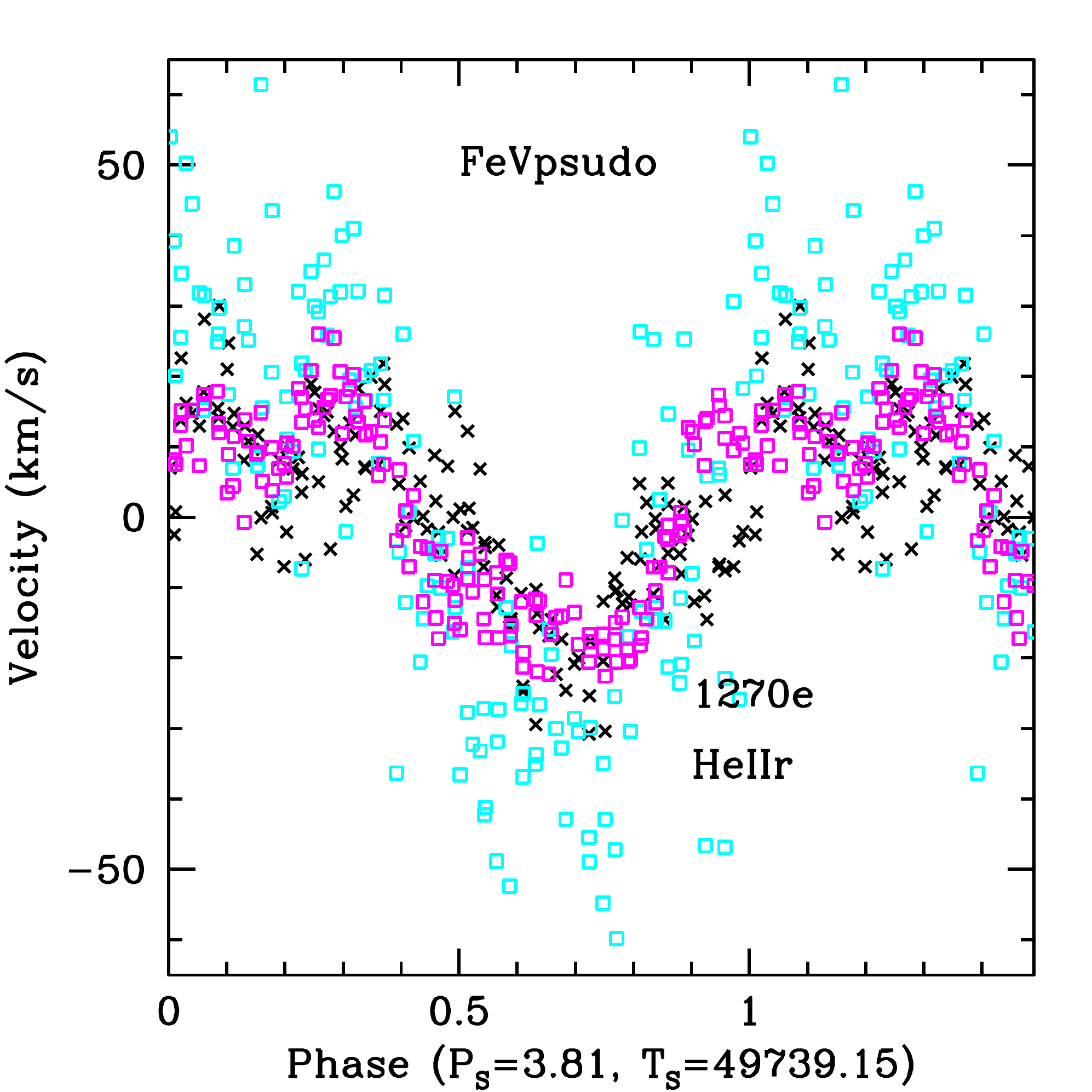}
\includegraphics[width=0.19\linewidth]{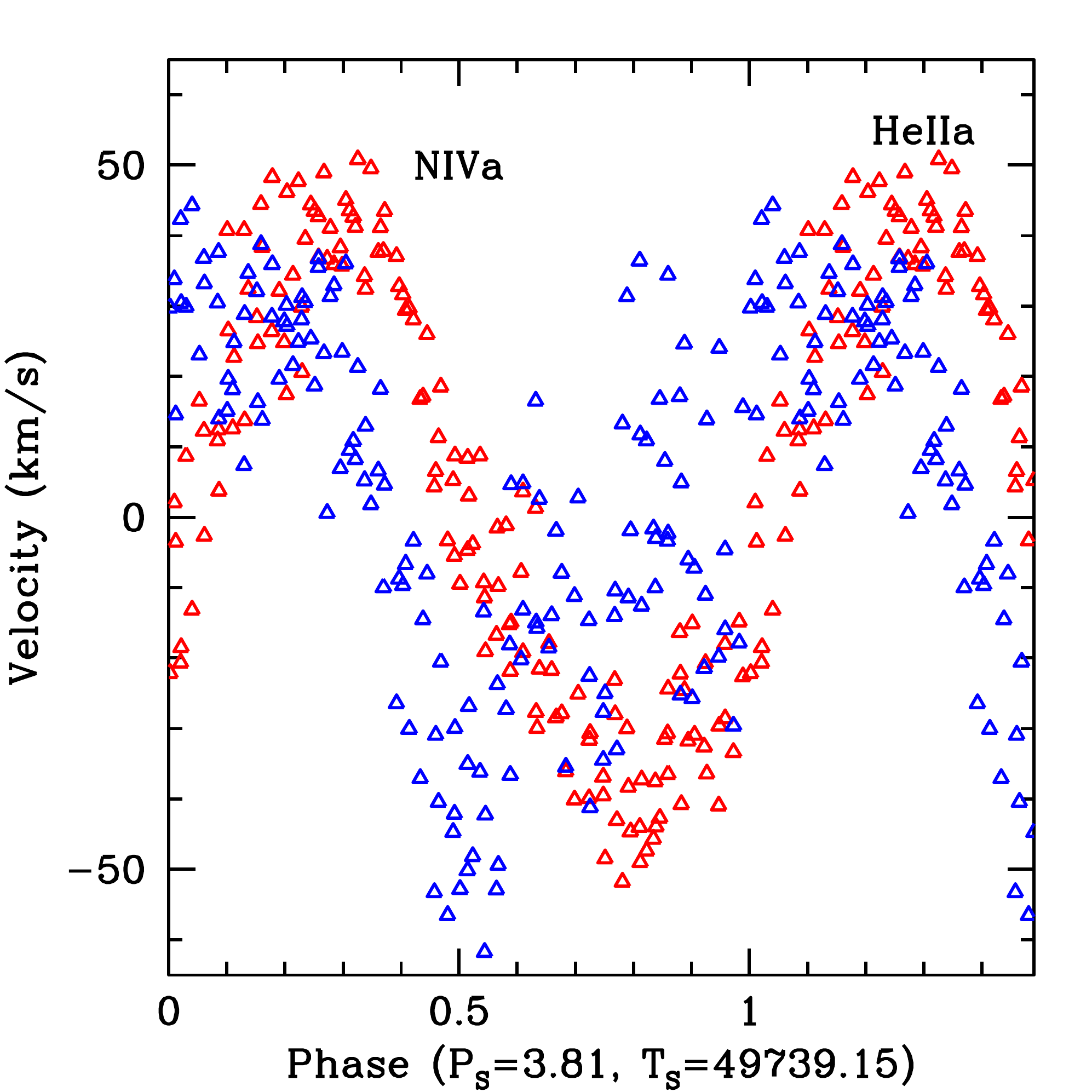}
\includegraphics[width=0.19\linewidth]{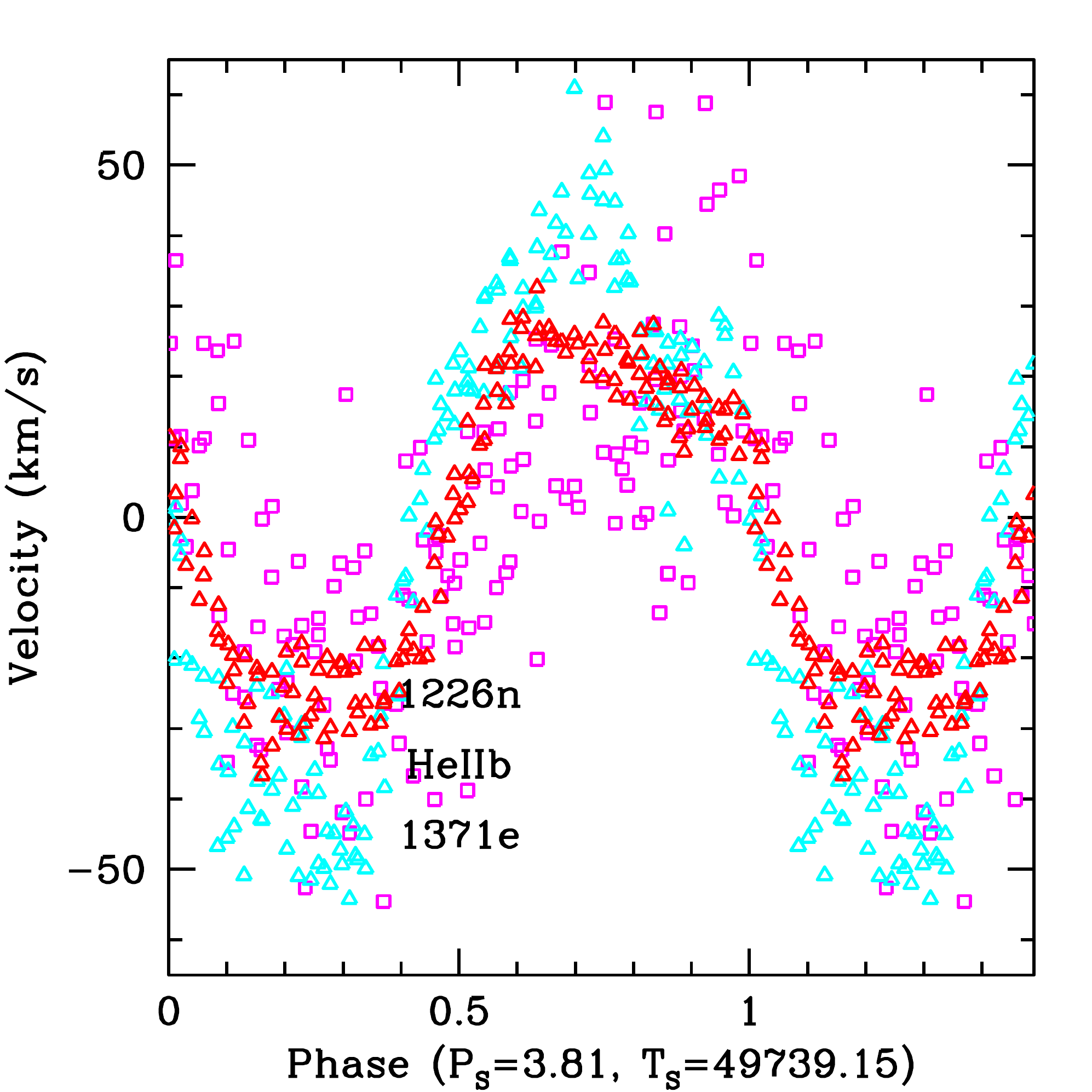}
\includegraphics[width=0.19\linewidth]{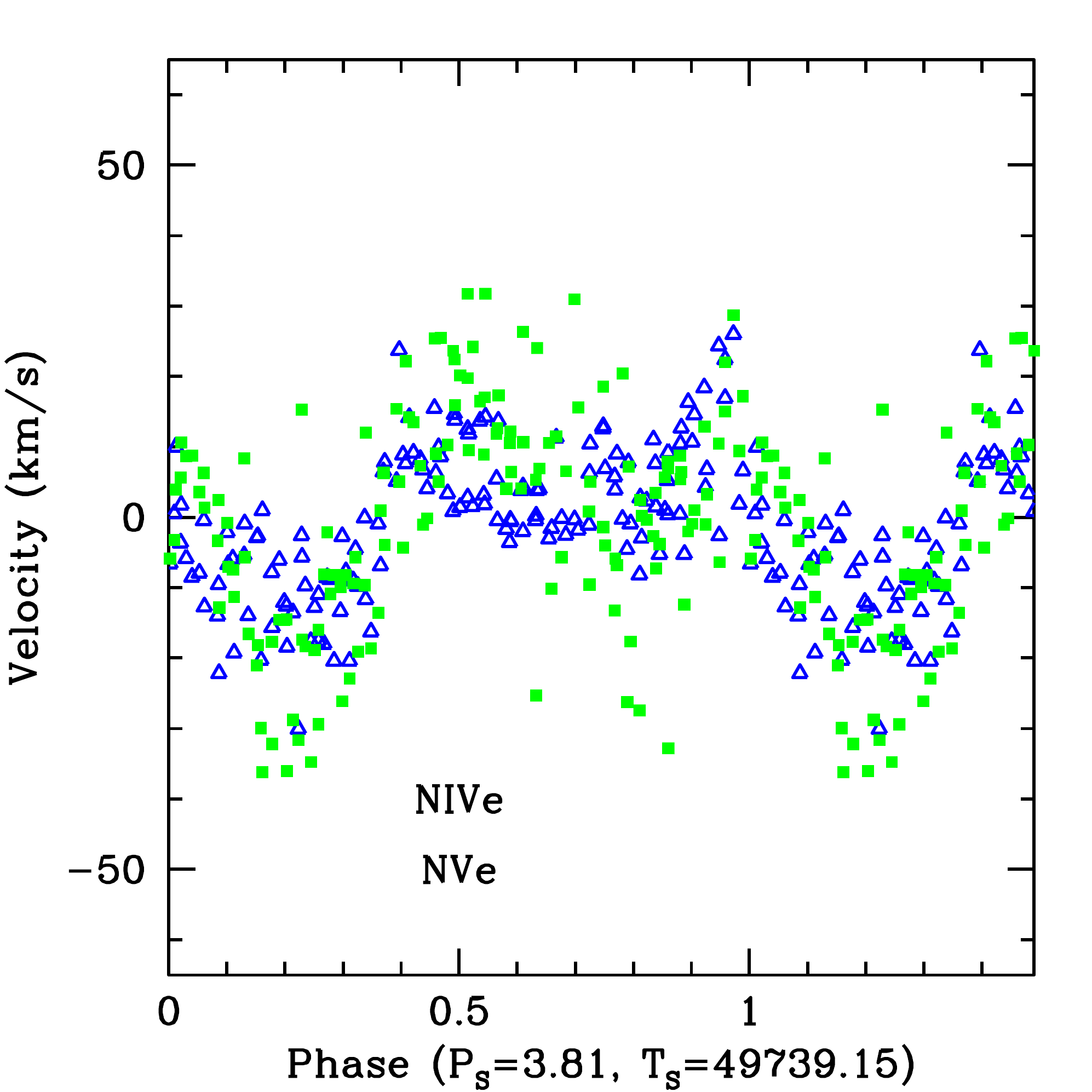}
\includegraphics[width=0.19\linewidth]{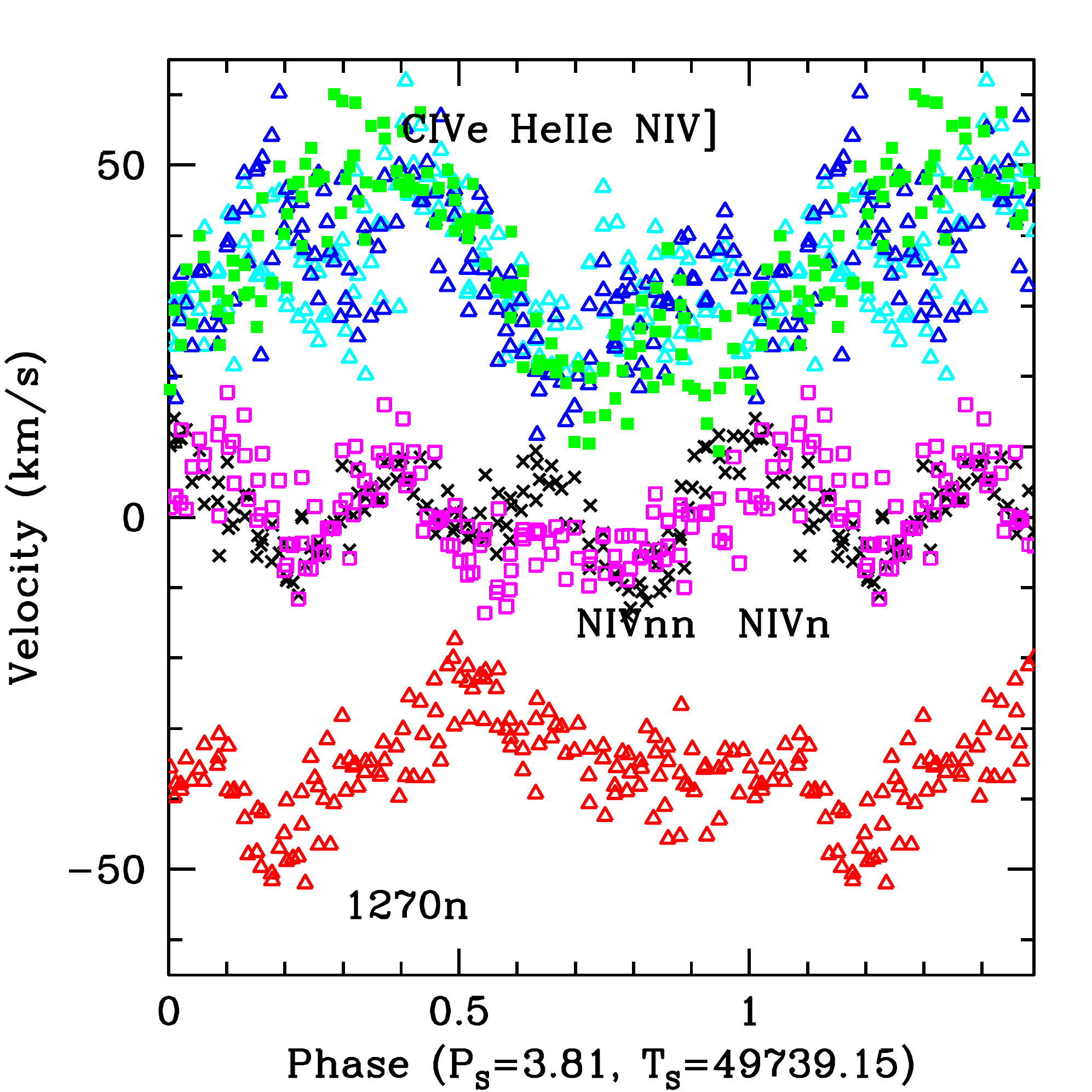}
\includegraphics[width=0.19\linewidth]{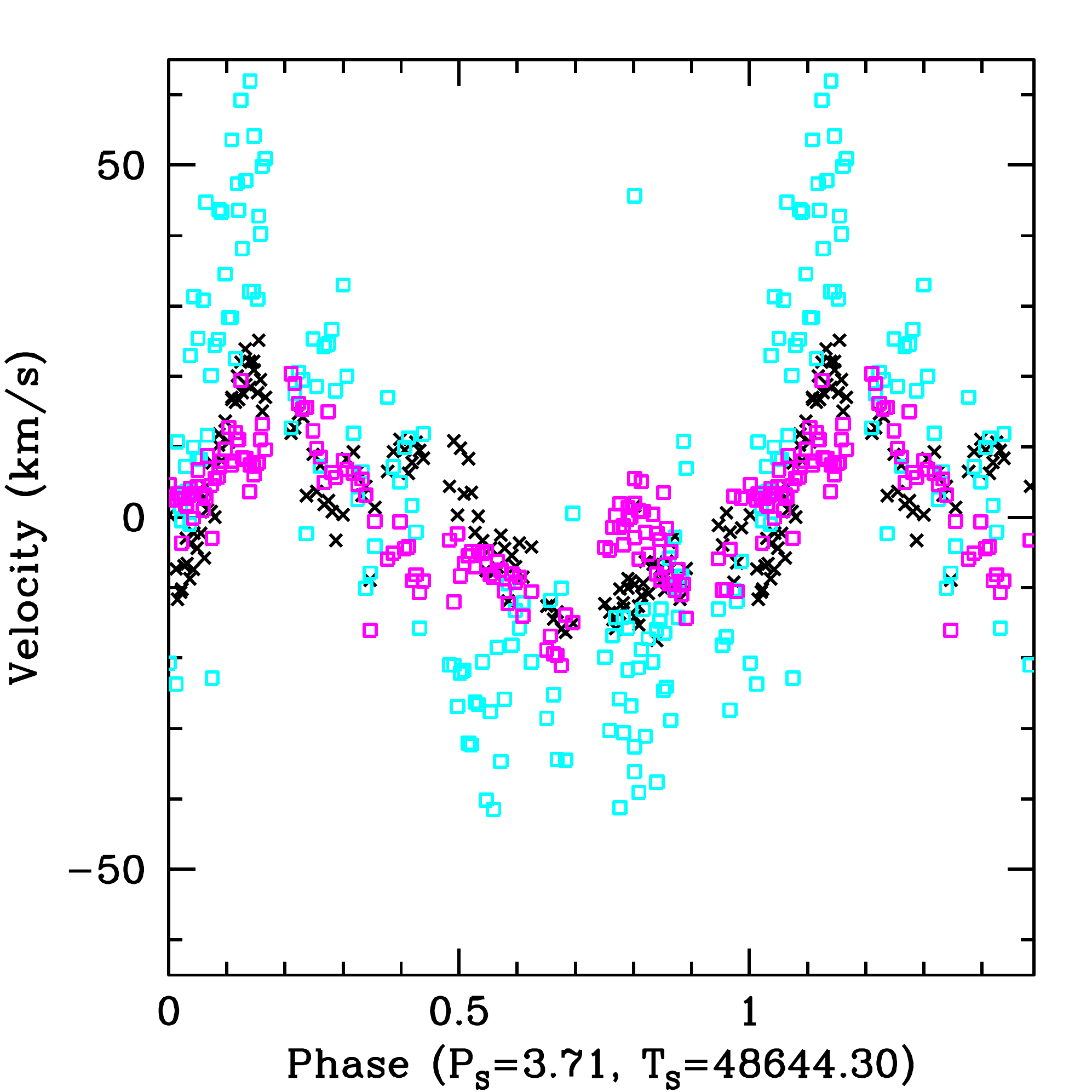}
\includegraphics[width=0.19\linewidth]{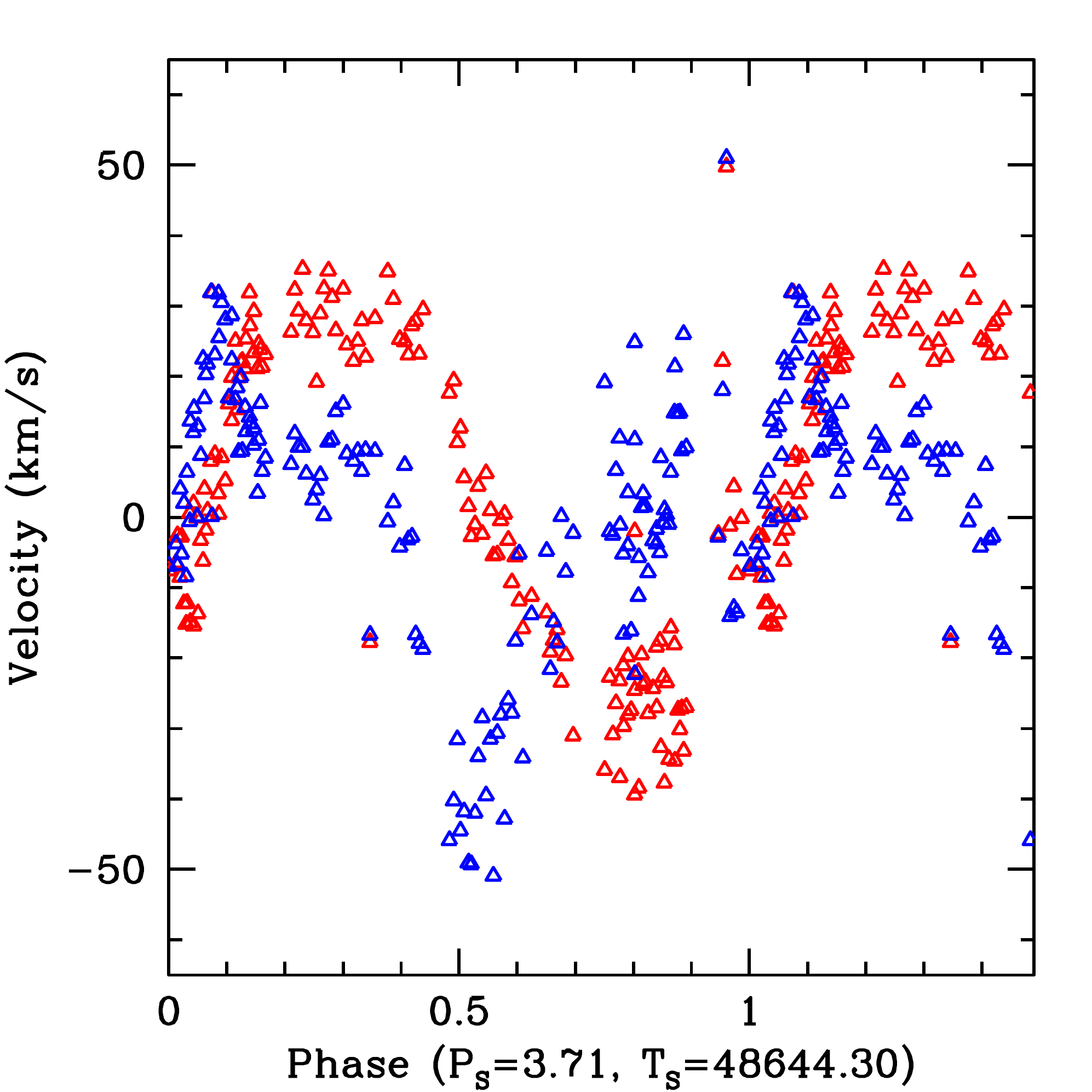}
\includegraphics[width=0.19\linewidth]{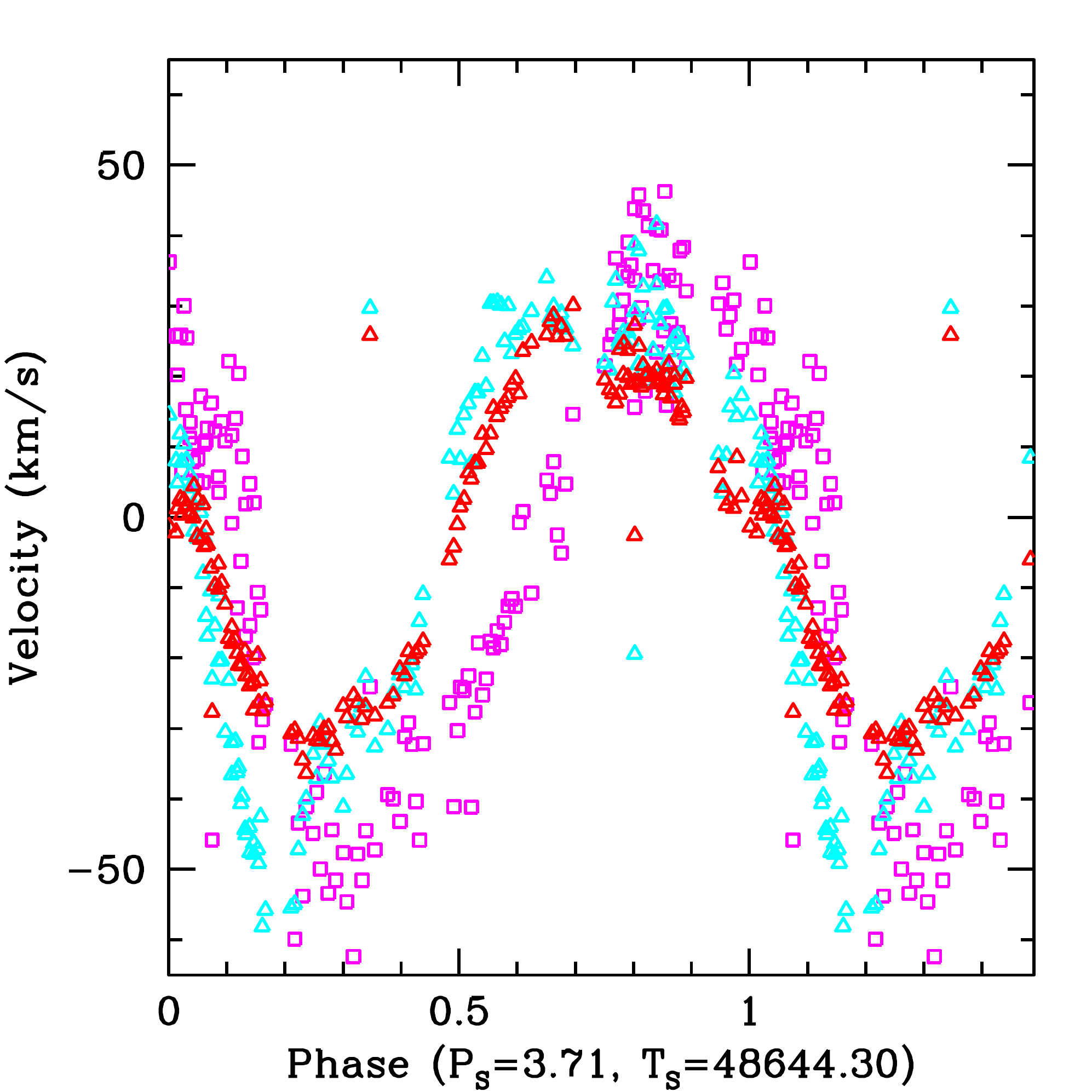}
\includegraphics[width=0.19\linewidth]{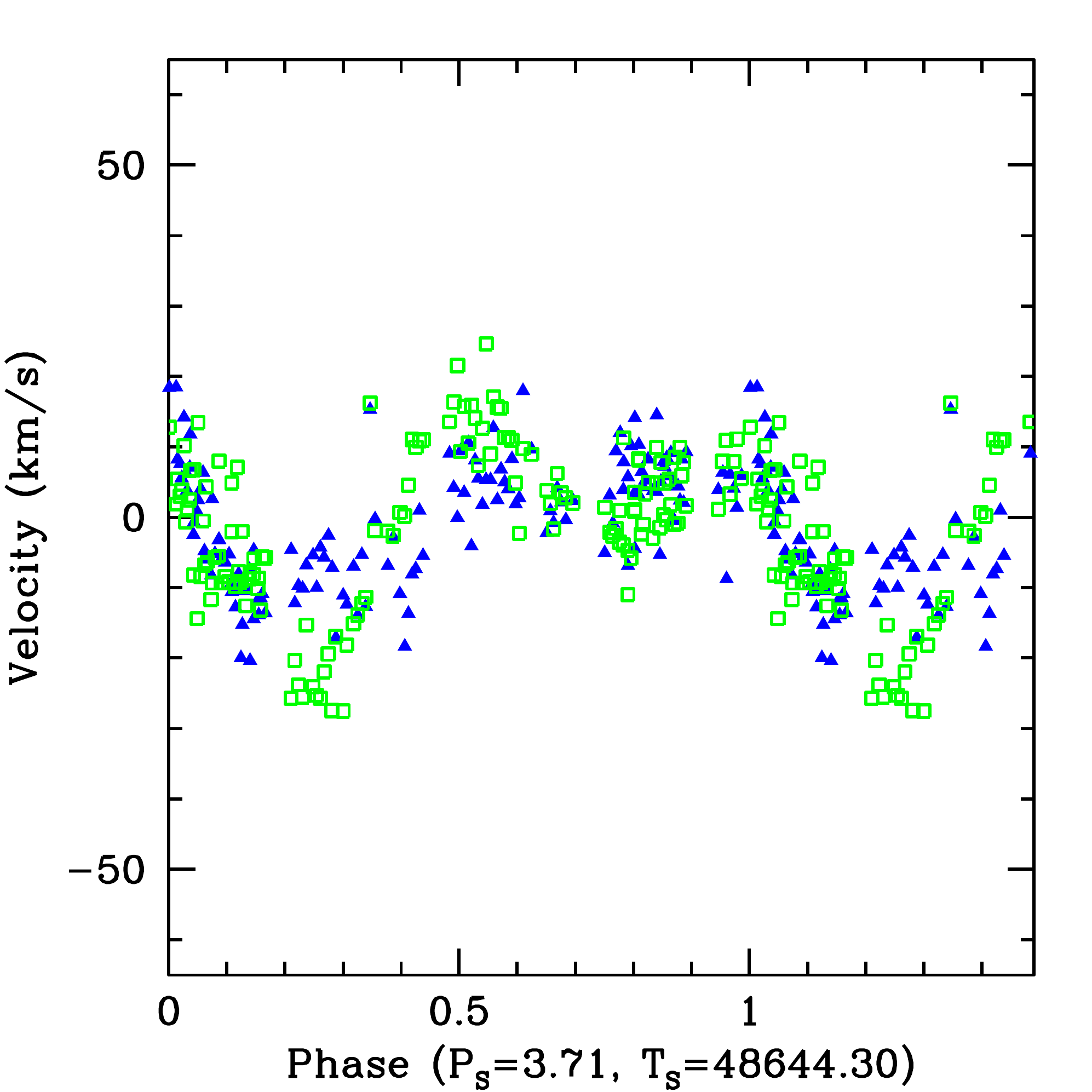}
\includegraphics[width=0.19\linewidth]{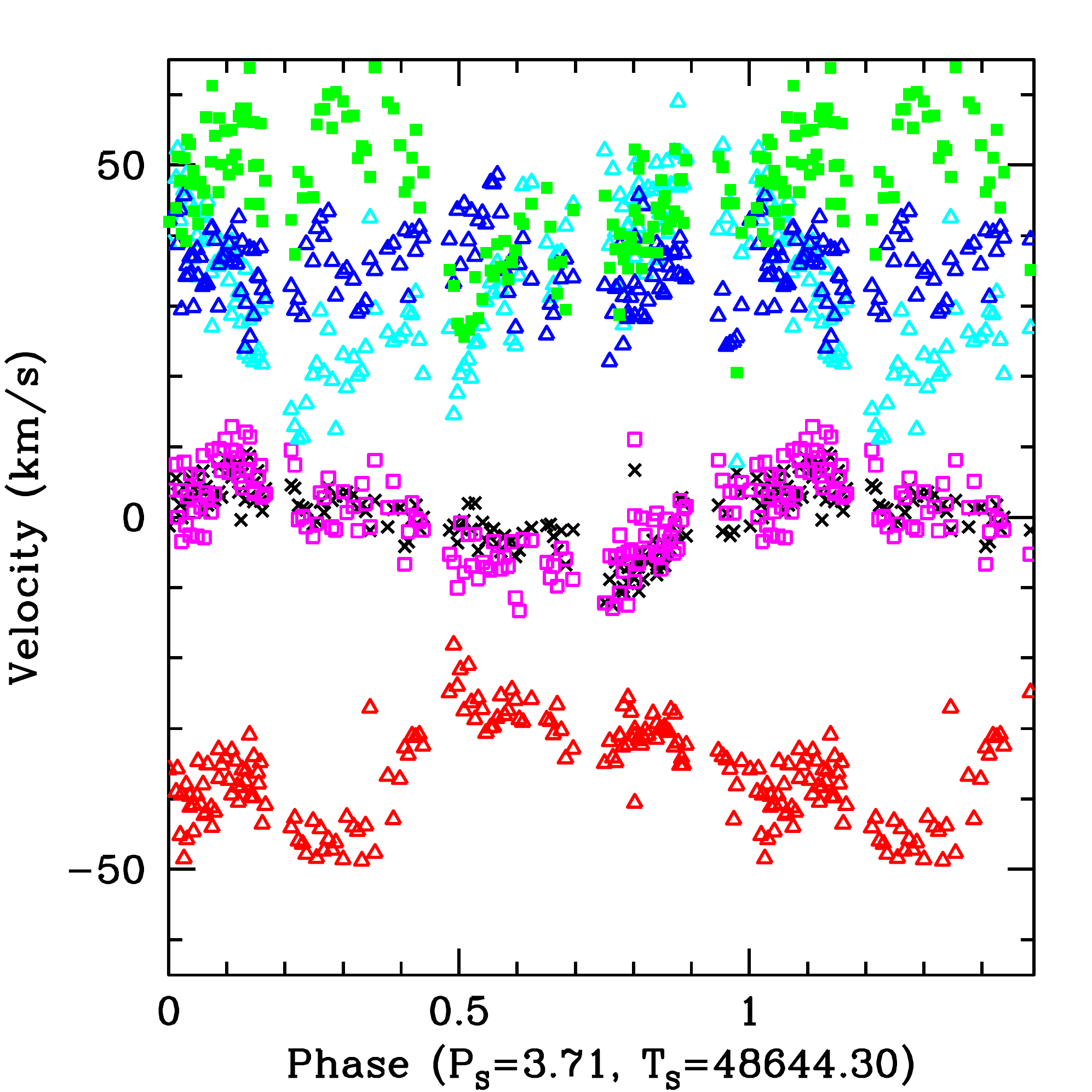}
\includegraphics[width=0.19\linewidth]{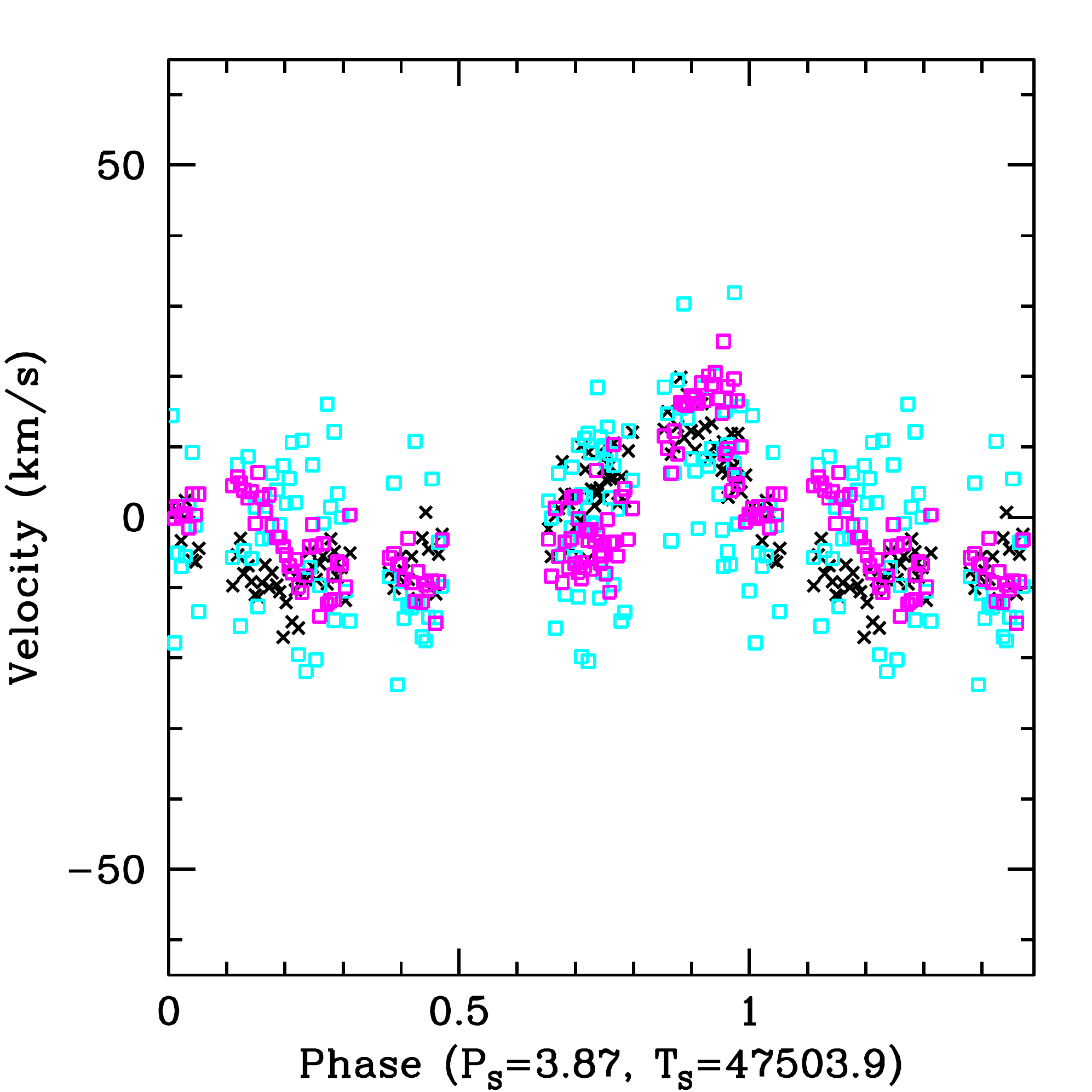}
\includegraphics[width=0.19\linewidth]{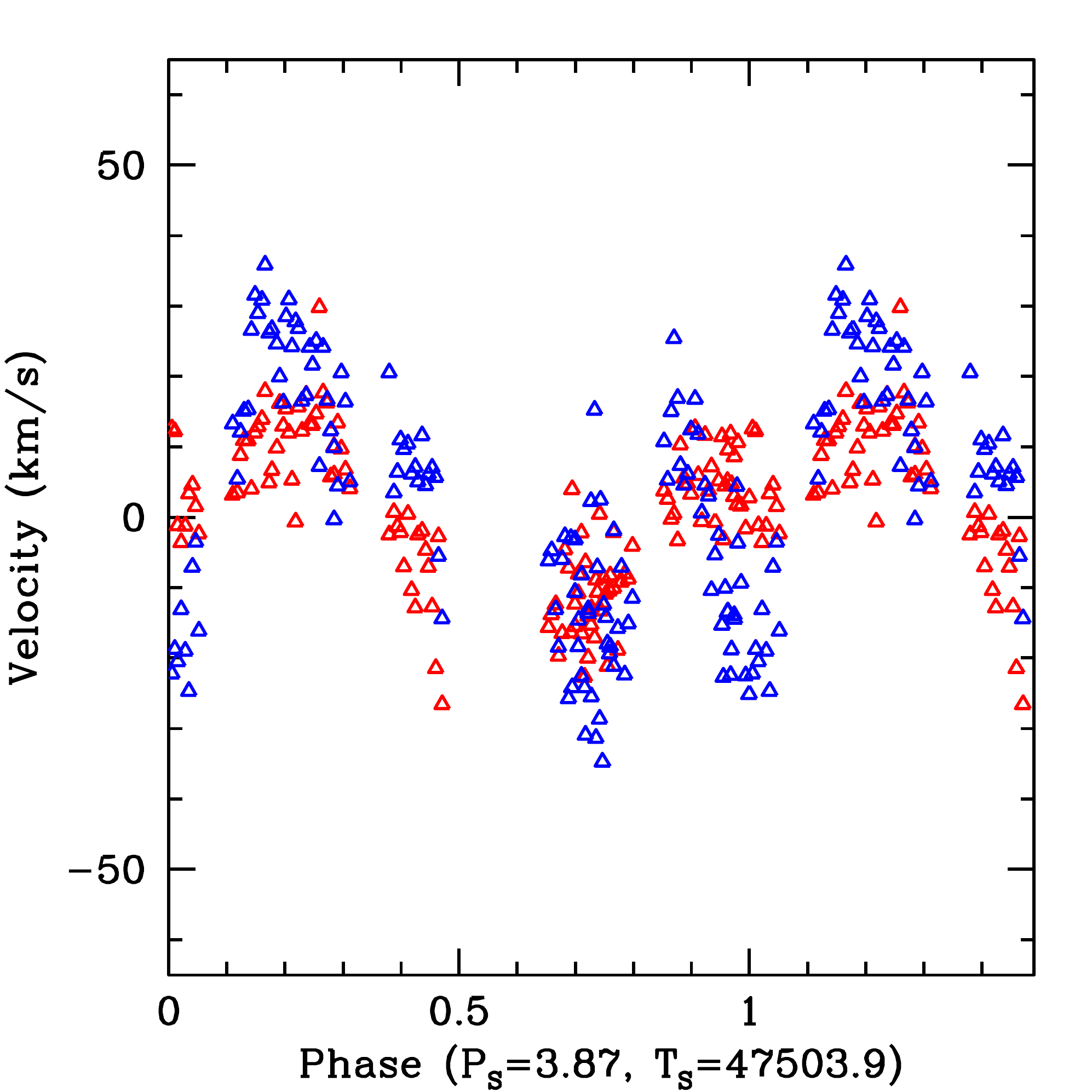}
\includegraphics[width=0.19\linewidth]{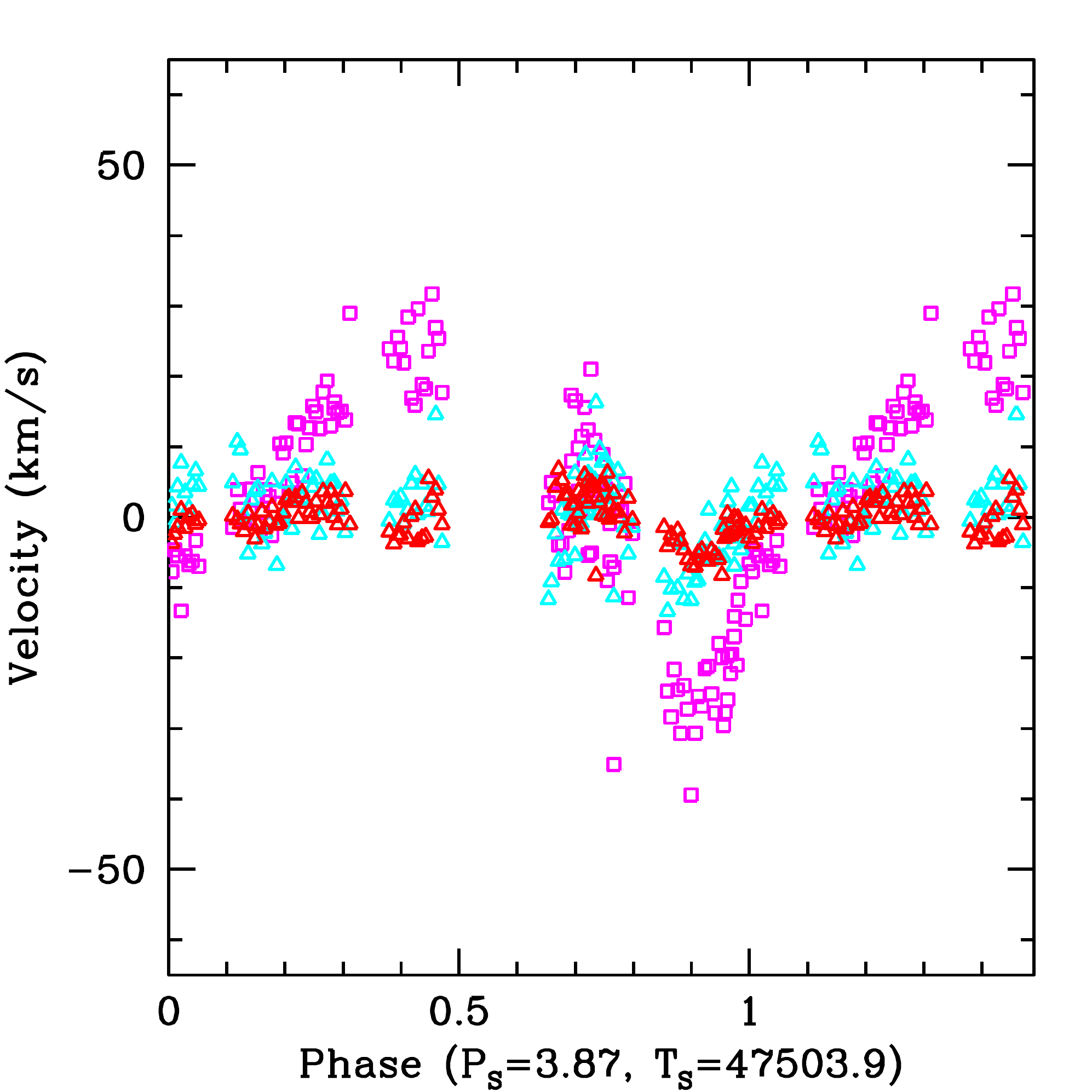}
\includegraphics[width=0.19\linewidth]{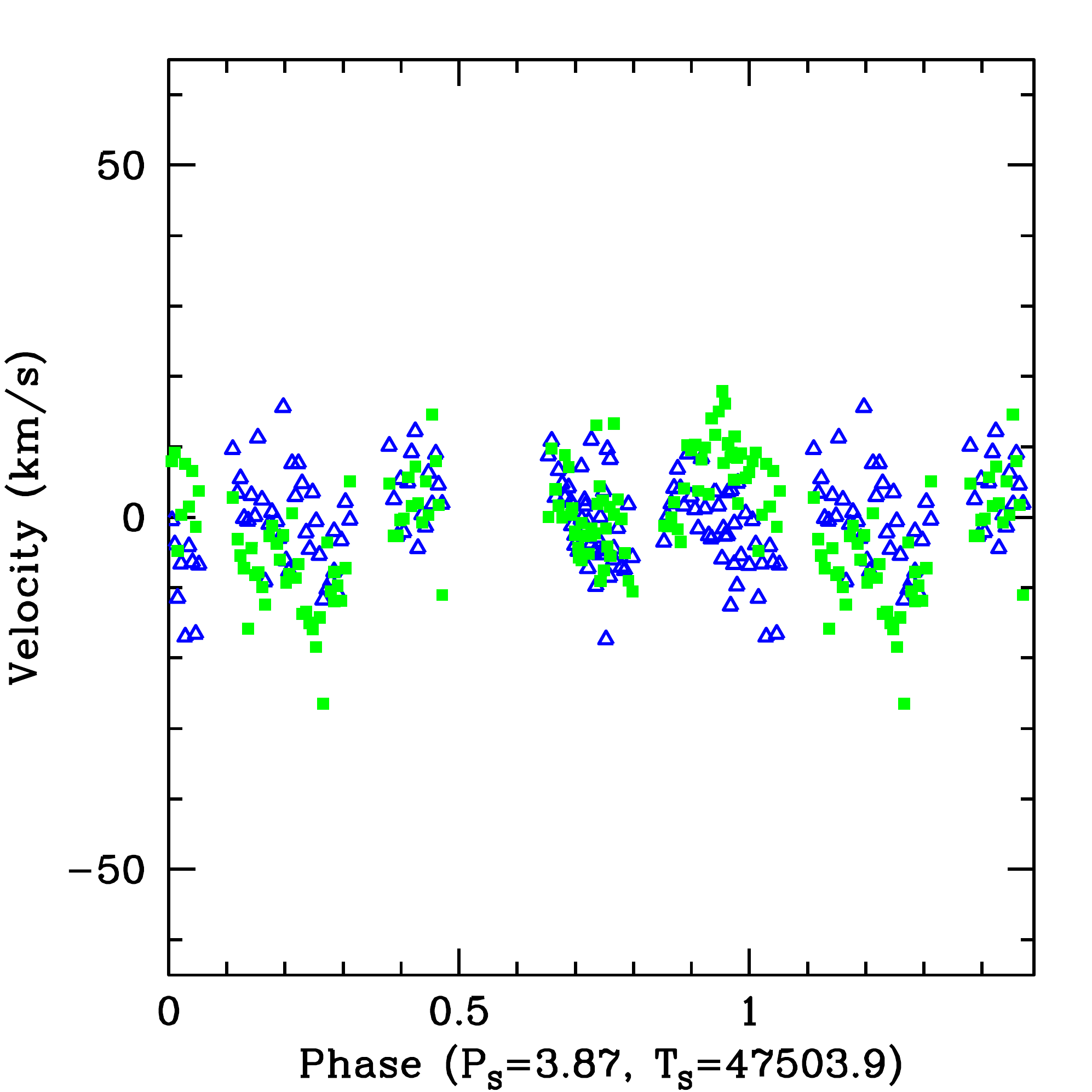}
\includegraphics[width=0.19\linewidth]{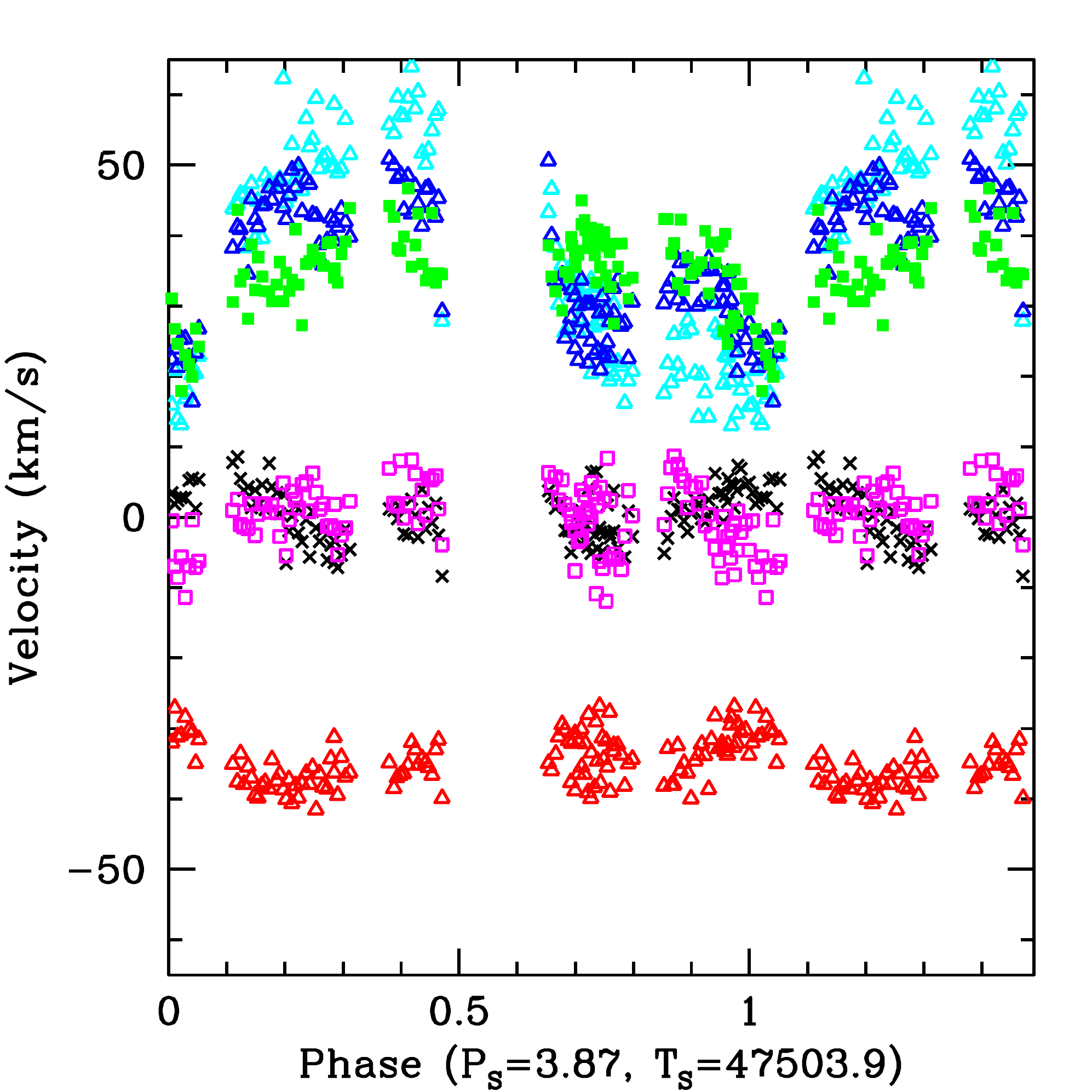}
\caption{Velocities of lines in IUE spectra folded in phase with (P$_S$,T$_S$) as listed in Table \ref{table_periods}. {\bf First row:} Set 1995; {\bf Second row:} Set 1992; {\bf Third row:} Set 1988. The spectral features that are plotted are labeled in the first row according to the naming convention given in Table A.1 (column 3).  They correspond to the following:
{\bf Column 1:} 1270e (black),HeIIr (magenta), FeVpsudo (cyan);
{\bf Column 2:}  HeIIa (red), NIVa (blue);
{\bf Column 3:} 1226n (magenta), 1371e (cyan), HeIIb (red);
{\bf Column 4:} NIVe (blue), NVe (green);
{\bf Column 5:} NIVnn (black), NIVn (magenta), CIVe+35km/s (cyan), HeIIe +35km/s (blue), NIV]+35km/s (green), 1270n-35km/s (red)
   }
\label{vels_group1}%
\end{figure*}

\begin{figure*}
\centering
\includegraphics[width=0.32\linewidth]{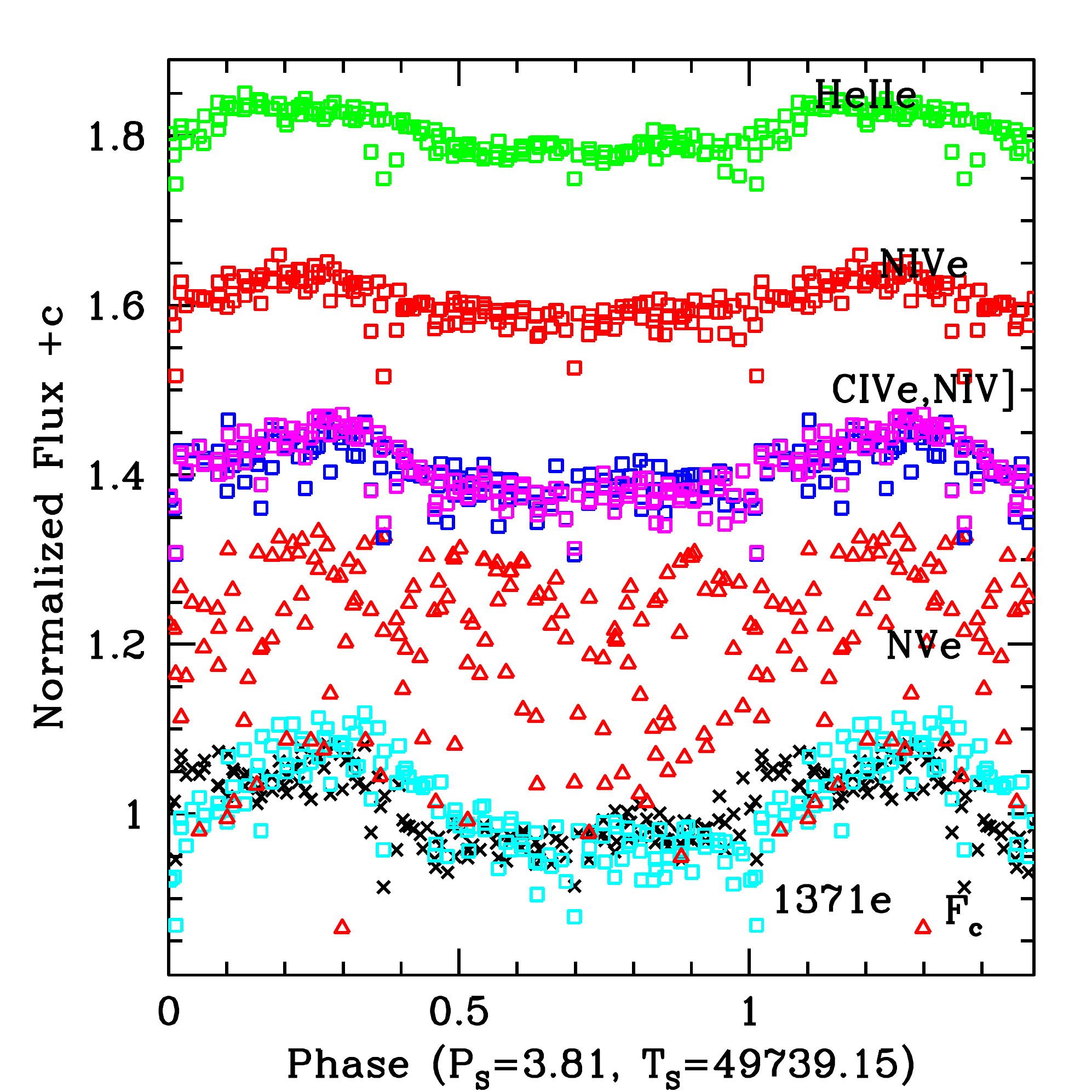}
\includegraphics[width=0.32\linewidth]{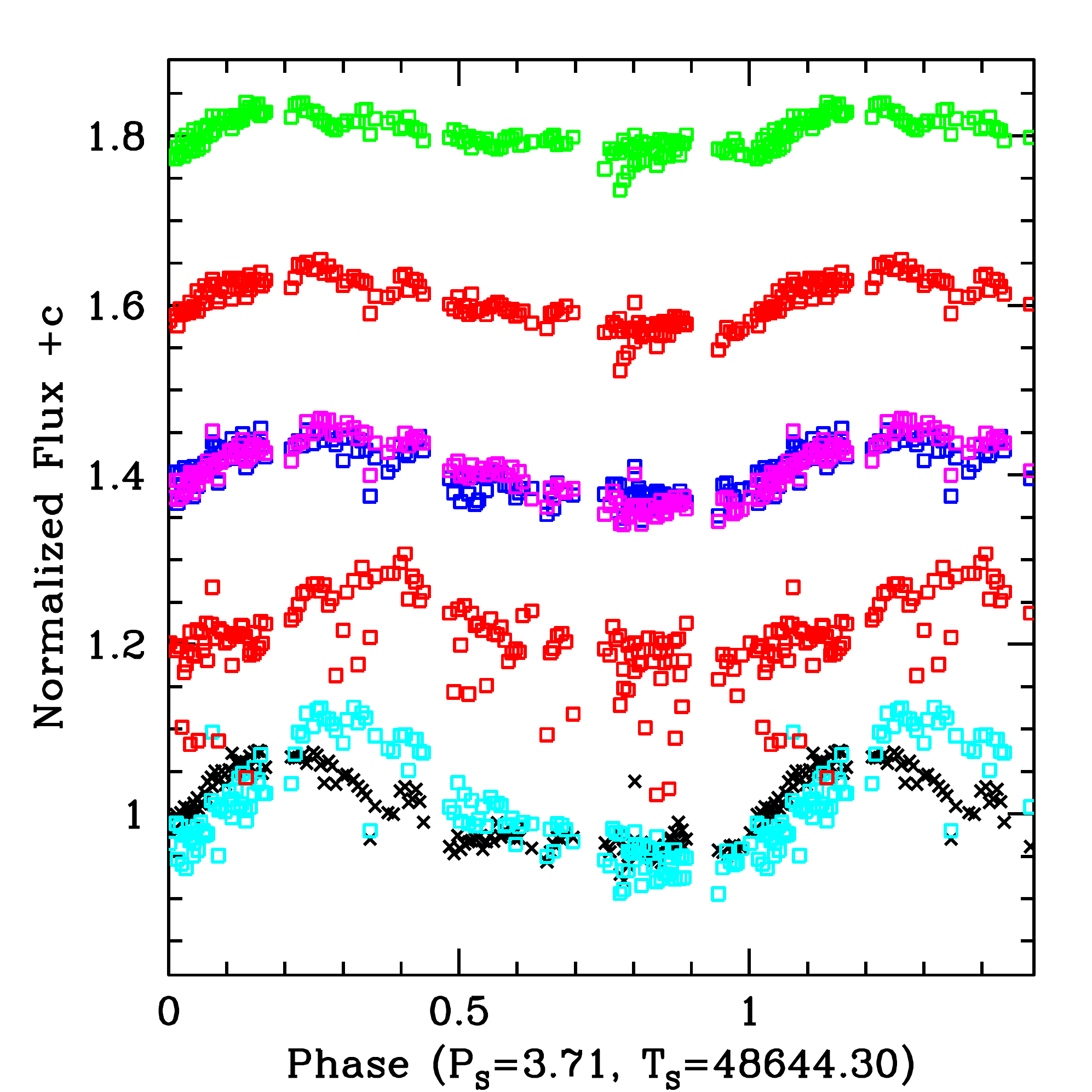}
\includegraphics[width=0.32\linewidth]{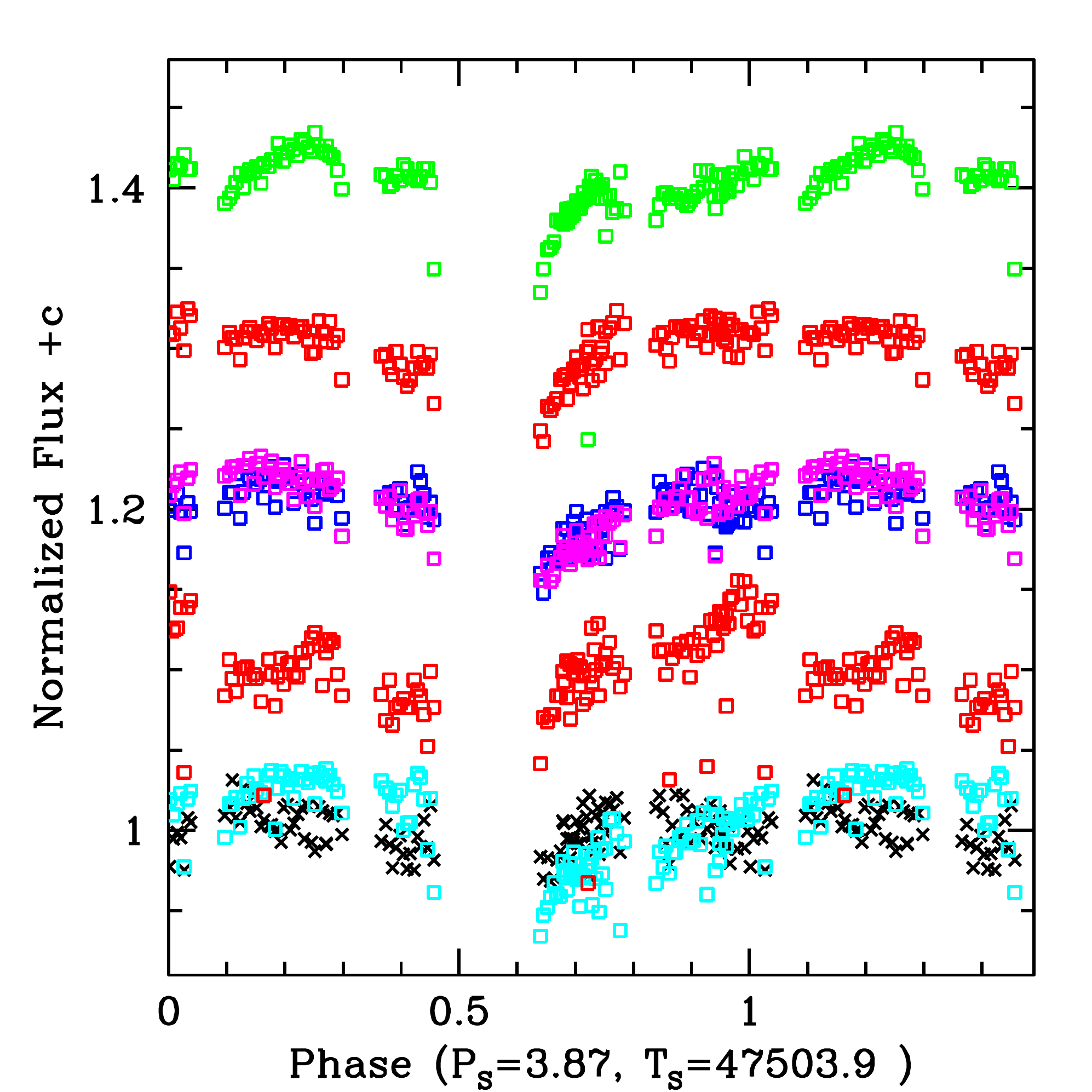}
\caption{UV fluxes as a function of orbital phase for the epochs 1995 (left), 1992 (middle) and 1988 (right).  Each line flux is normalized by the average value of the line for the given epoch.  
The significantly weaker modulation in 1988 compared to 1992 and 1995 is evident, as is the large cycle-to-cycle dispersion over time in the \ion{N}{V} emission flux of 1995.}
               \label{fluxes_phases}%
\end{figure*}

 \begin{figure}
 \centering
 \includegraphics[width=0.95\linewidth]{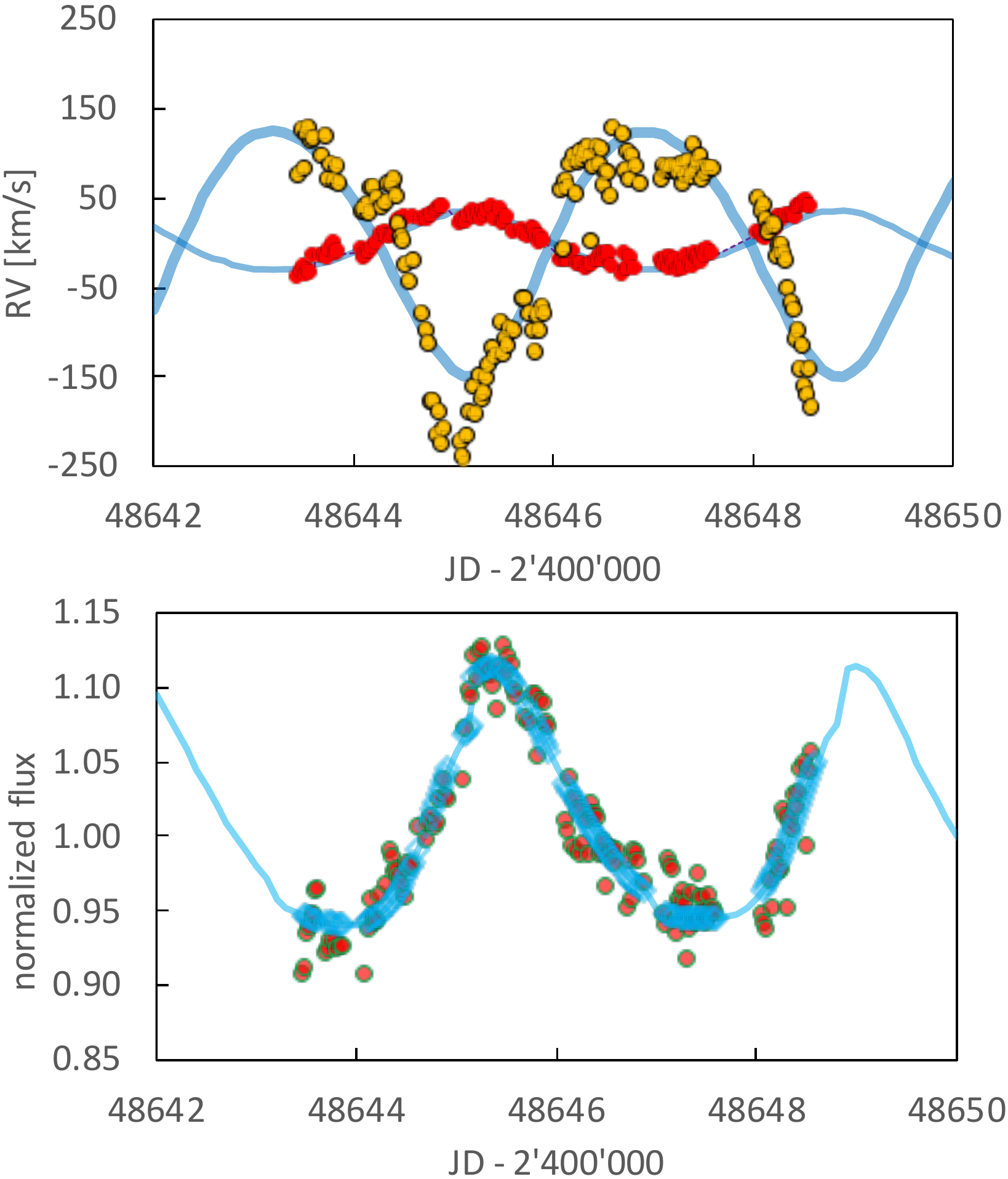}
\caption{Fit to the 1992  (red dots) and  
G1376 RVs (orange with black circle) (top panel) and the fluxes in the 1371e-band  (bottom panel).}
 \label{fits_1992}%
 \end{figure}

\begin{figure}
 \centering
 \includegraphics[width=0.9\linewidth]{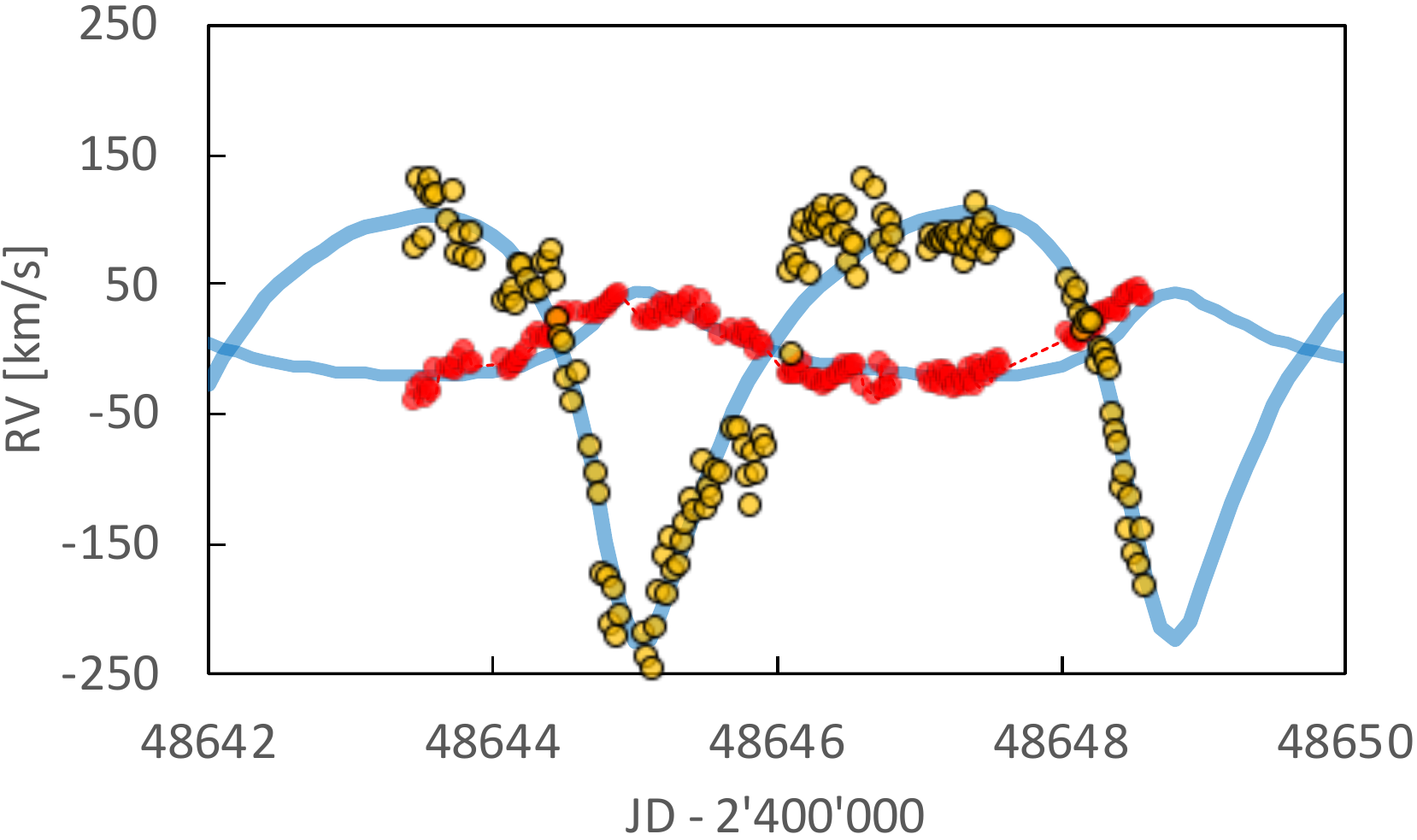}
\caption{Same as Fig. A.3, but here is the fit to the 1992  
G1376 RVs (orange with black circle) with an eccentricity $e=0.4$.}
 \label{fits_e0.4_1992}%
 \end{figure}
 
 \begin{figure}
 \centering
 \includegraphics[width=0.95\linewidth]{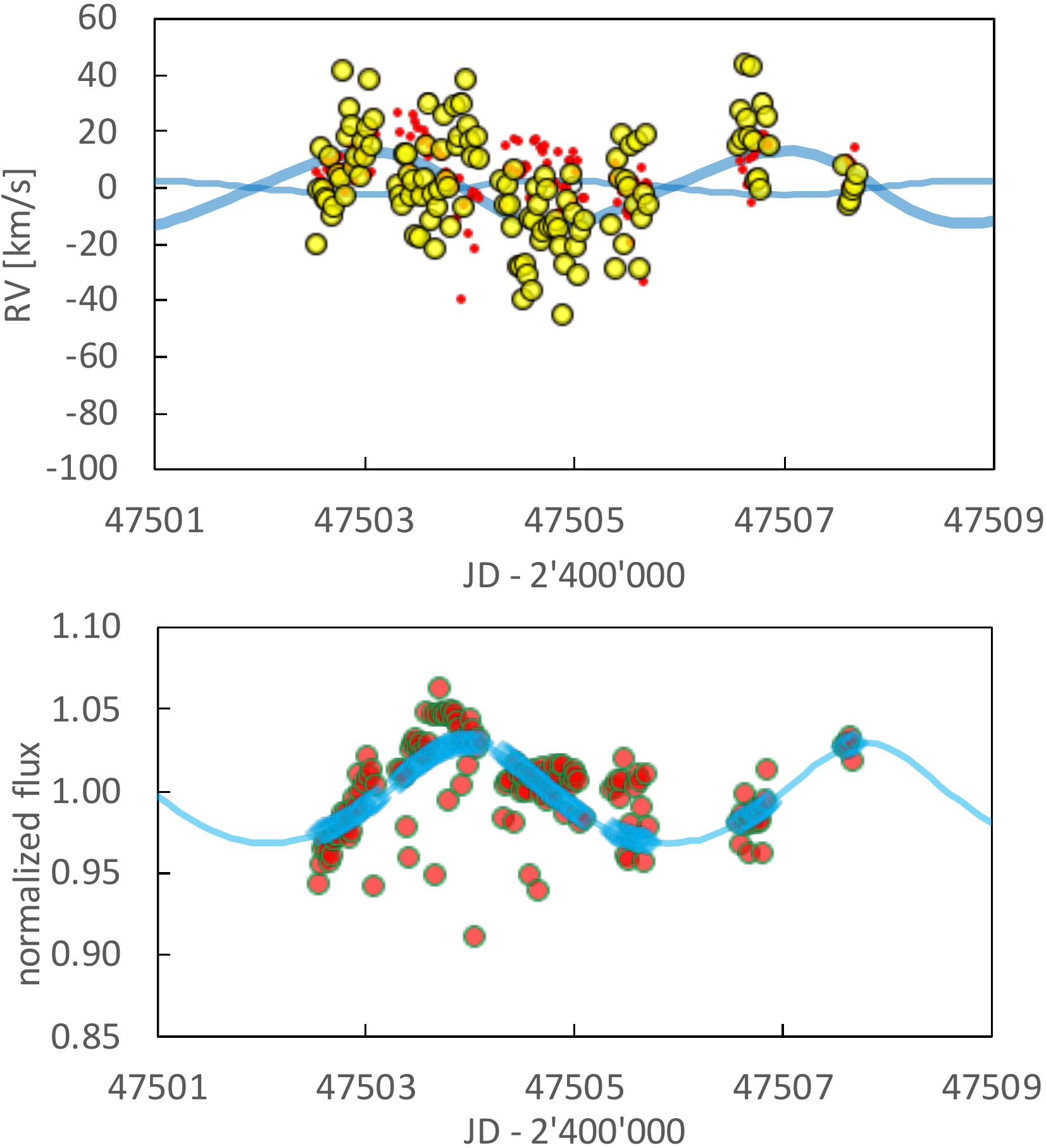}
 \caption{Fit to the 1988 IUE FeV+FeVI (red dots) and Gaussian-measured RVs (yellow with black circle) (top panel) and the fluxes in FeVIpsudo 1270-band and  (bottom panel).  
   }
 \label{fits_1988}%
 \end{figure}

\end{document}